# Beyond the Algorithm: Professional Experiences and Perceptions of AI Bias

Micarah Malone-Gawu

Submitted to the Faculty of the School of Computer and Information Sciences

in Partial Fulfillment of the

Requirements for the Degree of

Doctor of Philosophy Information Technology

University of the Cumberlands

**Approval for Recommendation**

This dissertation is approved for recommendation to the faculty and administration of the University of the Cumberlands.

Dissertation Chair:

______________________________

Alex Lazo, Ph.D.

Dissertation Evaluators:

______________________________

Alan Dennis, D.C.S.

______________________________

Mickey Hendon, Ph.D.

**Signatures are on file at the University.**

**Acknowledgments**

This dissertation marks the culmination of an incredible journey, and I am deeply grateful to all who walked beside me, guided me, and lifted me throughout this process.

First, I wish to express my deepest appreciation to my dissertation chair, Dr. Alex Lazo, for your steady guidance, thoughtful feedback, and belief in the importance of this research. Your support, patience, and encouragement were essential to the completion of this work. I am also profoundly grateful to my committee members, Dr. Alan Dennis and Dr. Mickey Hendon, for their time, expertise, and critical insights. Your perspectives challenged me to think deeply and elevated the quality of this dissertation.

To my husband, Evans, thank you for being my rock. Your love, encouragement, and quiet strength kept me grounded through every long night and challenging draft. I could not have done this without you.

To my family, friends, and colleagues, thank you for your support, prayers, and unwavering belief in me. A special thank you to those who listened, proofread, checked in, and reminded me of my purpose, your presence meant more than words can express.

Finally, this work is dedicated to all those working toward fairness, transparency, and equity in technology. May this dissertation be a small but meaningful contribution to building a more just and ethical future.

With heartfelt gratitude,

Micarah J. Malone-Gawu

## Abstract

The purpose of this qualitative multi-case study was to examine how social bias emerges, is perceived, and can be mitigated within artificial intelligence and machine learning systems by practitioners directly involved in their design, development, and governance. Although examples from healthcare, criminal justice, employment, and education were used to illustrate domains where automated systems shape everyday life, the study focused on the lived experiences and professional insights of AI practitioners rather than sector-specific populations. Guided by Intersectionality Theory and Cognitive Science, the study employed an interpretivist approach, utilizing semi-structured interviews with nine practitioners, supplemented by document analysis and triangulated case material to enrich contextual understanding. Findings showed that algorithmic bias arises from historical inequities, exclusionary design assumptions, and organizational pressures that prioritize speed and efficiency over ethical reflection. Participants emphasized that technical corrections alone cannot ensure fairness; instead, equitable AI requires structural accountability, diverse participation, and sustained cognitive awareness during the development lifecycle. Many described limited enforcement of ethical standards and organizational cultures that inconsistently support responsible practice. The study concludes that human-centered and socially grounded AI development depends on embedding ethics early in the design process, strengthening governance frameworks, and cultivating institutional environments that encourage reflective decision-making. These insights contribute to ongoing conversations on responsible AI and offer practical guidance for organizations seeking to design systems that are transparent, accountable, and aligned with the communities they affect.

**Table of Contents**

**List of Tables**

# Chapter One

# Introduction

## Overview

Artificial Intelligence (AI) and Machine Learning (ML) technologies are now deeply embedded in the daily lives of millions, influencing decision-making across critical sectors such as healthcare, education, criminal justice, and employment. While often framed as neutral tools that enhance efficiency and precision, AI systems are fundamentally sociotechnical constructs shaped by human choices, cultural assumptions, and structural power dynamics (Benjamin, 2019; Binns, 2018; Eubanks, 2018; Suchman, 2007). Consequently, these systems not only automate but also amplify long-standing patterns of social inequity and bias (Barocas et al., 2019).

Recent findings from the HAI AI Index Report (2025) reveal a 56% increase in AI-related incidents, emphasizing persistent gaps in fairness, transparency, and regulatory oversight. In addition, large language models (LLMs) exhibit emergent behaviors such as conformity and silence on bias-related topics, reinforcing dominant narratives under the guise of neutrality (Zhong et al., 2025). These patterns mirror the Spiral of Silence phenomenon, where agents align with perceived majorities, suppressing dissenting viewpoints even in the absence of emotional drivers (Communication Theory, n.d.).

This study critically investigates the societal consequences of algorithmic bias and explores strategies recommended by professionals across AI-related domains to mitigate harm and advance equitable system design. Rather than conducting a narrow, sector-specific analysis, the study draws upon high-impact domains as illustrative backdrops for broader cross-sectoral insights. These contexts highlight the tangible implications of AI bias while centering the perspectives of developers, data scientists, designers (Holstein et al., 2019; Raji et al., 2020).

What distinguishes this qualitative study is its interdisciplinary theoretical grounding. Intersectionality Theory serves as the central analytical lens, offering a framework to examine how overlapping social identities such as race, gender, class, and disability interact to produce distinct and unequal experiences with algorithmic systems (Bowleg, 2012; Crenshaw, 1991). This framework reveals how AI bias compounds systemic marginalization, particularly for individuals situated at the crossroads of multiple forms of oppression (Buolamwini & Gebru, 2018).

Complementing this lens, Cognitive Science is applied to explore how individuals perceive, interpret, and trust AI systems. This perspective elucidates the psychological mechanisms by which biased outputs are often accepted uncritically, especially when LLMs remain silent or present information in ways that appear objective (Buçinca et al., 2021; Choung et al., 2022). Findings from TUTORBench indicate that large language models rely entirely on statistical pattern recognition rather than adaptive explanation, emotional calibration, or student-level personalization, which are capabilities essential for the ethical and effective use of AI in education (Liang, Patel, Srinivasa, & Zhao, 2025).

The study employs a qualitative design drawn from semi-structured interviews with AI practitioners and subject matter experts who directly contribute to algorithmic decision-making. These interviews are triangulated with documentary analysis of public reports, policy briefs, institutional publications, and scholarly literature (Barabas et al., 2020; Whittaker et al., 2021). This approach enables a nuanced examination of both technical and sociocultural dimensions of bias, centering professional expertise and experiential insights.

This introductory chapter presents the study's background, significance, and guiding research questions. It also outlines the theoretical framework, highlights key assumptions and limitations, and defines essential terminology. By framing algorithmic bias as a multifaceted sociotechnical issue, this chapter establishes the foundation for a research inquiry that aims not only to document the problem but also to inform ethically grounded, interdisciplinary responses that promote understanding, trust, and community in AI system development (Binns, 2018; Raji & Buolamwini, 2019).

Chapter One introduced the study by outlining its background, purpose, significance, limitations, and assumptions. It also presented the research questions and defined key terms relevant to the investigation. The concepts introduced in this chapter were further expanded in the chapters that followed. Chapter Two provided a comprehensive review of existing literature related to algorithmic bias, fairness frameworks, and ethical AI governance. Chapter Three described the methodological framework and procedures used to conduct the research. Chapter Four presented the findings derived from the analysis of qualitative data in response to the study's research questions. Finally, Chapter Five discussed the findings in relation to existing literature, identified implications for theory and practice, and offered recommendations for future research.

**Background and Problem Statement**

It is not known how AI and machine learning practitioners experience, interpret, and navigate bias within the systems they design and deploy, nor how their ethical reasoning shapes the pursuit of fairness and accountability in real-world contexts. While prior research has explored algorithmic bias from technical and policy perspectives, the human and organizational dimensions of these challenges remain underexamined. Understanding practitioners lived experiences is critical to bridging the gap between theoretical principles of fairness and their practical application in design environments. This study addresses that gap by exploring how social hierarchies, institutional pressures, and cognitive processes intersect to influence ethical judgment and decision-making among AI professionals. Chapter Three, therefore, outlines the qualitative methodological approach used to investigate these experiences, detailing the research design, participant selection, data collection, and analytic strategies that guide this inquiry.

Artificial intelligence (AI) and machine learning (ML) systems are increasingly embedded within the infrastructures that shape daily decision-making in healthcare, education, employment, and criminal justice (Benjamin, 2019; Eubanks, 2018). Marketed as technologies that enhance efficiency, scalability, and objectivity, these systems often conceal the social and institutional forces that shape their design and deployment. Contemporary research demonstrates that AI and ML systems do not operate neutrally; they frequently reproduce and, in some cases, amplify existing social inequities. Racial bias in facial recognition technologies (Buolamwini & Gebru, 2018), inequitable diagnostic tools in healthcare (Obermeyer et al., 2019), and discriminatory hiring and risk assessment algorithms (Angwin et al., 2016; Raji et al., 2020) illustrate that algorithmic bias is not a rare technical malfunction, but a recurring sociotechnical pattern embedded within data, design, and institutional structures. Biased systems therefore do more than create isolated errors; they perpetuate systemic harm that shapes opportunity, economic stability, and public trust in technology.

Even as awareness of algorithmic bias increases, inequities continue to emerge in systems explicitly designed to promote fairness. Models trained on historical or incomplete data often replicate the inequalities of the environments from which they are derived (Crawford, 2021; Mitchell et al., 2022). For instance, an AI system designed to optimize healthcare resource allocation may disadvantage underrepresented patients if the training data reflects unequal access to care. Likewise, predictive policing tools trained on historical arrest data reproduce racialized surveillance patterns that stem from decades of disproportionate policing. These patterns of bias are rarely intentional; they reflect cognitive and structural blind spots within the design and deployment process that intertwine human reasoning, institutional norms, and broader social hierarchies.

Despite the proliferation of fairness metrics, auditing tools, and interpretability techniques, technical interventions alone cannot address the human and organizational dynamics that sustain bias (Holstein et al., 2019). Fairness in AI ultimately depends on human judgment on the decisions of those who define objectives, select datasets, and evaluate results. These decisions are shaped by competing demands such as efficiency and equity, and by cognitive heuristics that influence reasoning under pressure. For example, confirmation bias may lead a developer to favor models that reinforce prior assumptions, while institutional incentives may reward rapid deployment over ethical reflection. Understanding how practitioners experience and navigate these tensions is essential for identifying where fairness principles are weakened or sustained in practice.

In parallel with technical efforts, policymakers have begun introducing governance frameworks to guide responsible AI development. The European Union’s AI Act represents the first comprehensive regulatory framework establishing enforceable standards for transparency, accountability, and human oversight. It provides measurable guidance for organizations developing or deploying AI systems, setting a global benchmark for ethical governance. In contrast, the United States has not yet adopted a comparable federal mandate. The Blueprint for an AI Bill of Rights, released by the White House Office of Science and Technology Policy (OSTP, 2022) under the Biden Administration, articulates five guiding principles; safe and effective systems, protection from algorithmic discrimination, data privacy, notice and explanation, and human alternatives. However, the blueprint remains an advisory document rather than enforceable law.

Experts, including former U.S. Deputy Chief Technology Officer Nicole Wong, have noted that the absence of a binding national framework leaves companies largely responsible for self-regulation, resulting in inconsistent implementation and fragmented accountability structures (Massachusetts Daily Collegian, 2025). While the blueprint represents a step toward articulating public expectations for trustworthy AI, it lacks the enforcement mechanisms necessary to ensure compliance across industries. Practitioners working in this regulatory vacuum described uncertainty about how to interpret ethical expectations and align them with organizational goals. Without a U.S. equivalent to the EU AI Act, ethical commitments remain largely voluntary, highlighting the need for policies that translate high-level principles into enforceable standards and practical accountability.

Addressing these challenges requires understanding not only the structural forces that perpetuate bias but also the cognitive processes through which practitioners perceive and act upon ethical dilemmas. Intersectionality Theory and Cognitive Science together provide a framework for analyzing how social hierarchies and cognitive patterns shape ethical reasoning and decision-making in AI development. Intersectionality highlights how systems of power and identity, such as race, gender, class, and ability, intersect to structure both human experience and institutional behavior. Cognitive Science complements this by examining how mental models, heuristics, and reasoning processes influence perception and judgment. When applied together, these frameworks illuminate how social inequities are encoded through the cognitive and organizational routines of those designing AI systems.

Despite the growing scholarly attention to responsible AI, there remains limited empirical research that centers the lived experiences of practitioners who develop and maintain these systems. Much of the existing literature assumes that awareness of fairness principles naturally translates into ethical practice. However, evidence suggests that awareness alone is insufficient without institutional support, clear policy guidance, and mechanisms for accountability (Morse et al., 2022). Practitioners frequently report uncertainty about how to apply abstract ethical principles to complex, time-sensitive projects and frustration with the lack of standardized oversight (Pant et al., 2024).

The problem addressed in this study is that current discourse on algorithmic bias rarely includes the perspectives of the professionals most directly responsible for AI and ML system design and implementation. Existing research has largely focused on technical or policy remedies but has not adequately captured how practitioners perceive, experience, and respond to issues of fairness and bias in their everyday work. Consequently, the human and organizational factors that shape how bias is identified, rationalized, or mitigated remain underexplored.

This study seeks to address that gap by examining the lived experiences, ethical reasoning, and decision-making processes of practitioners who work with AI and ML systems. Informed by Intersectionality Theory and Cognitive Science, I examine how social hierarchies, institutional norms, and cognitive processes shape practitioners' behaviors and ethical decision-making. By centering practitioner voices, this study aims to generate empirically anchored insights that inform the development of equitable, accountable, and socially responsible AI systems.

## Purpose of the Study

The purpose of this qualitative, multi-case study was to examine how practitioners who build, manage, and regulate artificial intelligence (AI) systems understand the origins of algorithmic bias, interpret its societal impacts, and identify strategies for advancing equity in AI design and deployment. Guided by Intersectionality Theory and Cognitive Science, the study examined how human cognition, cultural context, and structural power dynamics influence the development and application of AI in high-stakes domains. This work addressed the problem that AI systems, often treated as neutral tools, reproduce existing inequities through biased data infrastructures, cognitive shortcuts, and institutional priorities that favor efficiency over fairness (Crawford, 2021).

The research questions aligned directly with this purpose by examining the emergence of bias, its societal implications, and practitioner-recommended mitigation strategies across healthcare, education, criminal justice, and employment. Together, these questions generated a comprehensive understanding of AI bias as both a technical and sociocultural phenomenon. The study contributed to the goals of the doctoral program by advancing theoretical and applied knowledge of responsible AI and offering an applied framework that connects cognition, conscience, and context to equitable system design.

**Significance of the Study**

This study aims to complement existing policy discussions on AI ethics by grounding its analysis in the lived experiences of those closest to technological development. Although many organizations have established internal fairness and accountability frameworks, practitioners described these initiatives as fragmented, unevenly applied, and frequently disconnected from the practical realities of design and implementation. In the absence of a governmental or regulatory mandate, ethical principles remain largely voluntary and subject to interpretation within each organization's culture, resources, and priorities. Participants emphasized that this absence of formal oversight creates a patchwork approach to ethics, where individual teams or companies determine their own definitions of fairness, transparency, and accountability.

Several practitioners expressed that standardized oversight or global benchmarks, such as those proposed under the European Union's AI Act, would offer the consistency and direction currently lacking across the field. Many viewed policy intervention not as a limitation but as a safeguard capable of aligning technical practice with social responsibility. Without such guidance, practitioners described facing competing pressures between ethical ideals and business imperatives, often having to justify fairness considerations in environments driven by deadlines, budgets, and performance metrics. These tensions revealed that initial intentions to design responsibly become diluted over time due to ambiguous policies and rapid development cycles.

Within this context, practitioners must interpret abstract ethical principles while managing unclear requirements and evolving project goals. Their decisions are shaped as much by cognitive shortcuts and organizational norms as by formal ethical reasoning, reflecting the intersection of social hierarchies, institutional culture, and human cognition central to this study's theoretical framing. By examining how fairness is understood and enacted within real-world workflows, this research offers an empirically grounded account of how bias emerges through technical choices, data assumptions, and institutional structures. In doing so, it bridges the gap between ethical aspiration and operational reality, demonstrating that sustainable equality in AI requires both practitioner agency and regulatory coherence.

## Research Questions

This qualitative study explored the experiences of professionals who design, develop, and implement artificial intelligence (AI) and machine learning (ML) systems and was guided by the following research questions:

RQ1: How do biases emerge in AI/ML systems, and what are their root causes?

RQ2: What are the societal impacts of AI bias across healthcare, education, criminal justice, and employment?

RQ3: What strategies do professionals recommend for mitigating bias and advancing equity in AI system design and use?

**Theoretical Framework**

This study was guided by an integrated theoretical framework combining Intersectionality Theory and Cognitive Science. These theories provided the conceptual foundation for examining how algorithmic bias develops and operates within AI systems. Intersectionality Theory informed the analysis of structural and institutional dimensions of bias, while Cognitive Science informed the analysis of the cognitive mechanisms involved in AI design and evaluation. This combined framework supported the study's qualitative design by offering two complementary perspectives on the origins and expressions of bias in sociotechnical systems.

Intersectionality Theory originates from critical legal studies and Black feminist scholarship, examining how overlapping systems of power, such as race, gender, class, and ability, shape social outcomes (Bilge & Collins, 2016; Crenshaw, 1989). Within the context of AI, this theory provides a lens for understanding how datasets, modeling practices, and deployment environments reflect historical and institutional patterns of inequality. The theory, therefore, provides a framework for understanding how algorithmic outputs may originate from or reinforce broader social hierarchies.

Cognitive Science examines mental processes, including attention, memory, perception,

reasoning, and decision-making. In this study, the framework was used to understand how these cognitive processes shape the development and deployment of AI systems. Research in cognitive psychology and human–computer interaction highlights phenomena such as automation bias, trust calibration, and confirmation bias, which influence how practitioners interpret and respond to algorithmic recommendations. Cognitive Science provides a basis for analyzing how heuristics and mental models affect technical decisions, evaluation practices, and interactions with automated systems.

Integration of the Frameworks allowed the study to examine both structural and cognitive contributors to algorithmic bias. Intersectionality Theory describes macro-level conditions that influence data infrastructures and institutional practices, while Cognitive Science describes micro-level cognitive processes that influence technical judgment and system interpretation. These theories supported the examination of how bias may emerge across multiple stages of the AI lifecycle, including data selection, model development, evaluation, and deployment. This dual framework guided the organization of the study's research questions, informed the development of the interview protocol, and provided the conceptual structure for analyzing practitioner narratives.

## Limitations

Like all research, this study has limitations that should be acknowledged. Recognizing these boundaries is essential for interpreting the findings appropriately and for identifying directions for future inquiry into the ethical design, development, and governance of artificial intelligence (AI) systems. The limitations arise from methodological, theoretical, and practical considerations related to the study's scope, data collection, and analytical focus.

A key limitation involves the use of purposive sampling and a relatively small participant pool. Participants were selected based on their professional experience with AI and machine learning (ML) systems rather than demographic or sectoral representativeness. This ensured inclusion of individuals with direct engagement in AI design, deployment, and governance while intentionally seeking demographic and cultural diversity. The resulting range of racial, gender, and cultural identities enriched the analysis by revealing how intersecting social positions shape ethical reasoning in AI practice. However, certain sectors. particularly those with more established commercial applications, may still be overrepresented. Professionals from technology and private-sector roles may have emphasized different ethical challenges than those working in public or social-impact contexts. Future research should expand participation across regions and industries to capture how organizational and governance structures interact with intersectional identities to shape decision-making.

Another limitation involves the interpretive nature of qualitative research. As an interpretivist study, the findings are shaped by the researcher's background, values, and prior experience with AI ethics and governance. The researcher's perspective inevitably influenced what was noticed, emphasized, and questioned during analysis. Reflexive practices such as memoing, bracketing, and peer debriefing were used to recognize and manage these influences, but complete neutrality is possible. The researcher's commitment to fairness and equity informed how data were interpreted and understood. Rather than viewing this as a source of bias, it was approached as a lens that shaped meaning through conscious reflection and transparency. Consistent with Denzin and Lincoln's (2011) concept of situated knowledge, the study acknowledges that understanding emerges from the interaction between participant experiences and the researcher's perspective, creating an interpretation that is both contextual and grounded in practice.

The use of virtual, semi-structured interviews introduces additional methodological constraints. Conducting sessions via digital platforms such as Zoom or Teams allowed broad geographic reach but influenced interactional depth. Differences in participants' comfort with technology, internet reliability, and privacy conditions may have shaped disclosure, particularly when discussing sensitive topics such as ethics or organizational culture. Social desirability bias also remains possible, as participants might have portrayed their practices or organizations in favorable ways. While rapport-building and confidentiality measures mitigated these effects, they may have influenced the openness and detail of some responses.

The study's temporal scope presents another limitation. The interviews capture practitioner perspectives at a specific point in time, reflecting the state of AI governance discourse and public awareness during the study period. As AI regulation and societal expectations evolve rapidly, perceptions of fairness and accountability may shift. Longitudinal or follow-up studies could assess how practitioner understanding and organizational responses change as responsible AI frameworks mature and become institutionalized.

The reliance on self-reported data is an additional constraint inherent in interview-based designs. While participants provided detailed accounts based on professional experience, these narratives represent perceptions rather than direct observation of development processes. Consequently, findings reflect how practitioners conceptualize and rationalize their work rather than how those practices unfold in real time. Combining interviews with ethnographic observation, document analysis, or technical audits in future studies would enhance triangulation and capture how ethical reasoning translates into practice.

The theoretical framework itself imposes interpretive boundaries. Integrating Intersectionality Theory and Cognitive Science foregrounds social hierarchy and human reasoning but necessarily narrows the analytic scope to these domains. Alternative frameworks such as Actor-Network Theory, Critical Data Studies, or Organizational Behavior could yield complementary insights. The intersectional-cognitive lens emphasizes relationships between identity, cognition, and institutional context but does not encompass the full complexity of AI ethics. Future research could expand this framework to include economic, political, or technological factors that shape how fairness is defined and operationalized.

Another limitation is the study's focus on practitioner perspectives rather than system-level audits. The research does not evaluate algorithmic performance directly but examines how individuals responsible for AI development perceive and address bias. While this focus provides crucial insight into human and organizational factors, it limits generalization about technical outcomes. Mixed methods approach combining qualitative interviews with quantitative fairness audits could yield more comprehensive understanding of where and how bias arises.

Geographic concentration presents a further boundary. Most participants were based in regions where AI governance infrastructures are well established, such as North America and Western Europe. Practitioners in the Global South or emerging AI economies may encounter distinct ethical and regulatory challenges influenced by local infrastructure, cultural norms, and policy environments. Expanding geographic scope would enable comparative analysis and deepen understanding of how local contexts mediate global AI ethics discourses.

Finally, as with most qualitative studies, transferability rather than generalizability defines the scope of application. The findings should be viewed as conceptually transferable to similar organizational settings that emphasize technical innovation while confronting emerging ethical mandates. The study's value lies in its depth of understanding and theoretical insight rather than statistical generalization.

In summary, these limitations delineate the interpretive boundaries of the research without diminishing its scholarly or practical contributions. The study offers an empirically grounded account of how AI practitioners experience and navigate ethical challenges. Acknowledging these constraints reinforces transparency and reflexivity, key principles of qualitative inquiry and provides a foundation for future studies to expand methodological scope, participant representation, and analytic frameworks in the pursuit of responsible AI.

**Assumptions**

This study is derived from several foundational assumptions that guide its design, purpose, and theoretical orientation. It assumes that artificial intelligence (AI) and machine learning (ML) systems are not neutral or objective artifacts but are embedded within the social, cultural, and institutional contexts in which they are created. Scholars emphasize that technological systems inevitably mirror the values, biases, and institutional priorities of their creators (Benjamin, 2019; Eubanks, 2018). This premise informs the study's focus on the human and organizational dimensions of AI, highlighting how decisions made during design, deployment, and evaluation influence the fairness, transparency, and ethical character of these systems.

A central assumption is that bias in AI systems does not arise solely from technical flaws such as incomplete datasets or suboptimal algorithms. Instead, bias reflects broader social and structural inequities replicated through data, design, and institutional processes. Research demonstrates that algorithmic bias frequently arises from historical exclusion, cultural assumptions, and power imbalances embedded within data infrastructures (Buolamwini &

Gebru, 2018; Noble, 2018). These findings suggest that technical interventions alone cannot fully resolve bias without addressing the deeper social forces that sustain inequity. Achieving fairness in AI, therefore, requires not only improved data quality or algorithmic design but also organizational and structural transformation in how responsibility, accountability, and inclusion are conceptualized.

The study assumes that AI systems operate within existing systems of inequality and can amplify, legitimize, or obscure them depending on how they are designed and governed. Social hierarchies and institutional practices shape every stage of AI development—from problem formulation and data collection to deployment and evaluation. Intersectionality Theory provides a conceptual foundation for understanding how overlapping forms of oppression such as racism, sexism, ableism, and classism intersect to produce compounded disadvantage (Bilge & Collins, 2016; Crenshaw, 1991). Algorithmic bias thus disproportionately affects individuals and communities at these intersections, often in ways that remain invisible to developers. Grounding analysis in intersectionality shows that bias in AI is not an abstract technical issue but a lived social reality with uneven consequences.

Another key assumption is that practitioners who design, build, and deploy AI systems hold essential knowledge for understanding how bias is produced and mitigated in practice. Research indicates that developers, data scientists, and project managers operate within ecosystems where ethical reasoning, organizational constraints, and technical problem-solving continually intersect (Barocas & Passi, 2019; Holstein et al., 2019). Their experiences reveal how fairness principles are interpreted, negotiated, or diluted in everyday practice. This research assumes that practitioner perspectives are indispensable for understanding how ethics is enacted in real organizational settings.

The study also assumes that participants will engage thoughtfully and candidly during interviews. Because discussions of bias and ethics can involve sensitive reflections on personal or organizational practices, guarded responses are possible. Nonetheless, qualitative inquiry recognizes that even moderated or coded language can reveal valuable insights into institutional norms and professional culture (Denzin & Lincoln, 2011). Through confidentiality and rapport-building, participants are expected to provide reflective accounts that illuminate their reasoning and strategies for addressing bias, even when expressed diplomatically.

A methodological assumption is that qualitative interviews are well-suited to exploring how practitioners interpret and respond to complex ethical issues. Qualitative approaches allow for deep contextual understanding and exploration of meaning-making processes that quantitative methods cannot capture (Creswell & Poth, 2018). Meaning related to fairness, bias, and accountability emerges through dialogue and situated experience. Practitioners' reflections provide insight into how ethical reasoning unfolds within the constraints of time, resources, and institutional priorities. Semi-structured interviews, therefore, serve as an effective means for uncovering the cognitive, emotional, and organizational dynamics that shape AI development.

The study further assumes that bias operates simultaneously at structural and cognitive levels and that addressing it requires attention to both. Intersectionality Theory frames bias as an outcome of social hierarchy and institutional power, while Cognitive Science explains how mental models, heuristics, and cognitive shortcuts influence human judgment in technical contexts (Buçinca et al., 2021; Choung et al., 2022). Together, these perspectives acknowledge that even well-intentioned practitioners work within perceptual and organizational constraints that can perpetuate inequity. Understanding bias requires examining how cognition and structure interact in practice.

Another assumption is that technical systems reflect organizational culture and values. Decisions about data selection, model optimization, and performance metrics are influenced not only by engineering considerations but also by business priorities, regulatory pressures, and workplace norms (Green, 2021; Selbst et al., 2019). Ethical outcomes in AI are therefore inseparable from the institutional context. This assumption reinforces the study's focus on organizations as critical sites for understanding how fairness and accountability are defined and operationalized.

The study also assumes that practitioners' cognitive and ethical awareness can be strengthened through interdisciplinary frameworks and reflexive practices. Cognitive Science suggests that individuals can mitigate bias through feedback, reflection, and ethical reasoning tools, while Intersectionality Theory emphasizes inclusion of marginalized perspectives in design and governance. Combining these approaches offers pathways for cultivating self-awareness and institutional learning within AI development teams.

From an epistemological standpoint, the study assumes that knowledge about AI ethics is constructed through interaction and interpretation rather than objective measurement. Consistent with interpretivist traditions (Denzin & Lincoln, 2011), meaning emerges from the co-construction of understanding between researcher and participant. The findings, therefore, represent an interpretive synthesis shaped by both participant narratives and theoretical framing, rather than universal truths.

Finally, the study assumes that the dual-theoretical framework provides a robust and complementary foundation for analysis. Intersectionality explains how social identities shape experiences of bias, while Cognitive Science elucidates how individuals perceive, interpret, and act upon algorithmic information. Together, these frameworks support a comprehensive understanding of bias as simultaneously structural and cognitive. Ethical AI requires not only technical solutions but also cultural and organizational transformation, as scholars have demonstrated (Raji et al., 2020; Whittaker et al., 2021). The integration of theory and practitioner insight provides a foundation for generating actionable strategies that bridge ethical ideals and professional practice.

Taken together, these assumptions establish the foundation for a study that is critically engaged, theoretically grounded, and methodologically rigorous. They clarify that this research is not premised on the neutrality of technology but on the recognition that all systems are human systems shaped by the values and limitations of those who design them. These assumptions also define the study's interpretive boundaries, which are revisited in Chapter Five when discussing implications, contributions, and directions for future research on equitable and responsible AI.

**Definitions**

The operational definitions of the key terms used in this study are as follows:

*Algorithmic Bias*: Systematic and measurable inequities produced by algorithmic systems when historical, structural, or cognitive biases become encoded in data, model assumptions, or deployment contexts (Barocas & Selbst, 2016; Obermeyer et al., 2019).

*Algorithmic Governance*: Institutional and regulatory mechanisms—including audits, transparency requirements, and oversight structures—that govern how AI systems are built, deployed, and held accountable (Batool et al., 2024; Novelli et al., 2024).

*Automation Bias*: A tendency for humans to over trust or defer to algorithmic recommendations, even in the presence of contradictory contextual information (Mosier & Skitka, 2018).

*Bias Variance Tradeoff*: A modeling principle describing the tension between underfitting (high bias) and overfitting (high variance) with implications for fairness across demographic groups (Brownlee, 2016; Buijsman, 2024).

*Cognitive Bias*: Predictable patterns of human judgment influenced by heuristics, mental shortcuts, or perceptual errors that affect system design, problem framing, and interpretation of model output (Buçinca et al., 2021; Kahneman, 2011).

*Data Feminism*: An approach that examines how power and inequality shape data practices, arguing for positionality, reflexivity, and inclusion in the creation and interpretation of datasets (D'Ignazio & Klein, 2020).

*Data Representativeness*: The extent to which datasets adequately reflect demographic, geographic, and phenotypic diversity in real-world populations (Adamson & Smith, 2018; Moons et al., 2025).

*Epistemic Injustice*: A condition in which marginalized communities are deprived of credibility, recognition, or representation within systems of knowledge production such as datasets and machine learning pipelines (Fricker, 2007; Noble, 2018).

*Explainability*: The ability to interpret, understand, or reconstruct an AI model's reasoning, data pathways, or decision logic (Lipton, 2016; Mitchell et al., 2019).

*Fairness Metrics*: Quantitative measures used to evaluate the equitability of model performance, including Equality of Opportunity, Average Odds Difference, and Demographic Parity (Almasoud & Idowu, 2025).

*Human-Centered AI*: A design philosophy emphasizing that algorithms should augment, support, and remain subordinate to human judgment, reasoning, and ethical agency (Amershi et al., 2019).

*Intersectionality*: A theoretical framework describing how overlapping identities such as race, gender, class, and geography produce compounded and structurally patterned forms of disadvantage (Collins, 2019; Crenshaw, 1991).

*Practitioner-Centered Ethics*: An approach that prioritizes the lived experience, constraints, and decision-making perspectives of AI developers, highlighting the gap between ethical principles and day-to-day implementation (Herzog & Blank, 2024; Pant et al., 2024).

*Predictive Policing*: The use of algorithmic tools to forecast crime or criminal behavior based on historical policing data that often reflects racially biased enforcement patterns (Almasoud & Idowu, 2025; Angwin et al., 2016).

*Proxy Variables*: Indirect indicators used in modeling that stand in for complex constructs, often encoding structural inequalities when proxies (such as cost or test scores) reflect access rather than need or ability (Lau, 2025; Obermeyer et al., 2019).

*Risk Assessment Algorithms*: Tools that estimate the likelihood of future events such as recidivism or health deterioration, often criticized for reinforcing disparities within criminal justice and healthcare systems (Angwin et al., 2016; Obermeyer et al., 2019).

*Sociotechnical System*: A system in which social, institutional, and technical components interact to jointly shape outcomes, revealing that technology is inseparable from human and organizational context (Winner, 1986).

*Structural Inequality*: Systemic and historically rooted disparities embedded in institutional practices, policies, and data infrastructures that become reproduced in algorithmic systems (Bridges, 2024; Noble, 2018).

*Transparency*: Organizational and technical practices that make system design, data sources, limitations, and model behavior visible for auditing and oversight (Cheong, 2024; O’Neil, 2016).

*Value-Sensitive Design*: A design framework that embeds moral and social values into engineering processes through stakeholder engagement and anticipatory ethical reflection (D'Ignazio & Klein, 2020).

**Summary**

Chapter One establishes the foundation for this dissertation by framing artificial intelligence (AI) and machine learning (ML) systems as sociotechnical constructs shaped by human judgment, institutional norms, and cultural values. Although often presented as neutral and objective, these systems are embedded within existing power structures that influence how they are designed and applied. When deployed in high-stakes domains such as healthcare, education, employment, and criminal justice, AI systems frequently reproduce and amplify social inequalities rather than mitigate them.

The chapter emphasizes that algorithmic bias is not a technical error, but a structural and systemic issue rooted in historical and organizational inequities. Bias arises from the cultural and institutional contexts that shape data collection, modeling decisions, and evaluation criteria. Addressing these challenges requires not only technical improvement but also critical attention to the social, cognitive, and ethical dimensions of AI development.

This study responds to that need by centering the experiences of AI practitioners such as developers, engineers, data scientists, and project managers whose decisions influence how fairness, accountability, and transparency are enacted in practice. Guided by three research questions, the study explores how practitioners conceptualize the origins and impacts of bias and what strategies they identify for advancing equity in AI systems. The research is framed through a dual-theoretical lens combining Intersectionality Theory, which examines how social hierarchies shape systemic inequity, and Cognitive Science, which investigates how human reasoning and heuristics influence ethical decision-making.

Methodologically, the study employs a qualitative, interpretivist approach using semi-structured interviews to capture practitioners lived experiences and meaning-making processes. This design allows for nuanced exploration of how ethical reasoning unfolds amid competing demands for efficiency, innovation, and compliance. The study assumes that bias is both structural and cognitive, that practitioners offer critical insights into its causes and mitigation, and that knowledge about ethics is co-constructed through interpretation and dialogue.

The significance of this research lies in its potential to bridge theory and practice in AI ethics. By examining how practitioners perceive and navigate ethical challenges, it contributes to a growing movement toward actionable fairness and accountability in AI governance. The findings aim to inform future approaches to ethics training, policy design, and organizational learning while advancing a more holistic understanding of algorithmic bias.

In summary, Chapter One defines the problem, purpose, significance, and theoretical grounding of the study, establishing that achieving fairness in AI requires both technical and cultural transformation. It positions ethical AI development as a human-centered endeavor that depends on the awareness and responsibility of those who create these systems. Chapter Two builds upon this foundation by reviewing the existing literature on algorithmic bias, fairness, and governance, situating this study within ongoing scholarly and practical debates about responsible AI.

# Chapter Two

## Review of Literature

### Introduction

This chapter examines how practitioners who work directly with artificial intelligence (AI) systems perceive bias, its origins, societal consequences, and the effectiveness of mitigation strategies. As AI technologies increasingly shape decision-making in sectors such as healthcare, criminal justice, education, and employment, concerns about fairness, accountability, and transparency have become more urgent. While much of the existing scholarship emphasizes technical solutions and statistical fairness metrics, this study centers on the perspectives of those closest to the data, design, and deployment processes. Their insights are essential for understanding how bias is recognized, interpreted, and addressed in real-world development environments.

The literature review was developed through a purposive and iterative process using academic databases and publicly available scholarly platforms. A comprehensive search of peer-reviewed and open-access sources was conducted through IEEE Xplore, Google Scholar, arXiv, Stanford Human-Centered AI Research publications, and the Journal of Artificial Intelligence Research (JAIR) to develop the foundation for this literature review. The search strategy focused on literature published within the past ten years (2015–2025) to ensure inclusion of both seminal and contemporary research on bias and fairness in artificial intelligence. Combinations of the following keywords and search phrases were used: “AI bias,” “algorithmic fairness,” “responsible AI,” “intersectionality in AI,” “machine learning ethics,” “AI governance,” and "cognitive bias in AI systems." Inclusion criteria required that sources directly examine bias, fairness, or ethical considerations in AI design, implementation, or governance, while exclusion criteria omitted non-scholarly commentaries and studies lacking methodological transparency. This search approach ensured that the selected literature reflects the theoretical foundations, technological advances, and applied strategies relevant to understanding and mitigating AI bias across high-stakes domains such as healthcare, education, and employment. As the review progressed, additional sources were identified through reference tracking and thematic exploration, allowing the review to remain responsive to emerging concepts while maintaining alignment with the study's purpose.

To establish the study's conceptual foundation, this chapter is organized into six sections. It begins with an overview of the theoretical frameworks, Intersectionality Theory and Cognitive Science, that inform the research design. The review then examines how structural inequities, and cognitive mechanisms collectively shape AI bias across multiple domains, supported by case studies illustrating how bias manifests in practice. Subsequent sections explore the historical development of algorithmic bias, technical sources of inequity in data and modeling, philosophical critiques of general AI, and ethical and governance frameworks. The chapter concludes with a review of practitioner-centered research and an analysis of gaps in the literature that this study seeks to address. Together, these discussions ground the study's analysis in a dual theoretical perspective that views AI bias as the product of both structural inequality and human cognition.

**Case Studies of Bias in AI Systems**

Bias in artificial intelligence (AI) systems is not a theoretical abstraction but a lived reality with tangible consequences across multiple sectors. From healthcare and criminal justice to employment and education, AI technologies have been introduced with the promise of efficiency, objectivity, and scalability. Yet, in practice, these systems often replicate and intensify existing social inequalities. This section presents a series of in-depth case studies that illustrate how bias manifests in AI systems, drawing on empirical evidence, practitioner insights, and interdisciplinary scholarship. Each case is examined through the dual lenses of intersectionality and cognitive science, which together provide a framework for understanding the structural and psychological dimensions of algorithmic harm.

Building on the dual theoretical framework of intersectionality and cognitive science, this section applies these perspectives to real-world contexts where artificial intelligence (AI) systems exert significant social and ethical influence. Case studies from healthcare, criminal justice, employment, and education illustrate how structural inequities and human cognitive processes intersect to shape algorithmic outcomes. Through these examples, the analysis demonstrates that bias is both socially embedded and cognitively enacted, showing that fairness challenges in AI development cannot be separated from the historical, institutional, and perceptual forces that sustain them. By grounding theory in empirical reality, the case studies highlight how practitioners' design decisions, organizational constraints, and mental models contribute to the persistence of algorithmic harm. Collectively, they establish the practical and ethical foundation for this study's qualitative exploration of how professionals understand and respond to bias in AI systems.

**Health**

AI systems in healthcare have been widely celebrated for their potential to revolutionize diagnosis, treatment planning, and patient monitoring. Proponents argue that these technologies can reduce human error, improve diagnostic accuracy, and expand access to medical care. However, emerging evidence demonstrates that these benefits are not distributed equally. The same algorithms that promise efficiency and precision often reproduce and amplify existing racial, gender, and socioeconomic disparities in health outcomes. The tension between innovation and inequity reaffirms a critical reality: when healthcare data reflect structural inequalities, AI systems inevitably learn and replicate them.

A foundational study examined a widely used healthcare risk prediction algorithm that identified patients for extra care management based on projected healthcare costs, illustrating this dilemma (Obermeyer et al., 2019). The system's developers assumed that higher spending equated to greater medical need. However, because Black patients historically receive fewer medical services and lower overall healthcare spending due to systemic barriers, the algorithm consistently underestimated their health risks. As a result, Black patients were less likely to be flagged for additional care compared to White patients with the same chronic conditions. The flaw was not a computational error but a structural one that reflected a deeply embedded assumption equating financial expenditure with access to quality care. Such algorithms institutionalize the racialized economics of health by translating social inequities into quantifiable data points that ultimately shape clinical decision-making (Bridges, K. W., Ekezie, W., & Kadan, M., 2024). This misalignment between algorithmic logic and human need reveals a broader problem in healthcare AI development. Developers often rely on proxies such as cost, readmission rates, or hospitalization frequency to approximate patient need. These proxies fail to account for the racialized and class-based disparities that shape access to healthcare in the first place. Such approaches often prioritize efficiency and scalability over equity, embedding inequality into digital infrastructure under the guise of neutrality (Lau, 2025). The BMJ PROBAST+AI framework (Moons et al., 2025) confirms that bias is prevalent across medical AI models due to poor data representativeness, lack of external validation, and the exclusion of social context in model design. In their review of AI-driven risk models, the authors found that nearly eighty percent of published tools were developed on homogeneous populations that failed to reflect demographic diversity, with only a small percentage including external validation across racial or geographic subgroups.

The problem of underrepresentation is particularly pronounced in dermatology. Artificial intelligence models for skin disease diagnosis have consistently shown lower accuracy for patients with darker skin tones because most training datasets are dominated by images of light-skinned individuals (Adamson & Smith, 2018). This imbalance is particularly concerning, given the global diversity of skin types and the clinical significance of early detection for conditions such as melanoma. When algorithms are trained primarily on lighter skin, they struggle to recognize subtle visual cues on darker skin, leading to delayed diagnoses and poorer outcomes for these populations. Recent reviews of dermatology AI systems have highlighted that publicly available image repositories often lack sufficient representation of Fitzpatrick skin types IV–VI, which correspond to medium to dark skin tones. This gap perpetuates systemic inequities, as patients from underrepresented groups are more likely to receive inaccurate assessments or require additional diagnostic procedures. Addressing these disparities requires deliberate efforts to curate inclusive datasets, implement bias detection protocols, and validate models across diverse populations. Such challenges illustrate how medical datasets frequently replicate existing patterns of exclusion in healthcare systems, embedding differences in access, utilization, and outcomes into the foundations of AI models.

From an intersectional perspective, these findings demonstrate how historical and structural inequalities in healthcare, ranging from insurance discrimination to geographic resource disparities, become digitally embedded in algorithmic systems. The data used to train medical AI are not abstract or neutral; they are artifacts of social hierarchies that determine whose illnesses are studied, whose pain is believed, and whose bodies are documented. Intersectionality exposes how overlapping dimensions of race, gender, and class produce compounded forms of invisibility within these datasets. For instance, low-income women of color are statistically less likely to appear in large-scale genomic or imaging datasets, making their experiences less legible to AI-driven diagnostic tools. This invisibility perpetuates cycles of underdiagnosis and delayed treatment, reinforcing structural patterns of exclusion under a technological veneer.

Cognitive science offers a complementary lens by highlighting the psychological and organizational mechanisms perpetuating these patterns. Developers and healthcare institutions often rely on cognitive shortcuts such as anchoring, where initial design assumptions, for example, using cost as a proxy for need, constrain later decision-making. They may also exhibit confirmation bias, seeking validation that their models align with established performance metrics rather than questioning the equity of those benchmarks. These biases are reinforced by institutional pressures that prioritize technical efficiency and regulatory approval over social accountability. As noted in the BMJ PROBAST+AI report (Moons et al., 2025), these cognitive and institutional tendencies create an ecosystem in which validity is defined by internal performance, not external fairness.

Intersectionality and cognitive science reveal that algorithmic bias in healthcare is not simply a byproduct of flawed data or coding errors. It emerges from the interaction of structural inequality and cognitive simplification, the human tendency to reduce complex social realities into measurable variables. When developers operationalize healthcare needs through cost or utilization, they unknowingly transform structural injustice into a technical problem. The resulting errors are not accidental but systemic, reflecting how inequality is translated into the language of computation.

The practical implications of these biases are profound. When AI systems misclassify patients, the harm extends beyond diagnostic error to resource allocation, access to care, and institutional trust. Patients from historically marginalized groups may receive delayed diagnoses, fewer preventive interventions, or less aggressive treatment plans. These outcomes reinforce a cycle of inequity that undermines the credibility of medical technology and the broader healthcare system. The BMJ PROBAST+AI framework (Moons et al., 2025) offers a potential model for accountability by recommending comprehensive fairness audits, demographic benchmarking, and participatory evaluation involving patient communities. Such strategies move bias assessment beyond statistical validation toward a model of ethical validation, ensuring that fairness is not an afterthought but a core design criterion.

Ultimately, the case of algorithmic bias in healthcare supports the central argument of this study. AI systems are not neutral tools but sociotechnical artifacts that reflect their creators' values, assumptions, and inequities and the institutions that deploy them. Intersectionality explains how systemic inequities are embedded in data, while cognitive science elucidates how cognitive heuristics and organizational pressures normalize those inequities as acceptable trade-offs. Together, these frameworks reveal that fairness in healthcare AI cannot be achieved through technical optimization alone but requires reimagining what counts as accuracy and need. AI can fulfill its promise as a tool for equitable healthcare rather than a mirror of historical injustice only through ethical reflexivity, inclusive data practices, and governance frameworks that integrate social context.

**Criminal Justice**

AI-driven decision-making in the criminal justice system illustrates how algorithmic tools trained on historical data can entrench rather than correct systemic inequities. Predictive models such as COMPAS (Correctional Offender Management Profiling for Alternative Sanctions) and PredPol (Predictive Policing) were initially designed to provide objective risk assessments, ostensibly reducing human subjectivity in sentencing, parole, and patrol allocation. However, empirical evidence demonstrates that these systems often replicate, and in some cases amplify, racial and demographic disparities that have long plagued the justice system. ProPublica's landmark analysis of COMPAS (Angwin et al., 2016) revealed that Black defendants were nearly twice as likely as White defendants to be incorrectly labeled as high-risk for recidivism, even when criminal history and socioeconomic background were nearly identical. Conversely, White defendants were more likely to be classified as low risk despite comparable or more severe records. This pattern of misclassification is not random but reflects the structural biases embedded in the data used to train these models, data shaped by decades of over-policing, unequal sentencing, and socioeconomic inequality.

The proprietary nature of these systems compounds their ethical implications. Because COMPAS and similar tools are privately owned, their inner workings remain inaccessible to public scrutiny. This dynamic has been described as a "code is law" paradigm, where opaque algorithms influence judicial outcomes without the procedural safeguards that govern traditional legal processes (Engel, Linhardt, & Schubert, 2025). The lack of transparency prevents defendants from contesting the accuracy of algorithmic assessments, raising serious concerns about fairness and due process. While sectors like healthcare have begun developing standardized auditing frameworks such as PROBAST+AI (Moons et al., 2025) to evaluate bias and performance, no comparable mechanisms exist for the justice system. This absence of structured accountability allows biased algorithms to operate unchecked, with their decisions carrying life-altering consequences for individuals and communities.

Recent research substantiates the complexity of this problem. The study Algorithmic Fairness in Predictive Policing offers a systematic review of fairness strategies in policing and recidivism algorithms, highlighting that while most research focuses on racial bias, other protected attributes, such as age, gender, and socioeconomic status, remain largely unaddressed (Almasoud & Idowu, 2025). Their empirical analysis of the Chicago Police Department's Strategic Subject List (SSL) dataset revealed significant age-related bias, with individuals under thirty being disproportionately classified as high-risk, even when they had no prior record of violent or narcotic offenses. The study introduced a mitigation strategy called *Conditional Score Recalibration (CSR)*, which reassessed risk scores using fairness metrics such as Equality of Opportunity Difference, Average Odds Difference, and Demographic Parity. It was demonstrated that incorporating criteria-based adjustments into predictive models can significantly improve fairness without compromising accuracy (Almasoud & Idowu, 2025). This empirical contribution reinforces the broader argument that algorithmic bias cannot be separated from data quality, representativeness, and the institutional practices that produce the data.

The broader implications of biased AI in criminal justice extend beyond individual sentencing errors. Predictive systems shape the allocation of resources, influence policing strategies, and affect public perceptions of fairness. When predictive tools like PredPol direct law enforcement to neighborhoods already subject to heavy surveillance, they create feedback loops reinforcing structural inequities. Increased police presence leads to more recorded offenses in targeted communities, which in turn justifies further policing, perpetuating a cycle of overcriminalization. Causal analysis supports this pattern, identifying police actions as the primary drivers of model discrimination (Almasoud & Idowu, 2025). Their study concluded that predictive policing models amplify the bias already present in police behavior, effectively automating patterns of discrimination rather than eliminating them.

Intersectionality provides a critical framework for understanding these outcomes. The datasets that feed predictive policing systems are not neutral records of criminal behavior but reflect the interplay of race, class, and geography in law enforcement. When arrest records and neighborhood crime statistics are treated as objective indicators, the systems inherit and perpetuate the hierarchies that shape those records. Intersectionality shows how Black and Brown communities, already marginalized by social policy and economic inequality, face compounded disadvantage: first through biased policing practices, and again through the algorithms that learn from them. This dual exposure transforms historical discrimination into a mathematical constant within modern governance systems.

Cognitive science complements this analysis by revealing how human judgment interacts with algorithmic systems. Decision-makers often exhibit automation bias, the tendency to over trust machine outputs, and confirmation bias, the inclination to interpret evidence in ways that confirm preexisting beliefs. Judges, parole officers, and policymakers may defer to algorithmic scores because they appear objective and data-driven, even when those scores reinforce existing stereotypes. The BMJ PROBAST+AI (Moons et al., 2025) report warns that overreliance on opaque models fosters a false sense of precision, replacing moral deliberation with procedural compliance. As a result, algorithms become not tools of justice but mechanisms that obscure accountability.

The combined insights from these frameworks reveal that predictive algorithms do not simply mirror social injustice; they mechanize it. Fairness in these systems is too often reduced to statistical parity, a numerical balancing act that fails to address more profound moral questions about legitimacy, equality, and harm. The illusion of neutrality allows institutions to uphold discriminatory outcomes under the guise of efficiency. This dynamic aligns with what has been termed the “accountability gap,” a mismatch between recognizing AI’s ethical risks and implementing governance structures to mitigate them (KPMG GenAI Survey, 2023). In the context of criminal justice, this gap translates into unchecked authority for systems that directly influence sentencing, parole, and surveillance.

Empirical studies have begun to challenge the false trade-off between fairness and accuracy. Applying fairness-aware strategies such as CSR and class balancing to biased datasets has been shown to improve equity across demographic groups and enhance model performance, increasing accuracy from 83 to 90 percent (Almasoud & Idowu, 2025). This finding contradicts the common assumption that fairness compromises efficiency. Instead, it suggests that the two are mutually reinforcing when ethical design and robust evaluation metrics are applied. Moreover, their inclusion of underexplored variables such as age and socioeconomic status broadens the discourse beyond race, encouraging a more comprehensive understanding of equity in algorithmic systems.

Reform within this domain requires an expansion of both ethical and procedural accountability. Models used in criminal justice should be open to independent audit, with full disclosure of their data sources, error rates, and demographic impact. Adapting fairness frameworks such as PROBAST+AI could help standardize the evaluation of predictive models, ensuring that fairness, transparency, and performance are jointly assessed rather than treated as competing objectives. Equally important is the reintegration of human judgment. Judges, analysts, and policymakers must be trained to interpret algorithmic outputs critically and question their legitimacy in light of social context and ethical responsibility.

Integrating empirical insights highlights a critical point: bias mitigation is both possible and essential for restoring public trust in algorithmic governance (Almasoud & Idowu, 2025). As predictive models continue to shape decisions about risk and justice, the challenge is to refine algorithms and reimagine the moral architecture within which they operate. Without transparent oversight, participatory design, and cross-disciplinary evaluation, predictive technologies will continue to transform social hierarchies into digital hierarchies, embedding inequity in the logic of justice itself.

Ultimately, the case of predictive algorithms in criminal justice demonstrates that technology alone cannot repair what society has failed to reconcile. The biases embedded in data are reflections of historical choices and institutional values. Intersectionality reveals who those choices have harmed, while cognitive science explains why those harms persist under the guise of rationality. They show that fairness is not a statistical condition but a social commitment. Until the justice system aligns its technological ambitions with this understanding, algorithmic tools will remain instruments of inequality in the language of progress.

**Employment**

Bias in AI-driven recruitment systems demonstrates how intersecting identities such as gender, race, and socioeconomic status affect visibility and opportunity in automated decision-making. Major corporations increasingly use recruitment platforms to screen résumés, evaluate performance potential, and recommend candidates for advancement. These systems are marketed as efficient and objective solutions to human bias, yet they often reproduce and legitimize the inequities they were intended to mitigate in practice. A notable example is Amazon's AI recruitment tool, which was trained on historical data reflecting the company's male-dominated hiring patterns. The algorithm penalized résumés that included references to women's colleges, sororities, or organizations such as "Women in Tech," effectively disadvantaging female applicants. This outcome was not a simple technical malfunction, but a reflection of the gendered assumptions and structural inequalities embedded in the data. Data-driven systems are not inherently neutral and can perpetuate historical exclusion if they are not critically examined at every stage of development and deployment (Barocas & Selbst, 2016).

The Amazon case illustrates a broader trend across the technology industry, where automation is increasingly used to streamline hiring without sufficient oversight of the ethical implications. When past hiring practices reflect systemic exclusion, algorithmic systems built upon those datasets will reproduce those inequities at scale. Bias enters through skewed data and the choice of performance metrics, defining what constitutes a "good" hire. If historical data privileges certain traits, schools, or professional backgrounds that correlate with dominant social groups, then these traits become encoded as desirable, perpetuating a cycle of homogeneity. Research has shown that commercial AI products often reflect the demographic makeup and cultural assumptions of the organizations that build them (Raji & Buolamwini, 2019). In recruitment, this means that models trained on biased data continue to favor candidates who mirror those already in power, reinforcing patterns of occupational segregation and systemic underrepresentation.

Gender bias in AI recruitment is further exacerbated by the lack of diversity within development teams and the opacity of algorithmic decision-making. Many hiring algorithms operate as black boxes, with proprietary code and undisclosed decision parameters that make it nearly impossible for candidates or regulators to contest discriminatory outcomes. The importance of explainable AI in promoting accountability has been emphasized, as systems lacking transparency create environments where bias can persist undetected under the pretense of objectivity (Mitchell et al., 2019). Without interpretability, bias becomes both invisible and unchallengeable, placing the burden of proof on those most likely to be harmed by its effects. In corporate contexts, this opacity is compounded by the pursuit of efficiency and cost reduction, where human oversight is minimized in favor of speed and scalability. The result is an algorithmic filter that prioritizes uniformity over equity, narrowing the pool of candidates based on criteria that appear neutral but are structurally discriminatory.

Intersectionality provides a lens for understanding how these systems disproportionately impact marginalized candidates along overlapping lines of race, gender, and class. Women of color, for instance, face a double bind in algorithmic hiring systems. Historical labor market discrimination means that their résumés are less likely to contain the credentials or professional trajectories that AI systems recognize as indicators of success. Moreover, models calibrated to dominant communication norms may penalize linguistic and cultural variations in self-presentation. A study on algorithmic grading and feedback found similar patterns in educational settings, where automated assessment tools systematically devalued nonstandard English or culturally specific phrasing (Young & Wajcam, 2023).

In recruitment, these same linguistic biases can disadvantage candidates whose expression of competence diverges from dominant professional codes. Intersectionality thus reveals that algorithmic bias is not monolithic; it compounds across social hierarchies, producing exclusion through multiple and intersecting vectors of disadvantage.

From the cognitive science perspective, these inequities are reinforced by the mental shortcuts and perceptual limitations that influence developers and end users. Developers may unconsciously encode their cognitive biases into system design through processes such as anchoring, where initial assumptions about desirable candidates influence feature selection and model tuning. Anchoring may lead to overreliance on historical indicators such as Ivy League degrees, continuous employment histories, or specific job titles that correlate with privileged groups. Confirmation bias further reinforces these tendencies when organizations interpret the algorithm's output to validate preexisting hiring preferences. Cognitive science helps explain why even well-intentioned practitioners can perpetuate discrimination: their reasoning is bound by organizational norms and limited feedback loops that reward efficiency over fairness. These cognitive distortions become institutionalized when embedded in automated systems that are treated as objective evaluators rather than extensions of human judgment.

The consequences of biased AI recruitment are far-reaching. At the individual level, applicants may be unfairly screened out of job opportunities, limiting their access to economic mobility and professional advancement. At the organizational level, bias undermines workforce diversity, stifles innovation, and erodes trust among employees and applicants. On a societal scale, the widespread adoption of biased recruitment systems risks deepening existing inequalities in employment, particularly for women, people of color, and individuals from lower socioeconomic backgrounds. A 2025 analysis by KPMG on leadership and generative AI adoption reported on this tension: while leaders view AI as a driver of productivity, only a small fraction report having clear governance frameworks for fairness, diversity, and inclusion. This gap between innovation and regulation highlights a growing accountability crisis in AI governance.

Researchers and advocacy groups have proposed strategies to address these challenges, emphasizing that fairness cannot be achieved through technical adjustments alone. Responsible AI requires embedding human judgment throughout the model lifecycle, ensuring that fairness is evaluated both statistically and socially (Holstein et al., 2019). This approach involves establishing diverse development teams, implementing algorithmic audits, and incorporating fairness impact assessments that consider historical and cultural context. Some organizations are experimenting with participatory design models that involve diverse stakeholders in reviewing training data and defining criteria for fairness. These approaches move beyond transparency toward shared governance, aligning with calls for algorithmic justice grounded in community engagement.

Ultimately, bias in AI-driven recruitment systems supports the broader theme of this study: technological neutrality is an illusion. Algorithms mirror the values, priorities, and inequities of the societies that create them. Intersectionality explains how exclusion is encoded through historical hierarchies of privilege, while cognitive science clarifies how human perception and decision-making biases reinforce these hierarchies within automated systems. Together, these frameworks reveal that fairness in AI recruitment is not a static technical goal but an ongoing ethical practice that requires vigilance, reflexivity, and inclusion. Without structural change in data and design, AI recruitment will continue to replicate inequality under the guise of innovation, transforming historical patterns of discrimination into automated norms of employability.

**Education**

The exact underlying mechanisms appear in education, where AI tools are designed to enhance fairness in admissions, assessment, and student support, and instead reproduce longstanding patterns of privilege and exclusion. Educational institutions have increasingly adopted AI systems to improve efficiency in evaluating applications, predicting student success, and personalizing instruction. However, the underlying data and design choices often reflect and reinforce preexisting inequities. College admissions algorithms trained on historical data frequently prioritize traditional achievement indicators such as high standardized test scores, attendance at elite secondary schools, and participation in extracurricular activities that require financial resources. These markers of privilege are deeply intertwined with race, class, and geography. As a result, students from marginalized or under-resourced communities are often evaluated as less competitive, not because of a lack of potential but because the algorithm learns to associate merit with the material advantages accompanying privilege.

A prominent case was documented in which an AI-driven college admissions tool disproportionately favored applicants from affluent families and elite preparatory schools while systematically deprioritizing students from lower-income backgrounds (Stevens & Burke, 2021). The algorithm's training data mirrored decades of systemic bias in college admissions, where access to advanced coursework, private tutoring, and extracurricular enrichment correlates strongly with wealth. Similarly, recent analyses of algorithmic admissions systems in the United Kingdom revealed that predictive models used to estimate student performance during pandemic-era exam cancellations downgraded students from public and low-income schools at significantly higher rates than their peers from private institutions (Young & Wajcam, 2023).

These examples demonstrate that what is often framed as an efficient innovation can, in practice, replicate historical hierarchies of access and opportunity.

From an intersectional perspective, these inequities cannot be viewed in isolation. Race, class, gender, and geography intersect to compound disadvantage for students who fall outside the normative parameters of success embedded in training data. For instance, students from rural areas or first-generation college applicants may have strong academic potential but lack access to the extracurricular or institutional markers valued by algorithms. Likewise, women and students of color may be systematically penalized when their academic or professional trajectories do not align with patterns historically associated with success in male-dominated or Eurocentric disciplines. Intersectionality shows how these overlapping inequities create cumulative barriers, transforming AI tools intended to democratize opportunity into gatekeeping mechanisms perpetuating exclusion.

Cognitive science provides a complementary explanation for why these failures persist despite bias awareness. Human evaluators, while imperfect, are capable of contextual reasoning that allows them to understand a student’s achievements relative to available opportunities, personal challenges, or community constraints. AI systems, in contrast, rely on statistical correlations and lack the capacity for such situated cognition. They cannot interpret the qualitative dimensions of effort, resilience, or creativity that often distinguish high-potential students from those who meet conventional criteria. Constrained by cognitive shortcuts such as anchoring and confirmation bias, developers frequently design systems that equate measurable outputs like test scores, grade point averages, or institutional rankings with merit. These heuristics simplify complex human potential into numeric representations, producing outputs that appear objective but are shaped by socially contingent definitions of worth.

This cognitive simplification is further compounded by institutional and cultural factors that valorize quantifiable metrics as the gold standard of fairness. Educational policymakers often adopt AI systems under the assumption that automation reduces subjectivity. However, as Vakali and Tantalaki (2024) note, attempting to eliminate human bias can result in overreliance on biased data and flawed models because objectivity becomes conflated with computability. Cognitive biases such as automation bias lead admissions officers and institutional leaders to trust algorithmic outcomes over human judgment, even when evidence of inequity is apparent. Once integrated into institutional processes, these systems gain normative power that shapes how merit, potential, and fairness are defined and operationalized.

The consequences of algorithmic bias in education extend far beyond individual admissions decisions. Over time, these systems influence the demographic composition of student bodies, the distribution of scholarships, and even long-term socioeconomic mobility. When algorithmic tools systematically filter out underrepresented students, higher education becomes less diverse and less reflective of society. The resulting feedback loop is self-reinforcing as graduates of elite institutions continue to dominate leadership roles, feeding new data that reaffirms exclusionary definitions of excellence. Scholars such as Ulnicane (2024) and King and Petty (2021) argue that these cycles of algorithmic selection contribute to a digital meritocracy in which privilege is masked as performance and systemic inequality is naturalized as meritocratic order.

Recent studies have addressed these concerns by proposing fairness-aware frameworks for educational AI. Decker et al. (2024) advocate for procedural fairness through public engagement, suggesting that diverse stakeholders, including students, educators, and community representatives, should be involved in defining the values and priorities embedded in AI-driven decision systems. Transparency alone is insufficient if the underlying data continue to reflect exclusionary histories. Instead, participatory governance and ethical auditing are required to ensure algorithmic systems align with educational equity goals. The AI Index Report (2025) recommends mandatory fairness assessments for all education-focused AI applications, including transparent reporting of demographic performance metrics and third-party audits of training data.

Viewed through the dual lenses of intersectionality and cognitive science, these findings demonstrate that bias in educational AI is not simply a matter of computational error but a product of the epistemic and institutional contexts that shape system design. Algorithms do not miscalculate potential; they calculate according to inherited definitions of merit. These definitions privilege quantifiable achievement over lived experience, excluding those whose brilliance is not easily captured by standardized metrics. Therefore, ethical integration of AI in education requires a fundamental rethinking of fairness and success. Equity must be defined in terms of statistical parity and recognition of the diverse pathways through which students demonstrate potential.

In conclusion, educational AI systems mirror the structural inequalities and cognitive assumptions of the societies that produce them. Intersectionality reveals how algorithmic tools magnify inequities across race, gender, and class, while cognitive science explains how human biases become computational logic. Together, these frameworks suggest that creating truly equitable educational technologies requires embedding contextual awareness, reflexivity, and inclusivity into every stage of system design and implementation. Without this shift, AI will continue to transform instruments of opportunity into instruments of exclusion, perpetuating privilege under the language of progress.

**Synthesis and Implications**

Across healthcare, criminal justice, employment, and education, these case studies collectively demonstrate that bias in AI systems is not an isolated technical malfunction but a deeply embedded sociotechnical phenomenon that emerges from the convergence of structural inequity, institutional culture, and human cognition. Each sector reveals a distinct manifestation of the same underlying pattern: systems designed to optimize efficiency and fairness instead of operationalizing historical hierarchies and cognitive shortcuts that perpetuate inequality. In healthcare, the assumption that cost equals care privileges economic access over clinical need. In criminal justice, predictive policing models transform racialized histories of surveillance into quantified measures of risk. In employment, hiring algorithms reproduce organizational biases disguised as neutral evaluation metrics. In education, admissions systems convert privilege into a predictive signal of merit. Taken together, these cases illustrate how algorithmic systems, though presented as objective and data-driven, serve as mirrors reflecting the biases of the societies that create them.

Viewed through the combined lenses of intersectionality and cognitive science, these examples make visible the mechanisms by which bias becomes institutionalized and self-sustaining. Intersectionality exposes how race, class, gender, and other social categories intersect to shape the design and impact of AI, showing that exclusion is not incidental but patterned and systemic. Cognitive science complements this perspective by explaining how human reasoning, constrained by heuristics such as anchoring, automation bias, and confirmation bias, reinforces the inequities designers aim to correct. Together, these frameworks reveal that the challenge of AI fairness is as epistemological as it is ethical: the systems we build reflect the limits of our understanding of justice, equity, and human complexity.

Addressing these challenges requires more than technical refinement; it demands a paradigm shift in how AI is conceived, developed, and governed. Ethical innovation must move beyond compliance-based frameworks toward approaches that embed social responsibility into every system design phase. Inclusive design processes can ensure that marginalized voices are meaningfully represented in defining the problems AI seeks to solve and the values it encodes. Transparent governance structures, including algorithmic audits, impact assessments, and participatory oversight, can transform accountability from a procedural formality into a substantive ethical commitment. Interdisciplinary collaboration—bridging computer science, social science, law, and ethics—enables a more holistic understanding of fairness that reflects technical accuracy and social consequence.

Community engagement is equally essential. The individuals and groups most affected by AI systems should not merely be subjects of study but active participants in shaping their design and deployment. Involving communities in decision-making fosters trust, contextual awareness, and ethical reflexivity, ensuring that AI technologies serve collective rather than corporate interests. As scholars such as Hagerty and Rubinov (2019) and Decker et al. (2024) emphasize, the legitimacy of AI governance depends on whether those impacted by technology have a voice in its creation and oversight.

The cases analyzed here provide a foundation for the broader inquiry at the heart of this study: understanding how practitioners themselves interpret bias, navigate ethical tensions, and imagine pathways toward equity in AI. They illustrate that technical skill alone is insufficient for responsible AI development. What is required is a shift in mindset—a recognition that fairness is not an endpoint but an ongoing practice of reflection, adaptation, and inclusion. As global dependence on algorithmic systems expands, the responsibility to align technological innovation with human values becomes an ethical and civic imperative.

Ultimately, the findings support that the future of AI will be determined not only by advancements in computation but by our collective willingness to confront the moral and cognitive foundations upon which those systems rest. Designing technically robust and socially responsible technologies requires acknowledging that data are never neutral, algorithms are never autonomous, and fairness cannot be programmed without empathy. Actual progress in AI will depend on creating systems that reflect the full complexity of human identity and cognition, capable of advancing justice rather than reproducing harm.

**Historical and Technical Sources of Algorithmic Bias**

Understanding algorithmic bias requires attention to both its historical roots and the technical mechanisms through which it is reproduced. Bias in artificial intelligence (AI) systems is not simply the result of flawed code or incomplete datasets, it reflects long-standing social, institutional, and epistemological inequalities that have shaped data infrastructures and computational logic. As the HAI AI Report and related scholarship emphasize, these biases are embedded in the very foundations of AI systems, from the data they are trained on to the assumptions that guide their design and deployment.

Historically, data collection practices have been shaped by exclusionary norms and unequal access to resources. In sectors such as healthcare, education, and criminal justice, the data used to train AI systems often reflects decades of systemic discrimination. For example, medical datasets frequently underrepresent Black, Indigenous, and low-income patients due to disparities in healthcare access and documentation. Algorithms trained on such data may therefore fail to recognize the health needs of marginalized populations, leading to underdiagnosis and unequal treatment (Obermeyer et al., 2019). These outcomes are not anomalies; they are predictable results of training models on data that encode historical inequities.

In the criminal justice system, predictive policing tools rely on crime data that is itself a product of over-policing in minority communities. The COMPAS algorithm, analyzed by ProPublica (Angwin et al., 2016), exemplifies how biased historical records produce discriminatory risk assessments. The system disproportionately labeled Black defendants as high-risk, even when controlling for prior offenses. This bias was not introduced by the algorithm alone but inherited from the data it was trained on. Data is shaped by decades of racial profiling, surveillance, and unequal enforcement. The HAI AI Report highlights that such feedback loops are common: biased inputs lead to biased outputs, which in turn reinforce the original patterns of exclusion.

From a technical standpoint, bias in AI systems emerges through several mechanisms. One of the most significant is the use of proxy variables (Lau, 2025); indirect indicators that stand in for more complex or sensitive attributes. In healthcare, for instance, using healthcare spending as a proxy for health need ignores the structural barriers that prevent some populations from accessing care. In education, standardized test scores serve as proxies for academic potential, despite their strong correlation with family income and access to preparation resources. These proxies simplify complex social realities and introduce bias by failing to account for the context in which data is generated.

Imbalance and exclusion in training datasets further perpetuate inequities. AI models trained primarily on majority populations often perform poorly for underrepresented groups. In dermatology, for example, datasets containing predominantly lighter skin tones lead to lower diagnostic accuracy for patients of color (Adamson & Smith, 2018). Similarly, employment algorithms trained on résumés from male-dominated industries penalize female applicants, as seen in Amazon's hiring tool (Dastin, 2018). These patterns are not technical mistakes but reflections of the demographic and cultural limitations of the data sources themselves.

Model architecture and evaluation practices can also amplify bias. Many systems are optimized for accuracy or efficiency without consideration of fairness, resulting in models that perform well on average but poorly for specific groups. Validation datasets often mirror the biases of training data, and performance metrics rarely include demographic stratification. Without deliberate evaluation across diverse populations, AI systems will continue to reproduce inequities (HAI AI Report, 2025).

Cognitive science adds another layer to this analysis by revealing how human decision-making shapes technical design. Developers rely on heuristics and mental models that reflect dominant cultural norms. These cognitive shortcuts influence how problems are framed, which data is prioritized, and how success is defined. For example, practitioners may assume that high test scores or elite educational backgrounds universally indicate merit, overlooking the structural advantages that produce these outcomes. These assumptions become embedded in algorithmic logic, producing systems that privilege dominant groups and marginalize others.

Viewed together, the historical, technical, and cognitive dimensions of bias are deeply intertwined. Data infrastructures reflect the legacies of exclusion, while design choices often fail to account for those legacies. Intersectionality illuminates how systemic inequities shape the data and institutions underlying AI, while cognitive science explains how human reasoning perpetuates those same inequities through unconscious bias and mental framing. Addressing bias therefore requires more than technical refinement, it demands a critical engagement with the social and cognitive structures that sustain it. As the HAI AI Report emphasizes, fairness in AI must be understood as a socio-technical challenge requiring interdisciplinary collaboration, inclusive design, and structural accountability.

In this study, these insights inform the analysis of how practitioners perceive and respond to bias in AI systems. By examining the historical, technical, and cognitive foundations of algorithmic bias, the research seeks to uncover the assumptions, constraints, and trade-offs that shape real-world development practices. This approach supports the broader goal of promoting AI systems that are not only technically robust but also socially just, systems that acknowledge and address the inequalities embedded in their foundations.

**The Bias-Variance Tradeoff in AI Fairness**

The bias-variance tradeoff is a foundational concept in machine learning with significant implications for the pursuit of fairness and equity in artificial intelligence (AI) systems. It encapsulates the tension between two types of model error: bias, which arises from overly simplistic assumptions that fail to capture data complexity, and variance, which results from excessive sensitivity to training data fluctuations. As Brownlee (2016) explains, this tradeoff is central to model generalization and predictive performance. Yet in the context of AI fairness, it extends beyond technical optimization and becomes a socio-technical dilemma with tangible consequences for marginalized communities.

High-bias models tend to generalize poorly across diverse populations, often encoding dominant group norms while overlooking the patterns and experiences of underrepresented groups. This results in systematic underperformance that effectively renders these populations invisible within the model's logic (Barocas & Selbst, 2016; Noble, 2018). For example, a high-bias model trained on data from predominantly white, affluent populations may fail to accurately assess risk or potential for individuals from Black or Indigenous communities, reinforcing disparities in education, healthcare, and criminal justice (Ajanaku, 2022; Mergen et al., 2025).

Conversely, high-variance models attempt to capture every nuance in training data, leading to erratic and unreliable outcomes, particularly for groups already underrepresented. In fairness-aware contexts, this overfitting can result in inconsistent treatment of similar individuals, undermining equity in high-stakes domains such as hiring or lending (IBM, 2025; Lee, 2025). The tradeoff therefore embodies a tension between predictive precision and social justice, where optimizing for one may compromise the other.

Recent research by Buijsman (2024) emphasizes that optimizing for fairness often requires sacrificing some degree of predictive accuracy, especially when fairness is defined through group parity or equalized odds. This introduces a normative dilemma: should designers prioritize statistical performance or explicitly account for structural inequities, even if doing so reduces accuracy for the majority group? Drawing on Rawlsian fairness, Buijsman proposes centering the needs of the most disadvantaged as a guiding principle for resolving these tradeoffs.

Cognitive science helps illuminate why this tradeoff is so difficult to navigate. Developers and data scientists rely on heuristics, mental models, and cognitive shortcuts that shape how they frame problems and interpret outcomes. These reasoning patterns often favor optimization metrics such as accuracy or loss minimization over contextual considerations of equity or harm. Unlike human decision-makers who can engage in moral deliberation and situated reasoning, AI systems operate through probabilistic inference and lack the capacity for contextual judgment (Clark, 2013; Sawyer, 2014). When human cognition privileges technical performance over social consequence, these priorities become embedded in algorithmic design itself.

To mitigate these challenges, researchers have advanced fairness-aware learning strategies that integrate social context into model development. Raftopoulos et al. (2025) categorize these into three primary approaches: pre-processing methods that rebalance datasets, in-processing techniques that embed fairness constraints into model training, and post-processing strategies that adjust outputs to reduce disparate impact. These techniques acknowledge that fairness is not a purely mathematical construct but a normative commitment that must inform every stage of the AI lifecycle.

Yet, fairness-aware models remain bound by the constraints of the bias-variance tradeoff. Increasing fairness often adds complexity, which can heighten variance and reduce generalizability. This presents particular risks for marginalized groups, who may face both inaccurate and inconsistent model predictions (Eduwik, 2025; IBM, 2025). As Pant et al. (2024) show, practitioners are aware of these tensions but struggle to operationalize fairness due to vague guidelines, limited institutional support, and competing performance metrics. This disconnect supports the need for frameworks that help practitioners reconcile ethical goals with technical constraints.

Intersectionality offers a critical lens for reimagining what constitutes an acceptable tradeoff. As Crenshaw (1991) and Himmelreich et al. (2024) argue, fairness cannot be divorced from the intersecting systems of power that shape experience. Models that treat race, gender, or class as independent variables risk missing the compounded effects of discrimination. Incorporating intersectionality requires granular subgroup analysis and fairness metrics that capture overlapping forms of inequality, an endeavor that further complicates the bias-variance balance but is essential for ethical AI design.

In practice, this means moving beyond accuracy as the singular benchmark for performance. Developers must adopt evaluation frameworks that prioritize equity, transparency, and accountability alongside predictive reliability. Integrating philosophical principles such as Rawlsian justice can guide these decisions by centering those most vulnerable to harm. As Westerstrand (2024) suggests, embedding justice principles throughout the AI lifecycle from data collection to deployment and auditing is vital to advancing equitable AI.

Ultimately, the bias-variance tradeoff in AI fairness is not a problem to be permanently solved but a dynamic tension that must be continuously negotiated. Understanding it through the combined perspectives of intersectionality and cognitive science highlights that fairness is both a structural and cognitive challenge requiring interdisciplinary collaboration, ethical reflection, and sustained community engagement. As AI systems become increasingly embedded in public and private decision-making, the stakes of this tradeoff will only deepen. Ensuring that these systems serve all members of society, particularly those historically marginalized, demands a redefinition of fairness itself, one that privileges justice, inclusivity, and accountability over mere optimization.

## Philosophical and Ethical Critiques of General AI

The development of general artificial intelligence (AGI) has prompted an extensive body of philosophical and ethical critique examining the assumptions, implications, and consequences of increasingly autonomous systems. These critiques span moral theory, sociotechnical analysis, and global justice, highlighting the difficulty of aligning AI with diverse and evolving human values.

Kakembo Aisha Annet (2025) contends that the dominant “common-sense ethics” approach to AI is inadequate for addressing the deeper moral dilemmas posed by AGI. She calls for a renewed engagement with classical ethical frameworks leveraging utilitarianism, deontology, virtue ethics, and social contract theory to guide AI development. Utilitarianism raises questions about what constitutes “beneficial” AI and whether such benefit can be objectively defined. Reinforcement learning, for instance, may generate agents that optimize outcomes in ways that conflict with moral reasoning. Deontological ethics, emphasizing rule-based conduct, reveals the difficulty of translating moral principles into machine-readable code. Virtue ethics redirects attention toward the moral character and integrity of AI developers, while social contract theory supports the need for transparency and accountability, suggesting that AI systems should be explainable and trustworthy in ways analogous to human experts (Annet, 2025).

Judith Simon and colleagues (2024) expand this discussion by exploring how AI technologies challenge democratic legitimacy, privacy, and inclusivity. Their analysis identifies four key domains: regulation and public legitimation, privacy in the data economy, solidarity and responsibility in design, and the need to reconceptualize AI ethics altogether. They warn that algorithmic systems embedded in public infrastructure risk undermining democratic values by shifting power from public institutions to private corporations. The increasing dependence on opaque algorithms in governance raises serious concerns about surveillance, discrimination, and the erosion of civic agency. Simon et al. (2024) advocate for participatory governance models and robust ethical frameworks that reflect pluralistic values and promote social justice.

David J. Gunkel (2018) offers a radical philosophical reframing through what he terms *relational ethics*, challenging the “properties approach” that assesses moral standing based on intelligence or autonomy. In *The Machine Question*, Gunkel argues that ethics should focus not on the intrinsic properties of machines but on the relationships and responsibilities that emerge in human–machine interactions. This relational turn reframes ethical inquiry around interdependence and mutual recognition, acknowledging that machines, though not sentient, occupy morally significant roles within social systems (Gunkel, 2018).

Ning Sun (2025) integrates philosophical and psychological insights to examine how AI reshapes human subjectivity. Drawing on Marx’s theory of alienation, Kant’s deontological ethics, and Heidegger’s concept of technological enframing, Sun argues that the rapid expansion of technological rationality has not been matched by ethical or emotional development. This mismatch produces a crisis of agency and meaning, as AI systems mediate decision-making, relationships, and emotional life. Sun proposes an approach guided by ethical reconstruction, psychological intervention, and institutional reform to reorient AI toward human-centered values and restore moral coherence in an increasingly algorithmic society.

Global perspectives further expand the ethical discourse. Hagerty and Rubinov (2019) critique the Western-centric orientation of mainstream AI ethics, showing that its principles often fail to capture the lived realities of the Global South. Their analysis demonstrates that AI frequently exacerbates inequality in low- and middle-income countries, where limited infrastructure and weak regulation compound technological dependency. They advocate for ethnographic research and culturally responsive governance frameworks to ensure that AI development promotes global rather than parochial notions of justice (Hagerty & Rubinov, 2019).

Taken together, these critiques highlight that the ethical challenges of AGI are neither purely technical nor solely philosophical. They are deeply relational, cognitive, and structural. Intersectionality provides insight into how ethical discourse must account for the unequal global and social conditions in which AI operates, while cognitive science highlights the limitations of human moral reasoning when translated into computational logic. Addressing AGI ethics thus requires reimagining agency, responsibility, and justice through frameworks that are ethically grounded, globally inclusive, and relationally attuned.

**Philosophical Limits of General AI**

While artificial intelligence has made significant strides in specialized domains, many scholars argue that the pursuit of General AI remains conceptually and technically flawed. Landgrebe and Smith (2023) present a critique rooted in the mathematical complexity of human cognition. They argue that even the general cognitive performance of animals, such as crows, cannot be replicated by machines. This claim is not based on philosophical speculation but reflects the limitations inherent in modeling complex systems through algorithmic logic.

Current GenAI models, including deep learning and neural networks, function primarily as pattern-recognition systems. These systems lack the ability to understand context, intentionality, or consciousness, which are essential components of human intelligence. This critique builds on earlier work by Hubert Dreyfus, who maintained that symbolic logic-based AI could never replicate human cognition because mental processes cannot be reduced to formal rule sets. Landgrebe and Smith extend this argument to modern machine learning, asserting that no algorithmic framework will ever fully simulate the depth and flexibility of human thought.

These limitations have practical implications. They suggest that GenAI may continue to excel in narrow, well-defined tasks but will struggle to achieve the adaptability and nuance required for human-level intelligence. This gap highlights the need for greater transparency in how AI systems are trained, evaluated, and deployed. Without clear documentation of data sources, modeling assumptions, and decision logic, it becomes difficult to assess where cognitive biases may be influencing outcomes.

Furthermore, cognitive theory suggests that bias mitigation must extend beyond technical solutions to include training developers to recognize and reflect on their own assumptions. Developers often operate under time constraints, organizational pressures, and implicit norms that shape how they interpret data and define fairness. Without intentional reflection, these mental models can reinforce cognitive shortcuts which influence design decisions in subtle but consequential ways.

Training programs that incorporate behavioral science, ethics, and inclusive design principles can help practitioners become more aware of how their cognitive processes shape algorithmic outcomes. For example, workshops that simulate ethical dilemmas or expose developers to diverse user perspectives can foster critical thinking and empathy. These interventions are most effective when embedded into the organizational culture, supported by leadership, and reinforced through ongoing dialogue and peer review.

Transparency also plays a vital role in this process. When developers document their assumptions, data sources, and decision rationales, it becomes easier to identify where bias may be entering the system. Transparent workflows enable external auditing, interdisciplinary collaboration, and stakeholder engagement, all of which contribute to more accountable and inclusive AI development.

Addressing these human factors is essential for building GenAI systems that are not only technically robust but also ethically grounded and socially responsive. By combining cognitive awareness with structural support, organizations can move toward AI practices that reflect the complexity of human judgment and the diversity of human experience.

**Governance and Accountability in AI Development**

The rapid proliferation of artificial intelligence (AI) technologies across sectors has intensified the call for robust governance and accountability mechanisms. As AI systems increasingly influence decision-making in domains such as healthcare, finance, education, and criminal justice, scholars have examined how governance frameworks and accountability structures can ensure that these systems are developed and deployed responsibly. Governance in AI refers to the policies, standards, and institutional arrangements that guide ethical development and oversight, while accountability ensures that actors including developers, organizations, and regulators are answerable for the outcomes and impacts of AI technologies. Together, these concepts represent foundational dimensions of responsible AI.

Batool, Zowghi, and Bano (2024) conducted a systematic literature review of responsible AI governance and developed a 3W1H model that explores four central questions: Who is accountable for AI systems, what elements are governed, when governance occurs during the AI lifecycle, and how governance is executed through frameworks, standards, and policies. Their analysis revealed that most existing approaches emphasize technical controls but overlook the human and process-oriented dimensions necessary for comprehensive oversight. This gap highlights the need for governance models that integrate ethical principles across every stage of the AI lifecycle from problem definition and design to implementation and monitoring.

Accountability has also been conceptualized as a relational and context-dependent construct. Novelli, Taddeo, and Floridi (2024) argue that accountability must encompass authority recognition, interrogation, and the limitation of power. Their proposed architecture identifies seven features context, range, agent, forum, standards, process, and implications alongside four key goals: compliance, reporting, oversight, and enforcement. These elements illustrate that accountability extends beyond compliance audits to include dynamic, reciprocal relationships among stakeholders. Accountability mechanisms, they contend, must evolve alongside technological and social changes to remain effective within rapidly shifting AI ecosystems.

Transparency and explainability are central to both governance and accountability efforts. Cheong (2024) emphasizes that opaque algorithmic systems undermine public trust and impede legal or ethical scrutiny. He categorizes transparency initiatives into four thematic areas: technical interpretability, legal disclosure requirements, ethical principles such as fairness and autonomy, and multi-stakeholder engagement. These strategies collectively aim to empower individuals affected by algorithmic decision-making to understand, contest, and seek redress for potentially harmful outcomes.

Global and regional policy frameworks have further advanced discussions of responsible AI governance. The OECD AI Principles (2019) stress inclusive growth, human rights, transparency, robustness, and accountability, while UNESCO's Recommendation on the Ethics of AI (2021) calls for human-centered development and democratic oversight. Within the regulatory domain, the European Union's AI Act, expected to take effect by 2026, introduces a risk-based approach that imposes strict requirements on high-risk AI systems and mandates transparency, human oversight, and independent conformity assessments (Leopard & Epstein, 2025). Collectively, these frameworks aim to promote ethical alignment and interoperability across national boundaries.

Despite notable progress, persistent challenges limit the impact of current governance initiatives. Jurisdictional fragmentation hampers cross-border enforcement, and corporate reliance on self-regulation often prioritizes innovation speed over ethical responsibility. Batool et al. (2024) observe that few governance frameworks address all phases of the AI lifecycle or specify clear accountability structures. Such fragmentation weakens oversight mechanisms and risks diminishing public confidence in AI systems.

To address these limitations, researchers advocate for multi-level governance that operates across organizational, national, and international layers. At the organizational level, companies are encouraged to establish internal ethics boards, conduct bias audits, and publish transparent accountability reports. At the national level, governments should define and enforce accountability standards, empower regulatory agencies, and ensure public participation. Internationally, collaboration among governing bodies is necessary to harmonize standards and minimize regulatory gaps.

Overall, governance and accountability are not merely technical challenges but deep ethical and political imperatives. They require integrative approaches that connect normative principles with enforceable oversight mechanisms. Literature collectively supports a persistent need for governance models that embed fairness, transparency, and accountability throughout the AI lifecycle. This study responds to that gap by introducing the Fairness by Design Framework, which conceptualizes fairness as a continuous socio-technical process embedded in organizational practice. Building on global governance models, this framework advances responsible AI through iterative design feedback, measurable fairness metrics, and transparent reporting mechanisms designed to sustain ethical accountability across development and deployment.

**Emerging Ethical Frameworks and Contemporary Theoretical Perspectives in AI Fairness**

Recent research highlights that as artificial intelligence systems evolve toward increasingly autonomous agents, conventional ethical frameworks may no longer be sufficient to guide their development or regulate their social impact. The emergence of large-scale, agentic systems capable of making independent decisions and interacting dynamically with their environments has shifted the moral landscape of AI. Gabriel, Keeling, Manzini, and Evans (2025) emphasize that as AI agents acquire the ability to perceive, reason, and act with limited human oversight, the boundaries of ethical responsibility become diffused. Their work calls for a new ethics that transcends traditional, anthropocentric models by recognizing the distributed nature of agency in human–machine ecosystems. This shift necessitates rethinking accountability and moral reasoning across the full lifecycle of AI from data collection to deployment and governance.

This reconceptualization aligns with broader debates in the AI fairness literature, which increasingly acknowledges that fairness is not solely a statistical or computational issue but a deeply social and relational one. As Gabriel et al. (2025) note, autonomous systems challenge long-standing assumptions about intentionality and moral authorship. If an AI agent can make complex decisions that materially affect individuals or groups, ethical evaluation must consider not only the algorithms themselves but also the institutional and cultural structures in which they operate. These developments highlight the need for an ethics strategy that is simultaneously technical, organizational, and social, capable of addressing the intertwined roles of data, design, and governance.

Intersectionality provides one such lens by grounding the interdependent nature of social categorizations and their impact on lived experience. Originating in Black feminist scholarship, Intersectionality posits that systems of oppression such as racism, sexism, classism, and ableism intersect to shape distinct and compounding forms of marginalization. In the context of AI fairness, Intersectionality highlights how algorithmic bias is rarely the result of isolated variables but instead emerges from overlapping structural inequities encoded in data and institutional practices. Rather than seeking a singular measure of fairness, Intersectionality calls for context-specific analyses that attend to how algorithmic systems affect people differently depending on their intersecting identities. This framework therefore reframes fairness as a question of equity and social consequence, not merely of statistical parity.

While Intersectionality provides a social-structural foundation for analyzing inequity, other emerging frameworks offer complementary approaches that operationalize ethics across design, implementation, and policy domains. Value-Sensitive Design (VSD), developed within the field of human–computer interaction, integrates moral and social values directly into the engineering process. It emphasizes proactive identification of stakeholders and anticipatory reflection on the societal implications of technological choices. By incorporating ethical deliberation during the initial stages of design rather than as a corrective measure after deployment, VSD provides a procedural path for translating values such as justice, autonomy, and inclusivity into system specifications. In practice, this approach aligns with intersectional principles by ensuring that marginalized perspectives are included in defining what “fairness” means for a particular context.

Data Feminism, proposed by D’Ignazio and Klein in 2020, extends these ideas by interrogating the epistemological foundations of data science itself. It challenges the presumption of neutrality in data collection and analysis, arguing that power is always present in decisions about what counts as data, who collects it, and how it is interpreted. From a Data Feminism perspective, fairness is achieved not through abstraction but through positional awareness recognizing that all data practices are situated within specific cultural, political, and historical contexts. This orientation resonates with Intersectionality in its commitment to amplifying underrepresented voices and revealing the structural forces that render certain populations more visible or invisible in algorithmic systems.

A third body of work, often described as Algorithmic Governance, situates ethical responsibility within institutional and policy mechanisms. Rather than focusing solely on the moral reasoning of designers or the technical accuracy of models, Algorithmic Governance treats fairness as a matter of organizational and regulatory architecture. It encompasses auditing protocols, public transparency measures, accountability mechanisms, and participatory oversight structures that define how AI systems operate within society. This perspective reframes fairness as a dynamic outcome of governance design focusing on how decisions are made, who is empowered to challenge them, and what recourse exists when harm occurs. For Intersectionality-informed research, this governance-oriented approach is essential because structural inequality cannot be mitigated solely through better algorithms; it requires systemic accountability and institutional reform.

These emerging frameworks illustrate an ongoing expansion in how scholars and practitioners conceptualize ethical AI. The shift is from principle-based ethics, which prescribes abstract values such as beneficence or justice, toward process-oriented ethics, which emphasizes how those values are enacted within technical and social systems. Intersectionality provides the critical grounding for understanding whose values are prioritized and whose experiences are marginalized, while frameworks such as Value-Sensitive Design, Data Feminism, and Algorithmic Governance offer methodological and procedural pathways for embedding those insights into practice.

Synthesizing these perspectives strengthens the analytical foundation of this study. Intersectionality ensures that the examination of AI bias remains anchored in real-world inequities and the lived experiences of those most affected by algorithmic decision-making. Value-Sensitive Design operationalizes this concern through stakeholder engagement and ethical deliberation. Data Feminism contributes reflexivity, urging recognition of positional power in data practices, and Algorithmic Governance extends accountability beyond individual actors to the institutional level. Collectively, they outline a comprehensive ethical landscape that views fairness as a continuous negotiation among social values, technical constraints, and governance structures. Within this pluralistic framework, AI ethics evolves from a compliance exercise into an iterative process of social responsibility, aligning technological development with broader commitments to equity and justice.

## Practitioner-Centered Research and Gaps in Literature

As artificial intelligence (AI) technologies become increasingly embedded in critical societal functions, the ethical dimensions of their development and deployment have attracted growing attention. Yet much of the discourse surrounding AI ethics remains dominated by abstract principles, top-down policy frameworks, and philosophical speculation. This has produced a widening gap between ethical theory and practical application. In response, scholars have called for practitioner-centered research studies that center on the experiences, perceptions, and challenges of those who design, build, and implement AI systems. Such research is essential to translate ethical ideals into practice and to illuminate how developers navigate competing pressures, institutional norms, and moral uncertainty in their daily work.

Pant et al. (2024) conducted a grounded theory literature review of 38 empirical studies examining AI practitioners' views on ethics. Their analysis yielded a taxonomy of five categories: awareness, perception, need, challenge, and approach. This framework captures the complexity of how practitioners engage with ethical concerns in real-world contexts. While many express awareness of principles such as fairness, transparency, and accountability, they often struggle to operationalize these values due to vague guidance, limited institutional support, and conflicting organizational incentives. The study concludes that ethical implementation requires more than compliance. It demands cultural transformation, interdisciplinary collaboration, and ongoing reflexivity within development teams.

A central issue emerging from the literature is the principles-to-practice gap emphasizing the disconnect between normative frameworks and the realities of AI development. Herzog and Blank (2024) critique approaches that treat ethics as a checklist or outsource it to external experts, arguing that such strategies reduce moral reflection to procedural compliance. Instead, they propose a systemic, collaborative model that embeds ethical deliberation throughout organizational ecosystems. Their framework emphasizes inclusive dialogue, shared responsibility, and reflexive infrastructure that enables practitioners to deliberate ethically amid technical and commercial pressures.

Despite a proliferation of ethical guidelines from the OECD, UNESCO, and national governments, empirical data on how these principles are interpreted and applied within organizations remains limited. Giarmoleo et al. (2024), in a systematic review of over 300 articles, found that ethical challenges in AI tend to cluster around two domains: system design and human–AI interaction. Yet few studies investigate how practitioners resolve these challenges in practice, underscoring the need for context-specific and experience-based research that connects policy ideals with daily development realities.

The fragmentation of ethical standards across jurisdictions further complicates implementation. Ricciardi Celsi and Zomaya (2025) note that AI governance diverges widely across regions: the European Union emphasizes rights-based regulation, the United States prioritizes innovation and self-regulation, and China integrates ideological imperatives into its frameworks. Such variation creates uncertainty for practitioners operating in multinational contexts. The authors advocate for transdisciplinary approaches that integrate ethical, legal, and technical expertise, enabling practitioners to reconcile divergent standards while maintaining ethical integrity.

The lack of ethical literacy and formal training among practitioners compounds these challenges. Khan et al. (2023) surveyed AI engineers and lawmakers across 20 countries and found that although transparency, accountability, and privacy were widely recognized as core principles, few practitioners had received formal education in ethics. This deficit limits their capacity for informed moral reasoning and contributes to inconsistent implementation. The study recommends the creation of capability maturity models and ethics-aware development frameworks to guide the systematic integration of ethical reflection into technical workflows.

Lin (2025) extends this critique by arguing that conventional principlism and technological solutionism fail to resonate with practitioners. He proposes a user-centered, realism-inspired ethics framework emphasizing contextual relevance, utility, and actionability.

This approach aligns ethical guidelines with the practical realities of AI design, encouraging ownership and engagement among developers. Such practitioner-led ethics initiatives echo broader calls in cognitive science for situated decision-making, an acknowledgment that moral judgment, like cognition itself, is context-dependent and shaped by environmental constraints.

These studies reveal that practitioner-centered ethics is critical to advancing responsible AI. It provides insight into the lived experiences of those closest to technological decision-making and identifies the structural, cognitive, and institutional barriers that hinder ethical implementation. Bridging the principles-to-practice gap will require bottom-up engagement, interdisciplinary collaboration, and co-created tools that embed fairness, accountability, and transparency across the AI lifecycle. Only by centering practitioner perspectives can AI systems be designed and deployed in ways that are not merely innovative but also equitable, reflective, and socially responsible.

## Conclusion

The integration of artificial intelligence (AI) into critical decision-making systems across healthcare, criminal justice, employment, and education marks a pivotal transformation in how institutions operate and how individuals experience power, access, and opportunity. While AI is often introduced with the promise of efficiency, scalability, and objectivity, the case studies and theoretical analyses presented in this dissertation reveal a more complex reality. Far from being neutral tools, AI systems frequently replicate, amplify, and entrench existing social inequities particularly when trained on historical data that reflect systemic discrimination and when deployed without sufficient ethical safeguards (Noble, 2018; Barocas & Selbst, 2016).

In healthcare, AI diagnostic tools demonstrate great potential to improve clinical outcomes and streamline decision-making. Yet, as shown in Case Study 1, these systems often underperform for patients from marginalized racial and ethnic groups due to underrepresentation in training datasets and a failure to account for social determinants of health. Intersectionality theory reveals that these disparities are not merely additive but stem from intersecting dimensions of race, gender, class, and geography (Lett & La Cava, 2023). A Black woman living in a medically underserved area, for instance, may face compounded risks invisible to systems trained primarily on data from white, urban populations. Cognitive science highlights a complementary limitation: AI lacks the capacity for contextual reasoning and cannot account for the embodied, lived experiences that shape patient outcomes (Clark, 2013; Sawyer, 2014).

In the criminal justice system, explored in Case Study 2, predictive policing tools and risk assessment algorithms such as COMPAS have been shown to disproportionately label Black defendants as high-risk and to direct enforcement toward communities already subjected to over-policing (Angwin et al., 2016; Malik, 2023). These outcomes stem from datasets shaped by decades of racially biased policing and sentencing practices. Intersectionality provides a lens to understand how these systems erase or distort the experiences of individuals who exist at the intersection of multiple marginalized identities such as Black transgender individuals or Indigenous women (Lee, 2025). From a cognitive science perspective, the absence of moral reasoning and contextual interpretation within these systems further constrains their capacity to render fair or just decisions (Riva et al., 2024).

Case Study 3 examined AI-driven hiring systems, which continue to reproduce structural inequities under the guise of automation. Research shows that algorithmic résumé screening tools often privilege white male candidates while disadvantaging Black women and other intersecting identities, even when qualifications are comparable (Wilson & Caliskan, 2025; An et al., 2025). These models rely on proxies for "success such as elite educational backgrounds or specific linguistic patterns that are deeply entangled with privilege. Intersectionality exposes how such proxies systematically exclude those outside dominant cultural norms, while cognitive science highlights the potential value of human evaluators who, when trained effectively, can recognize potential beyond rigid metrics (Howard, 2021).

In higher education, as illustrated in Case Study 4, AI systems applied to college admissions frequently emphasize standardized test scores, advanced coursework, and refined personal essays, which disproportionately benefit students from affluent and predominantly white backgrounds while reinforcing institutional hierarchies of prestige (Eubanks, 2018; Zwick, 2019). These systems fail to recognize the structural and contextual barriers faced by students from under-resourced schools, first-generation applicants, and students of color. Intersectionality clarifies how such disadvantages accumulate for individuals who experience multiple forms of exclusion, while cognitive science highlights the importance of contextual evaluation recognizing potential through lived experience rather than decontextualized data (Mangal & Pardos, 2024).

Across all four domains, a consistent pattern emerges: AI systems are only as fair and effective as the data, design choices, and institutional contexts that shape them. Technical solutions such as fairness-aware algorithms and explainable AI remain essential but insufficient. Achieving equity in AI requires embedding ethical reflection, transparency, and intersectional awareness throughout the development process. Integrating cognitive science principles into system design also ensures that algorithmic decision-making accounts for the nuances of human reasoning, uncertainty, and perception (Howard, 2021; Tan et al., 2025).

The bias-variance tradeoff, discussed in depth in this dissertation, epitomizes the tension between technical optimization and ethical responsibility. Efforts to minimize bias often increase variance, leading to inconsistent outcomes for underrepresented groups. This dynamic is not merely a statistical constraint but a moral question that demands redefinition of "acceptable performance" considering justice and equity. Frameworks such as Rawlsian fairness help center the most disadvantaged, while intersectional analysis ensures that their experiences are understood in their full social and structural complexity. The task is not to perfect algorithms, but to realign the goals of AI with ethical commitments and community values.

Governance and accountability must evolve in parallel. While policy instruments like the EU AI Act and the Algorithmic Accountability Act represent meaningful progress, ethical oversight cannot rest solely on regulation. Governance must be participatory and inclusive, co-created with the communities most affected by algorithmic systems. Embedding intersectional analysis and cognitive science throughout the AI lifecycle from data collection and model design to deployment and evaluation offers a path toward systems that are both equitable and empirically sound. Institutional accountability should include transparent auditing practices, interdisciplinary ethics boards, and meaningful oversight mechanisms.

Practitioner-centered research further demonstrates that ethical intentions often collide with systemic constraints such as unclear guidelines, insufficient institutional support, and conflicting performance metrics. Bridging the gap between ethical principles and technical practice requires interdisciplinary collaboration, ethics-aware development frameworks, and organizational cultures that prioritize accountability. Equipping practitioners with ethical literacy, decision-making tools, and supportive governance structures is essential for sustaining fairness in practice.

Future research must address ongoing gaps, including the lack of diversity in training datasets, the opacity of deep learning models, and the underrepresentation of intersectional identities in model evaluation. Advancements in explainable AI, fairness-aware learning, and hybrid human-AI decision systems will be essential for building technologies that are not only technically robust but also socially just. Continued research should also explore how cognitive science and intersectionality together can inform the design of systems capable of engaging with the complexity of human experience.

In conclusion, the use of AI in healthcare, criminal justice, employment, and education represents both an extraordinary opportunity and a profound ethical challenge. These systems can support more consistent and informed decision-making, but without intentional safeguards, they risk deepening existing inequalities. The findings of this dissertation highlight the intricate interplay between systemic structures and algorithmic outcomes. They call for a reimagining of AI development, one framed through the lens of intersectional analysis, informed by cognitive science, and accountable to the communities it serves. This reimagining requires robust governance, practitioner engagement, and a commitment to justice that transcends technical optimization. Chapter Three outlines the methodological approach used to examine how AI practitioners experience and interpret bias in their work.

# Chapter Three

## Procedures and Methodology

### Introduction

This chapter describes the research methodology and procedures used in this qualitative study. It outlines the research design, philosophical orientation, participant selection, data collection methods, and analytic strategies guiding the investigation. The purpose of this study was to explore how practitioners working with AI and ML systems understand the emergence of bias, interpret its societal impacts, and identify strategies for advancing equity in the design and deployment of these systems. The methodological decisions outlined in this chapter were designed to address the study's research questions and provide a structured, transparent approach for examining practitioners' experiences.

The study was grounded in a constructivist-interpretivist paradigm, which assumes that knowledge is shaped through social interaction and context. This orientation aimed to explore how practitioners assign meaning to fairness, accountability, and bias within their professional environments. A qualitative approach was selected because it enables a detailed examination of practitioner perspectives and the contextual factors that shaped decision-making in AI development. This chapter explains how this approach was operationalized through semi-structured interviews and document analysis, and how these methods were used to capture both individual experiences and organizational influences relevant to the research questions.

Intersectionality Theory and Cognitive Science served as guiding concepts in the development of interview questions and the design of the analytic plan. These frameworks informed the deductive components of the coding process but did not prescribe the findings. The chapter, therefore, focuses on the methodological procedures rather than on theoretical interpretation, detailing how data was collected, prepared, coded, and analyzed using a combination of inductive and deductive thematic techniques.

This chapter also describes the sampling strategy, recruitment procedures, and steps taken to ensure the ethical conduct of research, including informed consent, confidentiality, and secure data management. Procedures for establishing trustworthiness, such as triangulation, reflexive memoing, and member checking, are outlined to demonstrate methodological rigor. The chapter concludes by addressing practical limitations related to sampling, scope, and transferability.

Overall, this chapter provides a clear and systematic account of how the study was designed and conducted, establishing the methodological foundation for the findings presented in Chapter Four.

**Research Method and Paradigmatic Perspective**

This study was situated within a constructivist–interpretivist paradigm, which provided the philosophical foundation for its qualitative design. This paradigm assumes that meaning is socially constructed and shaped through experience, which aligns with the study's aim to explore how practitioners understand and describe bias, fairness, and accountability in their professional contexts. A constructivist orientation supports inquiry focused on subjective interpretation, while interpretivism emphasizes the importance of context, interaction, and participant perspectives in generating understanding.

This paradigmatic perspective informed the decision to use semi-structured interviews and document analysis as the primary methods of data collection. These methods were appropriate for eliciting detailed accounts of practitioners' experiences and for examining how participants interpreted ethical considerations in their work. The interpretivist emphasis on co-constructed meaning shaped the interview approach, allowing participants to elaborate on their reasoning, experiences, and decision-making processes. The constructivist stance supported thematic analysis by guiding attention to patterns in how participants constructed meaning related to algorithmic bias and fairness.

The paradigm also informed researcher reflexivity throughout the study. Constructivist–interpretivist inquiry acknowledges that the researcher plays an active role in interpreting data; therefore, reflexive memoing and ongoing attention to positionality were incorporated into the analytic process to enhance transparency. This philosophical orientation provided a consistent foundation for the methodological choices that follow, including participant selection, data collection procedures, and thematic analysis.

**Qualitative Research Approach**

This study is grounded in a constructivist–interpretivist research paradigm, which provides the philosophical foundation for its qualitative design. Constructivism states that knowledge is socially constructed and shaped by context, making it well-suited for examining how practitioners understand fairness, bias, and accountability in the development of AI (Creswell & Poth, 2018). These assumptions align with the purpose of this study, which seeks to explore how AI professionals interpret and make sense of ethical concerns in their everyday work, rather than measuring predefined variables or testing hypotheses.

Constructivism views reality as multiple and co-created through interaction. From this standpoint, practitioner experiences with AI systems are not treated as objective facts but as interpretations shaped by their organizational roles, professional histories, and social environments (Guba & Lincoln, 1994). This paradigm allows the study to examine how practitioners' understandings of bias and fairness emerge through their engagement with real-world projects, decision processes, and institutional expectations.

Interpretivism complements this perspective by emphasizing that meaning is created through dialogue and subjective interpretation (Schwandt, 2000). In this study, interpretivism is reflected in the use of semi-structured interviews that invite participants to describe their reasoning, experiences, and reflections in their own words. This orientation supports the goal of understanding how ethical considerations are defined, negotiated, and acted upon in practice, recognizing that these meanings vary across individuals and contexts.

The constructivist–interpretivist paradigm also informs the study's emphasis on reflexivity and relational engagement throughout the data collection and analysis process. Because the researcher serves as the primary instrument of interpretation, this paradigm requires awareness of how positionality and prior experience shape the analytic process. Reflexive memoing and iterative analysis were employed to maintain transparency and ensure interpretive rigor.

Overall, this paradigm provides an appropriate foundation for a qualitative inquiry into how AI practitioners understand and navigate algorithmic bias. It supports the study's focus on meaning-making, context, and subjective interpretation, and aligns directly with the methodological decisions described in subsequent sections, including sampling, interview design, and thematic analysis.

## Multi-Case Study Methodology

This study employed a qualitative multi-case study design to examine how professionals who design, deploy, or oversee artificial intelligence (AI) systems understand, experience, and respond to algorithmic bias. A multi-case approach was selected because it provides a structured way to explore a shared phenomenon across multiple contexts while allowing for depth and variation in practitioner experiences (Yin, 2018). This design aligns with the study's purpose and research questions by enabling analysis of practitioner accounts in relation to documented examples of bias in high-impact societal domains.

The cases in this study represent four sectors where algorithmic bias has been widely reported: healthcare, education, employment, and criminal justice. These cases were not developed through organizational fieldwork; instead, each case was constructed from publicly available reports, scholarly analyses, and existing documentation. Using secondary case materials allowed the study to situate practitioner narratives within real-world examples of algorithmic harm while maintaining consistency with a qualitative design focused on interpretation rather than causal inference.

The researcher collected from three sources: (a) semi-structured interviews with AI practitioners, which served as the primary data; (b) documentary materials, including regulatory guidance, institutional reports, and scholarly literature; and (c) secondary case studies derived from published accounts of algorithmic bias. Triangulation across these sources enhanced credibility and provided a basis for comparing practitioner interpretations with documented patterns of bias. Interviews offered direct insight into how professionals conceptualize fairness and navigate ethical tensions. Documentary analysis provided contextual information about governance expectations and institutional environments. The sector cases illustrated how algorithmic bias manifests in real applications.

Data analysis followed a two-stage structure. Within-case analysis was used to examine each sector individually, identifying the contextual and structural factors relevant to that domain. Cross-case analysis was then used to identify patterns, contrasts, and recurring features across cases. Interview transcripts were coded thematically using NVivo software, employing both inductive codes that emerged from the data and deductive codes informed by the study's theoretical frameworks of Intersectionality and Cognitive Science. Reflexive memoing and analytic journaling accompanied the coding process to ensure transparency and to document interpretive decisions.

The theoretical frameworks guided the analytic focus by directing attention to structural and cognitive dimensions of AI bias. Intersectionality provided a basis for examining how institutional and societal conditions shape the development and deployment of AI systems. Cognitive Science informed analysis of how practitioner reasoning, heuristics, and perception influenced their interpretations of fairness and accountability. These frameworks structured the deductive coding scheme and supported the interpretation of themes across cases.

The researcher maintained methodological rigor through multiple strategies, including data triangulation, member checking, an audit trail, and reflexive practice. Participants reviewed their interview transcripts to verify accuracy, and the analytic process was documented to support dependability and confirmability. A detailed description of cases and participant contexts enhances the transferability to similar sociotechnical environments.

Overall, the multi-case study design supported a comprehensive examination of how AI practitioners interpret algorithmic bias and how those interpretations intersect with evidence of bias across multiple sectors. This methodological structure aligns directly with the study's purpose and research questions by integrating practitioner experiences with documented examples of AI impact.

**Trustworthiness**

For this study, trustworthiness was established through four interrelated criteria: credibility, transferability, dependability, and confirmability. Each criterion was addressed through deliberate methodological strategies designed to maintain rigor, transparency, and alignment with the study’s multi-case design and interpretivist orientation.

**Credibility.** Credibility in this study was strengthened through prolonged engagement with the data, triangulation, member checking, and reflexive practices. Triangulation was achieved by integrating three data sources: semi-structured interviews, documentary analysis, and sector-based case materials, which allowed patterns to be cross-verified across multiple forms of evidence. Members checking further supported credibility by allowing participants to review and confirm the accuracy of their interview transcripts and clarify or expand their responses. Prolonged engagement with the data occurred through iterative coding cycles and repeated comparison of codes and themes against the raw data. Reflexive journaling accompanied all stages of analysis, enabling the researcher to document interpretive decisions and maintain awareness of potential bias. Together, these procedures ensured that the findings were grounded in participants’ intended meanings and consistently aligned with the evidence.

**Transferability.** Selecting cases from four sectors, healthcare, education, employment, and criminal justice, allowed the study to capture variation in the environments where algorithmic bias occurs. This contextual detail, combined with direct participant quotations and clear documentation of decision-making conditions, offers readers sufficient information to determine whether the findings may be relevant to their own sociotechnical settings.

**Dependability.** In this study, the researcher established dependability by ensuring that all methodological steps, from data collection to analysis, were implemented systematically and aligned with the study's purpose and its three research questions. This included maintaining clear documentation of how data were generated, managed, and interpreted, consistent with qualitative standards that emphasize procedural transparency and fidelity to method (Nowell et al., 2017).

To ensure dependability, the researcher maintained a comprehensive audit trail throughout the study. The audit trail included recruitment records, informed consent documentation, interview schedules, interview protocols, field notes, researcher memos, and all versions of coding frameworks. Each iteration of the interview guide and analytic structure was documented using version control, allowing for a clear record of how the researcher refined the instruments and procedures to ensure alignment with the study's purpose and research questions. This process ensured that every step in data collection and analysis was both traceable and replicable.

Consistency in data collection was maintained using a standardized, semi-structured interview protocol applied uniformly across all participants. The protocol directly reflected the study's research questions and theoretical foundations, providing a stable structure for eliciting comparable data across interviews while allowing participants to elaborate on their own experiences. Interview procedures, including opening scripts, probing strategies, and closing prompts, were consistently applied to minimize variability unrelated to the research aims.

Data management and analysis procedures were also stabilized to support dependability. NVivo software was used to organize transcripts, store codes, and document analytic decisions, ensuring that each coding action was recorded and linked back to the original data. The coding process followed a consistent sequence: initial open coding, application of deductive codes informed by the theoretical framework, categorization, and theme development. Reflexive memoing accompanied each stage to document analytic decisions, evolving interpretations, and links to the research questions. These memos provided a running record of how interpretations were formed and verified, supporting transparency and methodological clarity.

These practices ensured that data collection and analysis remained consistent, transparent,

and aligned with the methodological commitments described in Chapter Three. This level of documentation enables external reviewers to trace the logic of the study and confirm that findings emerged from a coherent and dependable research process.

**Confirmability.** Reflexivity played a central role in maintaining the confirmability of the findings. Throughout the research process, a reflexive journal was maintained to document methodological decisions, analytic reasoning, and reflections on how the researcher's positionality could influence interpretation. This journal served as a continuous self-audit, recording how interpretations developed across coding cycles and how those interpretations were checked against the study's research questions and theoretical framework.

Triangulation across the study's three data sources—semi-structured interviews, documentary materials, and sector-based case documentation—also strengthened confirmability by ensuring that themes were supported by multiple forms of evidence. NVivo software was used to maintain an organized and transparent record of all coding actions, including links between codes, analytic memos, and verbatim quotations. This created a traceable chain of evidence from raw data to final themes, ensuring that interpretations remained grounded in participant accounts and consistent with the analytic plan described in this chapter.

Confirmability was further enhanced through peer debriefing with the dissertation chair and methodologist at key stages of data analysis. These discussions provided external review of emerging themes, challenged interpretive assumptions, and ensured that conclusions were supported by the data and aligned with the study's research questions. Taken together, these strategies created a transparent and verifiable process that grounded the findings in the evidence collected, supporting confidence that the results reflect participant perspectives rather than researcher bias.

**Role of the Researcher**

Within this constructivist–interpretivist study, my role centered on ensuring that all stages of the research process remained aligned with the problem statement, purpose, and research questions established in Chapters One and Two. This required maintaining transparency in how data were collected, interpreted, and connected to the study's methodological framework.

The researcher's responsibilities included designing the study, developing the interview protocol, recruiting participants, conducting interviews, managing documentary materials, and completing the analytic procedures described in this chapter. At each stage, the researcher monitored how disciplinary background, professional experience in AI, and personal values could influence interpretation. To manage this influence, the researcher maintained a reflexive journal documenting methodological decisions, justifications for coding choices, and reflections on how their assumptions might shape interpretation. This practice served as a structured mechanism to ensure that findings remained grounded in participants' accounts rather than the researcher's presuppositions.

During data collection, the researcher adhered to a standardized semi-structured interview protocol to ensure consistency across participants while allowing flexibility for participants to elaborate on their experiences. This approach supported alignment with the research questions by ensuring that each interview addressed the core constructs of fairness, bias, and accountability. Their interactions with participants emphasized open-ended inquiry and neutral facilitation, minimizing the risk of leading questions or interpretive influence during data collection.

The researcher's role in documentary analysis also required reflexive attention. Policy reports, technical documentation, and case materials were analyzed using the same deductive–inductive structure applied to interview data. These sources were treated as contextual evidence supporting the multi-case design rather than as primary data. This distinction was documented in the analytic memos to preserve methodological fidelity and avoid conflating documentary interpretations with participant perspectives.

During the analytic phase, the researcher used NVivo to maintain a transparent record of coding, theme development, and cross-case comparisons. Deductive codes derived from Intersectionality Theory and Cognitive Science were applied systematically, and inductive codes were generated directly from participant language. Memoing practices documented how interpretations evolved and how decisions were linked back to the study's research questions and theoretical framework. Peer debriefing with the dissertation chair and methodologist provided additional external review to ensure that interpretations reflected the data and aligned with the study's purpose.

Ethical responsibilities were consistently upheld throughout the study, including obtaining informed consent, ensuring confidentiality protection, maintaining secure data management, and providing clear communication to participants regarding their rights. These practices supported procedural transparency and ensured that the relational aspects of qualitative research did not compromise methodological rigor.

Overall, the researcher's role in this study was to facilitate a research process that was reflexive, transparent, and aligned with the methodological commitments outlined in this chapter. By documenting decisions and monitoring the influence of positionality, the researcher aimed to ensure that the findings represent participants' experiences and perspectives while maintaining fidelity to the study's purpose, research questions, and qualitative design.

**Researcher Positionality**

In this study, positionality was monitored throughout all stages of the research process to maintain alignment with the research problem, purpose, and research questions, and to ensure fidelity to the qualitative multi-case design.

The researcher's academic and professional background in artificial intelligence, data governance, and project management informed their ability to understand the sociotechnical environments in which the participants work. This background provided familiarity with AI workflows and organizational structures; however, it also carried the risk of interpreting participant narratives through the assumptions and norms common in the technology field. To manage this, the researcher used structured reflexive practices, including a reflexive journal, analytic memos, and peer debriefing, to document where their assumptions might influence meaning-making, and to ensure that interpretations were grounded in the data rather than in prior professional experience.

During data collection, positionality was managed by adhering to a standardized, semi-structured interview protocol that was designed through field testing and expert review. The protocol ensured that all participants were asked questions aligned with the research questions while still allowing for open-ended elaboration. Maintaining consistency in how questions were posed minimized the researcher's background influence on the direction of the interviews. Reflexive notes were documented after each interview to record potential moments of bias, emotional reactions, or interpretive assumptions that required reconsideration during analysis.

Regarding the participants, a neutral and facilitative stance was maintained. Participants were treated as knowledgeable informants rather than subjects, aligning with the interpretivist paradigm's emphasis on co-constructed meaning. Awareness of power dynamics, particularly when interviewing participants with greater technical seniority or institutional authority, guided communication choices and helped ensure that responses were participant-driven rather than shaped by researcher expectations.

The researcher drew attention to issues of structural inequity and representation, central concerns in this study's application of Intersectionality Theory. While this perspective shaped interest in the topic, it was managed analytically by applying a consistent coding procedure, using both inductive and deductive coding, and grounding interpretations in explicit participant statements rather than personal assumptions. Theoretical commitments were noted in analytic memos to maintain awareness of how they informed themed development during coding and cross-case analysis.

Throughout the analysis, NVivo was used to create a transparent record of how codes, categories, and themes were developed, revised, and connected to data. Version-controlled coding files, memo trails, and peer debriefing sessions ensured that positional influences were documented and critically examined. These processes supported confirmability by demonstrating that findings emerged from systematic analysis rather than from researcher bias.

**Sampling Procedures and Data Collection Sources**

This section outlines the procedures used to identify study participants, obtain permissions, and collect the data necessary to address the research questions guiding this qualitative multi-case study. Because the purpose of the study was to understand how AI practitioners perceive the origins of bias, interpret its societal impacts, and describe strategies for equitable system design, data collection methods were intentionally selected to ensure direct alignment with RQ1–RQ3.

## Sampling Strategy

The target population for this study consisted of professionals who directly shape design, development, deployment, or governance of AI systems. Due to research questions focusing on how practitioners perceive the origins of algorithmic bias (RQ1), interpret its societal impacts (RQ2), and identify strategies for equitable system design (RQ3), participants were not required to work within the four sectors used for the secondary case materials. Instead, inclusion was based on their professional responsibility for decisions that influence model behavior, data practices, or ethical oversight.

Participants were recruited using purposive sampling to identify information-rich experts who could speak to bias in AI development (Patton, 2015). Snowball sampling supplemented this process by allowing participants to refer additional qualified individuals. Table 1 identifies the study's data source and collection methods. Recruitment was conducted through targeted outreach on LinkedIn, utilizing standardized messaging to ensure consistency and transparency. Participants were identified using Boolean search phrases such as "Responsible AI" or "Ethical AI" and contacted using an IRB-approved direct message. Inclusion criteria required participants to have (a) a minimum of three years of professional experience in AI development, deployment, or governance, and (b) direct or indirect influence on system design, data workflows, or algorithmic decision processes. Individuals with limited experience in interacting with AI systems as end-users were excluded. Prior to all recruitment activity, Institutional Review Board (IRB) approval was obtained.

A total of nine participants were recruited, which aligned with the study's goal of achieving thematic saturation. Saturation was reached when successive interviews no longer yielded new concepts or insights relevant to the research questions, consistent with the guidance for qualitative multi-case studies (Saunders et al., 2018). The final sample represented a diverse range of organizational contexts, including private-sector technology firms, academic institutions, and public-sector agencies. This variation supported the multi-case design by ensuring that practitioner perspectives could be interpreted across multiple institutional settings.

**Table 1**

*Data Sources and Collection Methods*

| **Data Source** | **Purpose** | **Type of Data Collected** | **Analytic Contribution** |
|---|---|---|---|
| Semi-structured Interviews | Capture firsthand practitioner experiences and interpretations of AI bias. | Verbal narratives, reflections, and perceptions were collected from professionals. | Supports thematic analysis of individual and cross-case experiences, contributing to understanding ethical reasoning in AI contexts. |
| Case Studies (Healthcare, Criminal Justice, Education, Employment) | Examine real-world examples of AI bias and institutional response across four high-impact sectors. | Archival records, published reports, policy analyses, and documented outcomes. | Provides contextual grounding for interpreting interview data and identifying sector-specific challenges in algorithmic fairness. |
| Documentary Analysis | Identify dominant narratives, governance | Organizational policies, public | Triangulates findings across sources and enhances |

| Data Source | Purpose | Type of Data Collected | Analytic Contribution |
|---|---|---|---|
| | structures, and technical standards influencing AI ethics. | datasets, and scholarly articles on AI accountability and fairness. | interpretive depth in relation to institutional practices. |

Semi-structured interviews served as the primary data collection method because they allowed participants to describe their professional experiences, perceptions of bias, and ethical decision-making processes in their own words. Interviews were conducted virtually via Microsoft Teams between October 6 and October 30, 2025, and lasted approximately 45–60 minutes. All interviews followed a standardized IRB-approved protocol to ensure alignment with the study's three research questions. The interview guide was field-tested with two professionals who met the inclusion criteria; their feedback led to refinements in question sequencing and clarity, thereby enhancing methodological fidelity. With participant consent, interviews were audio-recorded and transcribed verbatim by the researcher. Participants reviewed their transcripts as part of the member-checking process to ensure accuracy and completeness.

Documentary analysis was conducted concurrently to contextualize practitioner accounts and triangulate findings. Documents included regulatory guidance, organizational policy statements, scholarly literature, and technical reports. Inclusion criteria required that documents (a) address algorithmic bias, fairness, or governance, (b) pertain to one of the four sectors used in the study's case materials, or (c) demonstrate how institutions frame or respond to bias in AI. These sources provided sector-level and institutional context for RQ2 and RQ3.

The study also incorporated secondary case materials representing the healthcare, criminal justice, education, and employment sectors. These cases were derived from publicly available reports, prior scholarly analyses, technical audits, and investigative journalism. They did not serve as primary data sources; instead, they functioned as contextual cases within the multi-case study design. Their purpose was to illustrate documented manifestations of algorithmic bias and support cross-case triangulation with practitioner narratives. This structure ensured fidelity to the methodological design and alignment with the study's purpose and research questions.

Data retention and confidentiality procedures were followed in accordance with institutional policy. Interview recordings and transcripts were stored on an encrypted, password-protected device. Participants were assigned pseudonyms, and all identifiable information was removed from transcripts and reports. Audio recordings will be deleted upon completion of the dissertation, while de-identified transcripts will be stored securely for a period of three years, in accordance with university guidelines.

**Instrument Development**

This study employed a researcher-developed semi-structured interview protocol as the primary data collection instrument, supplemented by documentary and case study analysis guides. The instrument was designed to elicit in-depth reflections from professionals directly involved in the development, design, or governance of artificial intelligence (AI) systems. Its purpose was to explore how these practitioners perceived, interpreted, and addressed algorithmic bias within their professional contexts, aligning closely with the study's research questions and theoretical frameworks. Prior to use, the interview instrument underwent a formal field test and review process to ensure clarity, alignment, and content validity. Institutional Review Board (IRB) approval for the field test and instrument development was granted and is documented in Appendix C. Following successful field validation, full IRB approval for the research study, including participant recruitment and data collection procedures, was obtained and is included in Appendix A. The finalized version of the interview protocol is presented in Appendix D, and the documentary and case study analysis templates are provided in Appendices E and F, respectively.

The development of the instrument was informed by both the literature reviewed in Chapter Two and the theoretical foundations of Intersectionality and Cognitive Science. Intersectionality provided the conceptual grounding for exploring how overlapping social identities, institutional hierarchies, and systemic inequities influence professional experiences with AI bias (Bilge & Collins, 2016; Crenshaw, 1991). This framework guided the inclusion of questions addressing the social and organizational dynamics of AI development, particularly how practitioners perceive fairness, inclusion, and representational harm in data and model design. Cognitive Science complemented this framework by emphasizing the perceptual and reasoning processes involved in technical decision-making (Clark, 2013; Sawyer, 2014). It informed questions about how participants recognize, interpret, and respond to bias in algorithmic systems, particularly when such bias emerges from cognitive heuristics or decision shortcuts embedded in human-AI interaction.

A field test was conducted with two professionals who met the study's inclusion criteria. These pilot interviews evaluated the clarity, tone, and comprehensiveness of the instrument. Based on participant feedback, minor revisions were made to enhance flow and ensure that each question resonated with real-world professional experiences. Version control documentation was maintained throughout this process, capturing all modifications and their rationale to support transparency and replicability.

The interview instrument was constructed using best practices for qualitative research design (Creswell & Poth, 2018; Patton, 2015). Initial drafts of the protocol were derived from recurrent themes in the literature on AI ethics, fairness, and governance (e.g., Mitchell et al., 2021; Mittelstadt, 2019). These themes were mapped onto the study's three overarching research questions, ensuring conceptual alignment between theory and inquiry. The instrument was then subjected to an expert panel review, including a methodologist and a subject matter expert in AI ethics. Their feedback informed revisions to question sequencing, phrasing, and the inclusion of probing sub questions designed to elicit richer, more reflective responses.

In addition to the interview protocol, the study employed two supplemental instruments: a Document Analysis Guide and a Case Study Analysis Template. The document analysis guide provided structured criteria for selecting and coding institutional documents such as policy briefs, technical standards, and governance reports. The case study analysis template was used to extract and analyze information from four selected cases, one from each of the targeted sectors—healthcare, criminal justice, education, and employment. These instruments supported triangulation by contextualizing interview findings within documented real-world instances of AI bias and organizational response.

To ensure that the instruments were theoretically and empirically aligned, Table 2 was created linking each major research question to specific sections of the interview protocol and supporting data sources. This ensured comprehensive coverage of all inquiry domains and enhanced internal consistency across methods.

**Table 2:**

*Mapping Research Questions to Data Collection Instruments and Data Sources*

| **Research Question** | **Instrument / Data Source** | **Purpose of Instrument** | **Analytic Focus** |
|---|---|---|---|
| RQ1: How do professionals involved in AI development perceive the root causes of algorithmic bias? | Semi-Structured Interview Protocol (Appendix A) | Explore personal and organizational experiences shaping awareness of bias. | Thematic coding of narratives related to data quality, model design, and human decision-making. |
| RQ2: How do practitioners describe the societal implications of AI bias across high-impact sectors? | Case Study Analysis Template (Appendix C) | Examine real-world sectoral impacts (healthcare, criminal justice, education, employment). | Comparative analysis across four cases highlighting downstream effects of bias. |
| RQ3: How do professionals and institutions respond to and mitigate bias in AI systems? | Document Analysis Guide (Appendix B) | Review of policies, governance frameworks, and ethical guidelines. | Triangulation of participant reflections with institutional responses to fairness and accountability. |

See appendix for full table that provides a mapping of research questions to corresponding interview questions, demonstrating how each aligns conceptually with the study's aims and theoretical foundations

**Participant Recruitment**

Participants for this study were accessed through professional networks, online communities, and targeted outreach on professional platforms. The recruitment process was designed to ensure transparency, inclusiveness, and compliance with ethical research standards. The study received Institutional Review Board (IRB) approval prior to any recruitment or data collection activities, and all materials, including recruitment scripts, informed consent forms, and interview protocols—were reviewed and approved by the university's IRB. The official IRB Approval Letter and Informed Consent Form are included in Appendices A and B of this dissertation.

The recruitment process began with purposive sampling to identify individuals who met the study's inclusion criteria. Eligible participants were professionals with at least three years of direct experience in the development, governance, or ethical oversight of artificial intelligence (AI) systems. This included roles such as data scientists, AI developers, machine learning engineers, product managers, policy advisors, and ethics officers. Recruitment focused on individuals who influence AI design and deployment across various institutional contexts, including technology firms, academic research organizations, non-profit advocacy groups, and government agencies.

Potential participants were identified through a multi-step process. First, professional networks and public directories were searched using keywords such as "AI ethics," "machine learning engineer," "responsible AI," and "data governance." These searches were conducted within LinkedIn's advanced filters to locate qualified professionals across diverse sectors and geographic regions. Once potential participants were identified, they were contacted directly via LinkedIn's private messaging feature using a standardized recruitment script approved by the IRB. The script briefly introduced the researcher, described the purpose of the study, and invited the recipient to consider participating in an interview.

If an individual expressed interest, a follow-up email was sent containing the informed consent form and additional details about the study. The informed consent form outlined the research purpose, voluntary nature of participation, potential risks and benefits, confidentiality protections, and the right to withdraw at any point without penalty. Participants were asked to indicate their consent electronically by selecting "Yes" at the bottom of the form. Upon receiving consent, the researcher provided a scheduling link through Calendly that allowed participants to select an interview time at their convenience.

Recruitment continued until thematic saturation was achieved. A total of nine participants were enrolled in the study, representing diverse professional backgrounds and perspectives. Snowball sampling was used to supplement purposive recruitment. At the end of each interview, participants were invited to recommend additional professionals who might provide valuable insights based on the study's focus. These referrals were vetted using the same inclusion criteria to ensure that all participants possessed relevant expertise and experience.

No formal site permissions were required for this study, as all data collection occurred virtually and participants were engaged in their individual professional capacity rather than as representatives of their employing organizations. All interviews were conducted using Microsoft Teams, a secure platform that allowed for real-time communication and recording with participant consent. Each session lasted approximately 45 to 60 minutes, and follow-up interviews were scheduled as needed to clarify or expand upon prior responses.

Throughout the recruitment process, confidentiality and ethical standards were maintained rigorously. Participant names and organizational affiliations were anonymized immediately after consent and replaced with pseudonyms for all transcripts and written materials. Contact information and consent records were stored securely on a password-protected device accessible only to the researcher.

This structured recruitment and access process ensured transparency, replicability, and ethical integrity. By combining purposive and snowball sampling strategies and conducting all recruitment through secure professional channels, the study was able to engage qualified participants while safeguarding their privacy and autonomy. The approved IRB procedures and documentation, included in the appendices, provide full evidence of ethical compliance and participant protection.

**Data Collection**

Data for this qualitative multi-case study was collected from three primary sources: semi-structured interviews, documentary materials, and secondary case study evidence. These sources were selected to enable methodological triangulation, allowing the researcher to examine the phenomenon of algorithmic bias in artificial intelligence (AI) from multiple perspectives. The data collection process was carried out in accordance with the approved Institutional Review Board (IRB) protocol to ensure ethical integrity, methodological alignment, and transparency.

The primary data source consisted of semi-structured interviews with professionals who directly influence the development, governance, or ethical oversight of AI systems. The researcher developed a detailed interview guide (Appendix A) based on the study's research questions, theoretical frameworks, and the literature reviewed in Chapter Two. The guide included twelve core questions with corresponding probes designed to elicit rich, narrative responses about how practitioners perceive, experience, and address bias in AI systems. Questions were organized thematically to align with the study's three central research questions:

understanding the emergence of bias, exploring its societal impacts, and identifying strategies for mitigation and accountability.

Interviews were conducted virtually through Microsoft Teams, providing a secure and accessible environment for participants across geographic locations. Each session lasted approximately 45 to 60 minutes, with follow-up interviews conducted as needed for clarification or elaboration. Participants were informed that their participation was voluntary, and consent was obtained electronically prior to scheduling. All interviews were audio-recorded with explicit participant permission and subsequently transcribed verbatim by the researcher.

At the start of each interview, participants were asked to confirm their consent and were reminded of the confidentiality procedures in place. Basic demographic information was collected, including job roles, years of experience in AI-related work, and general industry or sector affiliation. This information was used solely for contextual analysis and is reported in aggregate form to preserve anonymity.

Following transcription, participants received a copy of their transcript via email for member checking. They were invited to review the document and provide corrections, clarifications, or elaborations as desired. This step enhanced credibility by ensuring that participant perspectives were accurately represented and that interpretations remained grounded in authentic narratives.

Documentary data were collected concurrently to contextualize and triangulate the insights gathered from interviews. Relevant documents included publicly available reports, technical standards, policy briefs, and organizational statements related to AI ethics and governance. These materials were accessed through credible sources such as ProPublica, the Stanford Human-Centered Artificial Intelligence (HAI) Annual Report, and case-specific reports published by recognized institutions and regulatory bodies.

Documents were selected using predefined inclusion criteria: relevance to AI bias or fairness, publication within the last five years, and direct connection to one of the four focal sectors (healthcare, education, criminal justice, and employment). Documents were analyzed using a structured document analysis guide (Appendix E), which included criteria for coding themes, identifying key discourses, and noting institutional responses to bias-related issues. This analysis provided contextual depth and allowed for comparison between practitioner narratives and institutional documentation of ethical AI practices.

The third data source consisted of four sector-specific case studies, each illustrating real-world examples of bias in AI implementation. One case was selected for each sector (healthcare, criminal justice, education, and employment) based on its public documentation, relevance to the study's focus, and potential to demonstrate the societal consequences of algorithmic bias. Examples included cases of algorithmic discrimination in hiring systems, predictive policing, automated grading software, and diagnostic tools.

Case study data were obtained from peer-reviewed journal articles, credible media investigations, and official reports from government and industry sources. The case study analysis template (Appendix C) guided the extraction of relevant information, including the background of the system, the nature of the bias identified, the affected populations, and subsequent policy or design responses. These cases provided empirical grounding for understanding how practitioner insights align or diverge from observable real-world outcomes.

All data, including interview recordings, transcripts, coded documents, and case materials, were securely stored on the researcher's password-protected computer. NVivo 14 software was used to organize and code all qualitative data. Each data source was categorized by type (interview, document, or case study) and linked to corresponding research questions within NVivo to facilitate cross-source analysis.

Participant identities were anonymized through the use of pseudonyms, and all identifying details were removed from transcripts and notes. Audio recordings will be permanently deleted after the completion of the dissertation, while de-identified transcripts and analysis files will be securely retained for three years in compliance with university policy.

The data collection process was designed to align with the study's multi-case qualitative methodology and to ensure credibility through triangulation of diverse sources. The integration of semi-structured interviews, documentary analysis, and sector-based case studies provided a comprehensive view of how bias in AI systems is experienced, contextualized, and addressed. The systematic procedures used, from recruitment to transcription and analysis, support transparency, replicability, and ethical rigor, enabling future researchers to replicate or extend this investigation within similar contexts.

**Data Analysis Procedures**

This study employed thematic analysis as the primary analytic method, supported by procedures recommended by Braun and Clarke (2006). Thematic analysis was chosen because it allows for systematic identification of patterns within qualitative data while maintaining flexibility to accommodate diverse data sources such as interviews, policy documents, and case studies. The analysis followed a multi-stage approach that combined inductive and deductive coding strategies. Inductive analysis allowed themes to emerge organically from participants' narratives, while deductive analysis was guided by the study's theoretical frameworks of intersectionality and cognitive science.

All qualitative data, including interview transcripts, documents, and case study materials, were uploaded into NVivo 14 software for organization and coding. NVivo facilitated the management of large volumes of textual data, allowing the researcher to categorize excerpts, identify relationships among codes, and track theme development.

**Step 1: Data Familiarization**

The first stage of analysis involved data familiarization. The researcher read each interview transcript and document multiple times to gain an in-depth understanding of the content. During this stage, reflective notes and analytic memos were recorded in a research journal to capture initial impressions, emerging questions, and potential connections between the different data sources.

**Step 2: Initial Coding**

In the second stage, open coding was conducted to identify key words, phrases, and concepts within the data. Coding was applied line-by-line for interview transcripts and paragraph-by-paragraph for documentary and case study materials. Each code represented a distinct idea, observation, or issue related to the study's central focus on bias in AI systems. This phase prioritized inclusivity, ensuring that all relevant data were captured without premature categorization.

**Step 3: Code Organization and Theme Development**

Once initial codes were developed, they were organized into broader categories to identify recurring ideas and patterns across participants and data sources. The researcher employed both within-case and cross-case analysis as recommended by Yin (2018) for multi-case study designs. Within-case analysis allowed the researcher to understand how bias and fairness were conceptualized within each sectoral case, while cross-case analysis identified similarities and differences across the four sectors. This comparative process helped reveal how systemic, institutional, and individual factors intersect in shaping experiences with AI bias.

**Step 4: Thematic Refinement and Triangulation**

In the fourth stage, codes were clustered into potential themes that represented significant patterns of meaning. These themes were reviewed and refined through iterative comparisons across interviews, documents, and case studies. Triangulation was used to validate themes and strengthen credibility by comparing insights from different data types and participant perspectives (Lincoln & Guba, 1985). For example, practitioner accounts of biased system design were cross-checked against documentary evidence of algorithmic audits and case studies illustrating real-world harm. This process ensured that themes reflected both experiential and contextual dimensions of the phenomenon.

**Step 5: Interpretation and Synthesis**

The final stage of analysis involved synthesizing the identified themes into a coherent narrative that addressed the study's three research questions. Interpretations were guided by the study's theoretical frameworks. Intersectionality provided a lens for understanding how systemic inequities influence AI development and decision-making, while cognitive science informed interpretations of how practitioners perceive, reason, and respond to algorithmic bias. The result was a multi-layer interpretation that connected individual experiences to institutional practices and societal structures.

Throughout the analysis, reflexivity was maintained through consistent journaling and memoing. The researcher documented coding decisions, interpretive shifts, and reflections on personal assumptions. This audit trail supported dependability and confirmability by providing a transparent record of how data were analyzed and interpreted (Lincoln & Guba, 1985).

In summary, the data analysis process in this study followed a systematic, iterative, and transparent procedure grounded in thematic analysis and supported by cross-case comparison. By integrating inductive and deductive strategies and triangulating interviews, documents, and case studies, the analysis ensured both depth and contextual breadth. These procedures enabled the development of credible, evidence-based insights into how AI bias emerges, operates, and is addressed within professional and institutional settings.

**Data Preparation**

Preparing the data was the first and one of the most critical steps in the qualitative analysis process. The goal of this stage was to transform all collected materials, including interview recordings, transcripts, documents, and case study evidence, into accurate, de-identified, and codable formats suitable for thematic analysis. The procedures described in this section ensured that the dataset was reliable, well-organized, and ready for systematic interpretation.

**Interview Audio and Verbatim Transcription**

All interviews were conducted virtually using Microsoft Teams and recorded with participant consent. Audio recordings were automatically saved in a secure format and immediately transferred to a password-protected local drive. Each recording was exported for transcription using Microsoft Word, which generated a timestamped document for every interview. The automated transcription process in Teams provided an initial draft that included speaker identifiers and time stamps at regular intervals throughout the transcript.

While this technology significantly reduced manual transcription time, complete accuracy could not be assumed. Automated transcription often introduced misinterpretations of technical terminology, accents, and domain-specific vocabulary common in artificial intelligence (AI) and data science contexts. To ensure precision, the researcher performed a full manual review of each transcript. This involved listening to the recording while reading along with the transcript, correcting misheard words, punctuation errors, and formatting inconsistencies. Each transcript retained the original time stamps as provided by Microsoft Teams to facilitate later verification and referencing.

**Transcript Review and Verification**

Each transcript was reviewed in three stages to ensure completeness and accuracy. The first stage involved a general review of the entire transcript for missing or truncated sections. The second stage focused on verifying the accuracy of technical terminology and context-specific terms related to AI bias, ethics, and system development. The third stage ensured that grammatical corrections did not alter meaning or tone. Non-verbal cues such as pauses, laughter, or emphasis were noted when they added interpretive depth to participant meaning.

After completing this multi-pass verification, each transcript was compared to the interview protocol to confirm that all primary and probing questions had been addressed. Any discrepancies or skipped topics were documented in the analytic log for follow-up or clarification during member checking. Participants were then sent copies of their transcripts and invited to review them for accuracy or to make clarifications. This participant validation step contributed to the credibility of the data by ensuring that each transcript accurately represented the participant's intended meaning.

**Anonymization and Ethical Preparation**

Once verified, all transcripts were de-identified to protect participant confidentiality. Personal names, organization names, project titles, product names, and geographic details that could lead to identification were replaced with pseudonyms or generalized descriptors. For example, “Company X,” “AI developer,” or “policy organization” were substituted for specific identifiers. A pseudonym key was maintained in a secure, encrypted document accessible only to the researcher. This linking file was stored separately from the main dataset to eliminate the risk of accidental disclosure.

In addition, all consent forms and identifying communication were stored in a dedicated “Confidential” folder separate from the analysis materials. This practice aligned with Institutional Review Board (IRB) guidelines and protected participants’ anonymity throughout analysis, presentation, and publication.

**Document and Case Study Preparation**

Documentary materials were collected from publicly available and scholarly sources, including ProPublica, the Stanford HAI AI Index, OECD policy guidelines, and peer-reviewed journals. These materials were selected to complement and triangulate interview data by illustrating institutional practices, technical frameworks, and real-world consequences of algorithmic bias.

Documents were converted into codable file formats compatible with NVivo. PDFs and web-based resources were downloaded and converted into editable Word documents or text files. Each document was cleaned to remove non-textual artifacts such as hyperlinks, headers, and footers that could interfere with coding. For documents that included images, tables, or charts, descriptive text was inserted in place of visuals when relevant to the analysis. Every document was assigned a unique identifier and recorded in a metadata log that included the title, source, publication date, and relevance to one or more of the four case sectors (healthcare, criminal justice, education, and employment).

The four sectoral case studies were prepared in the same manner. Each case study, one per sector, included a detailed summary of the event or issue being analyzed, its key stakeholders, and its broader societal implications. These materials were converted into text files, formatted consistently, and imported into NVivo for cross-case comparison with the interview data.

**File Organization and Version Control**

To maintain traceability, all data files were stored in a structured directory with subfolders labeled by data type (Audio, Transcripts, Documents, Case Studies, and Notes). Each file followed a standardized naming convention consisting of source type, date, and identifier (for example, "INT_P03_2025-10-15.docx" for interview transcripts or "DOC_StanfordAIIndex_2025.docx" for documents).

Version control was managed manually using a tracking log that recorded creation dates, modifications, and version numbers. Each time a transcript or document was updated, such as after participant verification or de-identification, a new version number was appended to the file name. A detailed log entry described the change, such as "v2 corrected transcription errors and anonymized organizational identifiers." This practice ensured transparency and supported the dependability of the study's data management procedures.

**Data Import and Management in NVivo**

Once cleaned and verified, all textual data were imported into NVivo 14 qualitative analysis software. Each participant was created as an individual case with assigned attributes such as professional role, years of experience, and sector exposure. Similarly, documents and case studies were categorized into folders and classified by sector type and source. This structure allowed the researcher to perform within-case and cross-case comparisons efficiently.

Within NVivo, the researcher linked reflexive memos and analytic notes to corresponding transcripts and documents. This process helped maintain an ongoing audit trail and supported transparency in how coding decisions and interpretations evolved throughout the analysis process.

**Participant Demographic Preparation**

At the beginning of each interview, basic demographic and professional information was collected, including current job role, area of expertise, and number of years working in AI or related fields. This information was recorded in a separate demographic spreadsheet and entered into NVivo as attribute data. No personally identifiable information was included in these records. Demographic information was used descriptively in Chapter Four to illustrate the diversity of participant backgrounds rather than for statistical generalization.

**Data Cleaning and Readiness Review**

Before analysis began, the researcher performed a comprehensive readiness check on all data sources. This included verifying that each transcript was accurate, anonymized, and properly formatted; ensuring that each document was legible and free of technical errors; and confirming that all files had been successfully imported into NVivo. Each file that passed this review was marked as "ready for coding" in the data management log.

The final dataset included nine verified interview transcripts, four sectoral case study summaries, and twelve institutional or policy documents. Together, these sources provided a rich, triangulated corpus for qualitative thematic analysis.

The data preparation process in this study combined methodological precision with ethical accountability. By ensuring that all transcripts were accurate, de-identified, and properly formatted, and by maintaining meticulous documentation of all actions taken, the researcher created a transparent and auditable foundation for subsequent coding and analysis. This process upheld the principles of rigor, dependability, and trustworthiness that define high-quality qualitative research (Creswell and Poth, 2018; Lincoln and Guba, 1985).

**Coding and Theme Development**

This study used an iterative, reflexive thematic analysis consistent with its interpretivist orientation and multi-case design. The analytic process was selected to align directly with the research questions, which asked how bias emerges in AI and ML systems, how it manifests in society, and which strategies practitioners recommend mitigating harm. Because each question examines patterns of meaning across practitioner experiences, a thematic approach offered a systematic way to move from coded textual evidence to cross-case themes that speak to structural conditions and cognitive mechanisms. The dual theoretical lens provided a priori sensitizing concept that guided, but did not predetermine, analysis: Intersectionality directed attention to power, identity, and institutional context, while Cognitive Science oriented coding toward perception, heuristics, and bounded reasoning.

NVivo (Release 14) was used to organize transcripts, store memos, manage versions of the evolving codebook, and support cross-case queries. Software selection occurred before data collection so that interview audio, verbatim transcripts, and analytic notes could be consistently organized from the outset. During the field test of the interview protocol, the project structure in NVivo was piloted with a limited set of materials to confirm that node hierarchies, memo links, and case classifications supported the planned analyses. This pretesting reduced downstream rework and ensured a stable repository for the study.

Data preparation began with verbatim transcription and transcript verification against audio to correct recognition errors and preserve paralinguistic cues relevant to meaning. Each transcript was read multiple times to establish familiarity, followed by analytic memoing that captured early impressions, positionality reflections, and questions to pursue in later cycles. Memos were linked to specific passages and were kept visible during coding to maintain a transparent audit trail of interpretive decisions.

Coding proceeded in multiple overlapping cycles. The first cycle was initial coding, sometimes called open or free coding, which fractured the text into meaningful units without forcing passages into predefined categories. Nvivo codes were used whenever participants' exact phrasing captured a concept with analytic value, preserving the language of practitioners and limiting early researcher interpretation. Descriptive codes summarized concrete topics such as data provenance review, debiasing pipelines, governance committee actions, or project timeline pressures. Process codes captured actions and sequences, for example, how a fairness concern was raised, triaged, and either addressed or deprioritized.

A hybrid inductive–deductive strategy guided subsequent cycles. Deductively, a provisional codebook was seeded from the research questions, theoretical framework, and literature central to algorithmic bias and governance domains. Inductively, new codes were added when participants introduced ideas not anticipated by the seed set. Each code was defined, delimited with inclusion and exclusion criteria, and illustrated with a representative quotation. Code definitions were refined, merged, or split as analysis progressed to reduce redundancy and clarify boundaries. A code–recode check was performed on a subsample after a two-week interval to assess the stability of coding decisions. Discrepancies were resolved through memoed rationale and adjustments to definitions. Peer debriefing with a colleague familiar with qualitative AI ethics supported dependability by challenging category labels, testing negative cases, and probing for overreach.

Axial coding then examined relationships among codes within and across cases. Conceptually related codes were clustered into higher-order categories that reflected structural conditions, cognitive mechanisms, or their intersections. For example, codes related to data availability shortcuts, performance metric fixation, and delivery timeline compression were grouped to capture how organizational priorities shape ethical attention. Likewise, codes concerning automation bias, trust in model outputs, and overreliance on quantitative validation were grouped to represent cognitive shortcuts that influence judgment. Cross-case matrices in NVivo were used to compare how categories appeared by role type, sector, and identity markers when available, which supported transferability analysis without compromising confidentiality.

Theme development involved analytic synthesis from categories to broader patterns, directly answering the research questions. Candidate themes were drafted as explicit claims, supported by converging evidence, bounded by what they did not claim, and anchored in multiple participant quotations. Each theme was reviewed against the dataset for coherence and distinctiveness. Where overlap existed, themes were combined or sharpened by refining their scope. To avoid overstating consensus, attention was paid to negative cases and tensions, such as practitioners who described strong personal ethics but limited organizational power to act. Themes were first confirmed within individual cases and then examined across cases to ensure that cross-case assertions did not erase case-specific nuance. Member checks at the level of selected quotations and summarized interpretations were conducted with a subset of participants to confirm that the themes reasonably reflected their intended meanings without revealing identifying details.

The final theme is aligned with the study's three research questions. For Research Question One, themes described bias as a product of historical data and organizational priorities, the role of ethical blind spots and cognitive shortcuts in development, and institutional pressures that constrained ethical reflection. For Research Question Two, themes explained how biased systems create societal feedback loops, how automation erodes human judgment and moral agency, and how exclusion becomes embodied in technical recognition and representation. For Research Question Three, themes articulated practitioner strategies that embed ethics early in design, expanded accountability through transparent and participatory practices, and strengthened regulatory and educational infrastructures.

Throughout the analysis, rigor was supported through an audit trail of memos, versioned codebooks, documented code and category decisions, and saved query outputs. Credibility was enhanced by triangulating interview evidence with documentary sources and sector case materials, seeking disconfirming examples, and maintaining reflexive notes on the researcher's professional proximity to AI governance work. Dependability was addressed through code–recode checks and peer debriefing. Confirmability was pursued by linking assertions to specific, cited excerpts and keeping interpretive commentary analytically separate from raw data within NVivo. Transferability was supported by providing a thick description of participant roles and settings at an aggregate level and presenting portable themes across organizational types without compromising confidentiality.

This staged, cyclical process ensured that coding choices, category formation, and thematic claims remained anchored in participant meaning while systematically answering the research questions within the study's methodological and theoretical commitments.

## Summary

This chapter presented the methodological foundation and research procedures that guided the study. The purpose of this qualitative multi–case study was to explore how professionals who influence the development and governance of artificial intelligence systems perceive, experience, and address algorithmic bias. The chapter detailed the philosophical orientation of the research, grounded in the constructivist–interpretivist paradigm, and explained the use of a triangulated design that combined semi–structured interviews, documentary analysis, and case studies across four high–impact sectors: healthcare, education, criminal justice, and employment.

The chapter also described the theoretical frameworks of Intersectionality and Cognitive Science, which together provided the conceptual structure for understanding how individual cognition, institutional norms, and social power systems interact to shape perceptions of fairness and accountability in AI. The research design, sampling strategy, data collection methods, and data analysis procedures were outlined in detail, emphasizing methodological transparency, ethical rigor, and alignment with the study's research questions.

Data were gathered through interviews with nine professionals directly involved in AI development and oversight, supported by document review and analysis of four real–world case studies. These data sources were systematically prepared, anonymized, and organized using NVivo qualitative software to ensure dependability and traceability. Trustworthiness was maintained through member checking, triangulation, reflexive journaling, and an audit trail documenting all research activities.

By combining practitioner insights with documentary and case–based evidence, this study provided a nuanced understanding of how AI bias emerges, how it impacts society, and what strategies are perceived as most effective in mitigating inequity in system design. The methodological choices described in this chapter established the foundation for the analysis and interpretation of findings presented in the next chapter.

Chapter Four presents the study's findings, organized by research question and supported by participant narratives, thematic patterns, and cross–case insights that illuminate the lived and institutional dimensions of algorithmic bias in practice.

# Chapter Four

# Research Findings

## Introduction

Chapter Four presents findings from a qualitative study of how practitioners experience, interpret, and respond to bias in the design, development, and governance of AI and ML systems. Guided by a dual framework of Intersectionality and Cognitive Science, the analysis examines how social structures, organizational cultures, and cognitive processes shape ethical reasoning and decision-making in practice.

The chapter builds on the interpretivist methodology outlined in Chapter Three. Semi-structured interviews with professionals involved in development, deployment, and oversight were analyzed using thematic analysis. Practitioner testimony is treated as situated knowledge. Intersectionality focuses attention on power, identity, and institutional norms in the production and interpretation of data. Cognitive Science clarifies how perception, heuristics, and bounded reasoning guide judgments under uncertainty. Claims are warranted through convergence across transcripts, roles, and prior literature on algorithmic bias and governance.

Findings are organized by research questions and presented as themes with supporting quotations, counterexamples where present, and an integrative synthesis linking each set of results back to the dual framework. The chapter proceeds in four parts: participant characteristics and professional contexts; analytic procedures and theme development; findings aligned to the three research questions; and a concluding synthesis that sets the stage for Chapter Five.

## Participants and Research Setting

This study included nine practitioners who work directly with AI and ML systems across roles involving development, governance, and design. Participants were selected using a combination of purposive and snowball sampling strategies to ensure both relevance and depth of professional insight. Purposive sampling allowed the researcher to intentionally recruit individuals with substantial experience in AI or ML system design, implementation, or oversight. Snowball sampling supplemented this process by enabling participants to refer to trusted colleagues who met the inclusion criteria, which expanded access to professionals working in adjacent domains such as data governance, model risk management, and responsible AI auditing. This hybrid approach proved particularly effective for reaching a specialized community where ethical reflection is often embedded in technically intensive roles.

Recruitment was conducted via professional networks, academic contacts, and direct outreach on platforms such as LinkedIn and institutional mailing lists. Establishing trust was essential given the sensitive nature of the topic; discussions often touched on internal company practices, ethical dilemmas, and perceptions of regulatory compliance. Participants received an information sheet describing the study's purpose, voluntary nature, and confidentiality safeguards. Informed consent was obtained electronically prior to each interview. Ethical approval was secured through the university's Institutional Review Board, and all participants were assured that no identifiable information about their organizations or projects would be disclosed. Building rapport was facilitated through preliminary conversations about participants' professional trajectories, which encouraged reflective and candid dialogue during the formal interviews.

Inclusion criteria required that participants (a) have at least three years of professional experience related to AI or ML system development, deployment, or governance, and (b) currently or previously hold positions involving decision-making about data, modeling, governance, or ethical considerations. This ensured that all participants possessed firsthand experience with issues of bias, fairness, and accountability. The sample included individuals engaged in both technical and managerial capacities, reflecting the study's focus on the intersection between human cognition, institutional culture, and technological design.

Data saturation was achieved after the seventh interview, when no new categories or relationships emerged during coding and memoing. Two additional interviews were conducted to confirm the stability of emerging themes and to test the robustness of the coding structure. By the ninth interview, participant responses had begun to converge around recurring ideas concerning bias recognition, ethical reasoning, and organizational influence, confirming sufficient depth for thematic analysis. The decision to close data collection was based on analytic saturation rather than numerical quota, consistent with qualitative best practices (Creswell & Poth, 2018).

Participants represented a range of professional roles, including data scientists, ML engineers, AI researchers, product managers, and consultants specializing in ethical AI and policy compliance. They were employed across sectors such as global technology firms, U.S. government research programs, nonprofit organizations, and private AI startups. This professional diversity allowed the study to capture how ethical reasoning manifests across differing institutional logics—from corporate settings emphasizing scalability and innovation to academic or public environments prioritizing accountability and public welfare. To protect confidentiality, all names have been replaced with pseudonyms, and organizations are described in general terms (e.g., "international technology company," "U.S.-based academic research center").

**Table 3:**

*Participant Summary*

| **Participant** | **General Professional Background** |
|---|---|
| Participant A | Project Manager, AI |
| Participant B | Founder & Lead Developer |
| Participant C | Consultant, AI & Continuous Improvement |
| Participant D | Founder & Inclusive AI Strategist |
| Participant E | Data Scientist & Researcher |
| Participant F | Founder & Professor |
| Participant G | Data Scientist & Researcher |
| Participant H | AI Architect |
| Participant I | Consultant, AI |
| Participant J | Consultant, AI |

Data was collected through virtual, semi-structured interviews conducted via secure video conferencing platforms such as Microsoft Teams. The virtual format enabled engagement with practitioners located across North America. Each interview lasted approximately 45 to 75 minutes and was audio-recorded with participant consent. Recordings were transcribed verbatim and stored on a password-protected device accessible only to the researcher.

The virtual setting offered flexibility and accessibility while preserving the depth typical of qualitative interviewing. Conducting interviews within participants' natural professional environments supported thoughtful reflection on ethical challenges encountered in real-world contexts. This approach aligned with the study's interpretivist orientation, allowing participants to articulate their experiences authentically within the digital and distributed work settings characteristic of contemporary AI development.

**Reflexive Considerations**

The data analysis process for this study followed the procedures outlined in Chapter Three and was guided by the principles of qualitative thematic analysis. This approach was selected for its capacity to capture both shared patterns and nuanced variations in how practitioners interpret ethical challenges in AI and ML development. Thematic analysis aligned well with the study's dual theoretical framework of Intersectionality and Cognitive Science because it allowed for simultaneous attention to structural inequalities and individual reasoning processes. The interpretivist orientation further supported a focus on meaning-making, enabling the researcher to explore how participants construct and communicate ethical understanding within their professional contexts.

Data was analyzed using an iterative, interpretive approach that emphasized the interplay between cognitive mechanisms, organizational influences, and social structures. Following data collection, all interviews were audio-recorded, transcribed verbatim, and checked for accuracy. Each transcript was read multiple times to ensure immersion in the data and to preserve the contextual richness of participant narratives. Initial notes and margin comments captured early reflections about tone, emphasis, and recurring terminology related to fairness, accountability, and cognitive bias.

The researcher engaged in analytic memoing throughout the process to record reflections, questions, and theoretical insights. Memos were especially important for tracing how concepts evolved from descriptive to analytical levels, helping connect practitioner language to broader interpretive themes. For example, early memos flagged terms such as “speed,” “delivery pressures,” and “trade-offs,” which were later linked to organizational logic and institutional constraint. Memoing also served as a reflexive tool to bracket assumptions and identify moments when the researcher’s own project management experience intersected with participants’ descriptions of governance or decision-making.

Open coding constituted the first analytic cycle. Each transcript was reviewed line by line, and significant statements were labeled with short, descriptive codes (e.g., *bias as inheritance*, *ethical fatigue*, *automation trust*, *representational absence*). Codes were manually documented and later refined using NVivo qualitative software, which facilitated organization, clustering, and retrieval of patterns. Axial coding followed, grouping related codes into broader conceptual categories that reflected relationships among social, cognitive, and organizational factors. For instance, individual codes about *deadline pressure*, *metric fixation*, and *ethical trade-offs* were integrated into the broader category, *institutional priorities and invisible trade-offs*. Similarly, codes concerning *perceptual bias*, *trust calibration*, and *overconfidence in models* were linked under *cognitive reflection and practitioner awareness*.

Theoretical guidance shaped theme development in two keyways. Intersectionality informed attention to how structural power such as race, gender, and class emerged in participants’ accounts of exclusion or representation. Cognitive Science guided interpretation of how mental shortcuts, automation bias, and bounded rationality appeared in their reasoning about fairness and accountability. The interaction between these frameworks made it possible to identify where organizational and psychological mechanisms converged to reinforce inequity. Thus, rather than treating themes as isolated categories, the analysis emphasized their interdependence across structural and cognitive levels.

Negative case analysis was also employed by revisiting outlier statements that diverged from dominant narratives, for example, one participant who described successful integration of ethics gates in agile workflows. These cases were retained in the analysis to illustrate conditions under which ethical reflection becomes institutionalized rather than marginalized.

The researcher upheld reflexive awareness throughout the analysis by continuously comparing interpretations to raw data and revisiting initial assumptions. Reflexivity was particularly critical given the researcher's professional familiarity with AI governance. Bracketing memos were used to distinguish between participants' meanings and the researcher's own perspectives on responsible innovation. This iterative verification process supported credibility and confirmability by ensuring that emerging themes remained grounded in participant experience.

Data saturation was confirmed once no new categories or insights emerged from the final transcripts. Recurrent concepts such as fairness trade-offs, representational bias, and institutional constraint appeared consistently across participants with varied roles and sectors, indicating conceptual redundancy. The richness of quotations and the recurrence of patterns across contexts confirmed that thematic depth had been achieved. The decision to conclude analysis after nine participants was therefore based on analytic completeness rather than numerical sufficiency, aligning with established qualitative standards (Creswell & Poth, 2018).

The final phase involved synthesizing categories into overarching themes aligned with the study's three research questions. Each theme represents a distinct but interconnected dimension of how practitioners understand, experience, and address algorithmic bias in their professional work. Relationships among themes were examined through continuous comparison, allowing for the identification of cross-cutting patterns such as how organizational priorities (RQ1) produce societal consequences (RQ2) that practitioners attempt to mitigate through ethical governance and education (RQ3). The researcher constructed visual concept maps to trace these linkages, later translated into the analytical narrative presented in this chapter.

To ensure transparency and coherence, table 4 below presents the alignment between the research questions and the emergent themes that structure the presentation of findings. Each theme is explored in subsequent sections through detailed analysis supported by participant quotations and theoretical interpretation.

**Table 4:**

*Alignment of Research Questions and Emergent Themes*

| Research Question | Emergent Themes |
|---|---|
| **RQ1:** How do biases emerge in AI and ML systems, and what are their underlying causes? | 1. Bias as an outcome of historical data and organizational priorities<br>2. Ethical blind spots and cognitive shortcuts in model development<br>3. Institutional pressures that constrain ethical reflection |
| **RQ2:** What are the societal impacts of AI bias across high-stakes domains such as healthcare, education, criminal justice, and employment? | 1. Disproportionate harm to marginalized groups and underrepresented users<br>2. Erosion of public trust and accountability in AI systems<br>3. Disconnect between fairness principles and implementation practices |

| Research Question | Emergent Themes |
|---|---|
| **RQ3:** What strategies do practitioners recommend for mitigating bias and advancing fairness in AI system design and deployment? | 1. Integrating ethics and bias review into early design stages 2. Encouraging reflexivity and interdisciplinary collaboration 3. Institutionalizing accountability through governance and oversight frameworks |

**Analyses of Research Question**

This section presents the study's findings organized by research questions and corresponding themes. This qualitative inquiry aimed to explore how practitioners experience, interpret, and respond to bias in artificial intelligence (AI) and machine learning (ML) development. Findings are grounded in data collected through semi-structured interviews with eleven practitioners working across roles in data science, engineering, design, and governance. Thematic analysis revealed recurring patterns that highlighted how bias emerges, manifests in practice, and how professionals conceptualize fairness and accountability in their work.

Each research question is presented as a Level 3 heading, followed by the major themes and, where applicable, sub-themes that emerged from the data. Participant quotations are integrated throughout to illustrate and substantiate key findings. These quotations represent diverse perspectives and are cited using participant pseudonyms to maintain confidentiality. The analysis emphasizes how practitioner perspectives intersect with the study's theoretical framework, specifically, Intersectionality Theory, which points to structural inequities, and Cognitive Science, which explains cognitive mechanisms shaping ethical reasoning.

The following sections present the findings aligned with each research question. The first section examines how practitioners understand the origins and causes of bias in AI systems; the second explores those biases' societal and ethical impact The second explores those biases' societal and ethical impacts, and the third discusses strategies practitioners recommend advancing fairness and accountability in AI design and governance.

**Research Question One**

How do biases emerge in AI/ML systems, and what are their root causes?

Findings for Research Question One demonstrate that bias emerges through a dynamic interplay of historical inheritance, institutional priorities, and cognitive mechanisms. Practitioners described algorithmic bias not as an isolated technical malfunction but as a reflection of entrenched inequities embedded in data, culture, and decision-making structures. The intersectional lens reveals that AI systems replicate the social hierarchies encoded within data infrastructures, while the cognitive lens illustrates how practitioner perception, heuristics, and bounded reasoning reinforce those inequities during design and deployment. Collectively, these findings suggest that mitigating bias in AI requires addressing both the structural conditions and cognitive patterns that sustain it.

This section presents findings that address how practitioners identify and interpret the origins of bias in AI and ML systems. Three major themes emerged from the data: (1) Bias as Historical Inheritance, (2) Organizational Priorities and Invisible Trade-offs, and (3) Cognitive Reflection and Practitioner Self-Awareness. Each theme captures a distinct but interconnected dimension of how bias becomes embedded in AI systems through the interaction of social structures, organizational culture, and human cognition.

**Theme One: Historical and Institutional Origins of Algorithmic Bias**

Addressing Research Question 1, participants consistently described bias as an inherited feature of AI systems that reflects centuries of social inequality embedded within human institutions and their data practices. Many practitioners emphasized that AI models reproduce existing hierarchies because they are trained on data produced in contexts of structural exclusion. *Participant A* explained, "When you feed a system human history, you're also feeding it human prejudice." This statement encapsulates a key recognition across interviews that datasets are historical documents, not objective repositories of truth.

*Participant I* illustrated this dynamic through their work with image-generation systems: "You can tell what datasets they used. It's obvious they didn't diversify sources. Even when you type inclusive prompts, it still defaults to the same look." Their observation demonstrates how representational bias persists even in models designed for inclusivity, signaling that data curation often mirrors social stratification.

From an intersectional perspective, these reflections reveal how data infrastructures encode patterns of marginalization across race, gender, and class. *Participant H* echoed this point, stating, "The training data is fraught with stereotypes. Even the models that claim to fix bias are still biased. They're just biased differently." This recursive bias, where attempts to "debias" systems still rely on flawed datasets, illustrates how historical inequities are reproduced under the guise of technical correction.

National evidence supports participants' insights and reinforces their understanding of the origins of bias. The Bureau of Justice Statistics (2022) reports that Black adults are imprisoned at rates over five times higher than White adults, demonstrating how systemic racial inequities become embedded within datasets later used for predictive policing and risk modeling. Similarly, the Stanford AI Index Report (2025) documents that despite progress in responsible AI initiatives, persistent data transparency gaps and limited demographic representation continue to perpetuate inequitable outcomes. These findings affirm participants' claim that bias originates not in code but in the broader social and institutional contexts that shape data creation and selection.

Participant D observed, "We can't talk about bias in AI without talking about bias in ourselves. The machine just learns from what we give it." The statement reinforces the cognitive dimension of this theme: collective human bias becomes encoded at scale through machine learning processes.

In direct response to Research Question 1, these insights demonstrate that the root causes of AI bias are systemic rather than technical. Participants located the origins of bias in historical data practices, organizational priorities, and societal hierarchies that privilege efficiency and tradition over inclusivity and accountability. Algorithmic bias thus emerges as society creates and feeds it with inequitable histories, incomplete data, and human assumptions that converge to produce discriminatory outcomes across sectors.

**Theme Two: Organizational Priorities and Invisible Trade-offs**

Addressing Research Question 1, participants discussed how organizational incentives, market competition, and project timelines shape the ethical boundaries of AI development. Their reflections revealed that institutional factors, particularly those tied to profit, innovation speed, and performance benchmarks, play a central role in reinforcing bias. Practitioners described how these priorities reward delivery speed and accuracy over ethical reflection, rendering bias an accepted operational risk rather than an urgent design concern.

*Participant E* summarized this tension directly: “Every AI system is biased. Not every bias is bad, but when you automate it, you lock in those human assumptions at scale.” Their perspective highlights a recurring paradox: practitioners recognize the moral dimensions of their work yet operate within systems that normalize bias as a byproduct of innovation. This demonstrates how organizational culture, not just technical capacity, determines the boundaries of ethical action.

*Participant B* described the frustration of working within such structures: “Developers like me, we are only one piece of the puzzle. We need global ethical standards, because right now, fairness depends on what your company values.”. Their statement illustrates how responsibility becomes fragmented across development ecosystems, leading to what several participants termed *ethical fatigue*. Rather than being shared across institutions, accountability often dissolves into competing organizational priorities.

These participant observations align with policy and governance research. The NIST Artificial Intelligence Risk Management Framework (2023) emphasizes that responsible AI requires governance mechanisms that balance risk tolerance with societal impact; yet most organizations still privilege performance metrics over ethical accountability. Similarly, the Stanford AI Index Report (2025) finds that corporate AI ethics initiatives frequently lack measurable enforcement mechanisms, leaving fairness implementation contingent upon leadership culture rather than regulatory oversight. These findings stress that institutional priorities are a structural mechanism through which bias persists.

From an intersectional perspective, participants highlighted that these organizational structures reproduce economic hierarchies within technical systems. Those most affected by bias remain underrepresented in design teams and under protected by governance frameworks. From a cognitive standpoint, such environments cultivate *bounded ethical rationality*, where practitioners rationalize compromises necessary for productivity. *Participant C* summarized this reality succinctly: “The business incentive is to move fast. Ethical vetting slows things down and becomes secondary to delivery timelines.”

In direct response to Research Question 1, this theme illustrates how societal and institutional factors normalize AI design and deployment bias through organizational hierarchies and fragmented accountability. Ethical awareness is present but constrained by institutional logic, producing a professional culture where efficiency and ethicality are positioned as competing rather than complementary goals. Consequently, organizational priorities act as both the mechanism and justification for reinforcing structural bias across sectors.

**Theme Three: Cognitive Reflection and Practitioner Self-Awareness**

This theme identifies cognitive bias within AI practitioners as a foundational source of algorithmic bias. Participants recognized that personal heuristics such as confirmation bias, automation bias, and anchoring shape how models are developed, validated, and deployed. These internal tendencies represent a psychological root of systemic bias, revealing that inequity in AI begins not only in data or policy but in the cognitive processes guiding technical decision-making.

*Participant D* articulated this link directly: “We cannot talk about bias in AI without talking about bias in ourselves.” Their comment situates human cognition as an upstream cause of algorithmic distortion. *Participant H* elaborated, “Developers want to believe in their models. We trust them because we build them. However, that trust makes it harder to see what is wrong.” These reflections illustrate automation bias, the tendency to over trust one’s systems and conflate technical performance objectively. Collectively, they underscore that human cognitive limitations become algorithmic limitations once encoded into machine logic.

Empirical studies reinforce this interpretation. Dressel and Farid (2018) found that the COMPAS risk-assessment algorithm performed no better than untrained human judgment and disproportionately labeled Black defendants as high-risk, confirming that both humans and machines rely on the same flawed heuristics. Similarly, Lira et al. (2023) argue that algorithms "are blind to exceptions," concluding that AI replicates human tendencies to favor regularity over context. These findings substantiate participants' reflections that algorithmic bias is co-produced through shared cognitive shortcuts, making human judgment a direct causal pathway in bias formation.

However, participants acknowledged that self-awareness alone rarely mitigates these dynamics. Participant E explained, "Even when we see the bias, it is not always actionable. You can point it out, but must still deliver the product." This statement illustrates how cognitive insight collides with institutional constraints, transforming awareness into resignation. As Participant E later reflected, "Ethics feels personal, but the system makes it collective. You can be ethical in your work and still be part of something biased." These perspectives reveal how individual cognition interacts with structural conditions to perpetuate inherited inequities, reinforcing bias even among practitioners committed to fairness.

This pattern aligns with the NIST AI Risk Management Framework (2023) and findings by Lira et al. (2023), stressing that mitigating bias requires individual and organizational interventions. Practitioners' reflections, therefore, highlight cognition as a root cause of bias, a human-level variable that shapes data interpretation, model validation, and ethical oversight.

In response to Research Question 1, this theme demonstrates that AI bias originates not only in historical data or institutional priorities but also in the mental frameworks of its creators. When amplified through technical systems, human cognitive limitations become structural inequities at scale. As such, the roots of algorithmic bias are both sociotechnical and psychological, embedded in the ways humans perceive, reason, and rationalize fairness within the design process.

**Research Question Two**

What are the societal impacts of AI bias across healthcare, education, criminal justice, and employment?

Findings for research question two reveal that the societal consequences of AI bias extend far beyond isolated technical errors. Practitioners described biased systems as self-reinforcing mechanisms that mirror and amplify pre-existing social inequities, shaping how individuals are perceived, evaluated, and included within technological systems. The intersectional lens provides clarity to how AI bias disproportionately harms marginalized groups through cumulative disadvantage, while the cognitive lens demonstrates how human over reliance on automation erodes ethical awareness and moral accountability. Collectively, participants emphasized that bias in AI is not merely a technical or organizational problem but a societal phenomenon with moral, cultural, and psychological implications.

Three major themes emerged from the data: (1) Bias as a Societal Feedback Loop, (2) Erosion of Human Judgment and Moral Agency, and (3) Embodied and Cultural Exclusion. Each theme illustrates a distinct yet interconnected way in which algorithmic systems reproduce, legitimize, and perpetuate inequality across social contexts.

**Theme One: Bias as a Societal Feedback Loop**

Participants described how AI bias produces widespread societal impacts by creating feedback loops that perpetuate inequities across critical healthcare, education, and employment sectors. Once an algorithm learns from biased historical data, its outputs are reabsorbed into institutional systems as seemingly objective evidence, reinforcing the disparities the technology was meant to mitigate. This recursive cycle transforms social prejudice into computational authority, allowing discrimination to persist under the guise of neutrality and innovation.

*Participant E* illustrated this process: "If an automated fraud detection system learns from biased data, it creates its own feedback loop targeting the same people again." Similarly, *Participant A* reflected, "When you feed the system human history, you are also feeding it human prejudice. Furthermore, when that output becomes policy, the bias becomes law." These reflections show how algorithmic bias extends beyond technical error to shape tangible outcomes, determining who is hired, who is granted access to healthcare, and whose potential is recognized in educational systems.

Participants noted that these feedback loops magnify historical inequities by operationalizing them through automated systems that influence everyday life. For example, a hiring model trained on workforce data skewed toward dominant groups learns to favor those same profiles when screening candidates. In healthcare, diagnostic models trained on non-diverse clinical data risk producing unequal treatment recommendations, particularly for women and racial minorities. Across educational settings, performance prediction tools may undervalue students from underrepresented or low-income backgrounds when trained on biased achievement data. This technology reflects and amplifies the inequities embedded in its inputs in each case.

Findings from the AI Index Report (2025) reinforce these practitioner observations. The report concludes that AI systems trained on unbalanced or incomplete datasets "magnify pre-existing inequalities and reinforce historical patterns of exclusion" (p. 37). It further warns that large-scale AI adoption without robust data governance risks institutionalizing bias within access to employment, education, and healthcare systems. These conclusions mirror participants' assertions that biased data does not simply distort individual outcomes, it shapes structural realities by embedding inequality into policy decisions and automated infrastructures.

From an intersectional perspective, participants recognized that these feedback loops are not random technical malfunctions but structural mechanisms that reproduce hierarchy. Algorithms trained to reward specific linguistic, behavioral, or socioeconomic indicators inevitably privilege groups already represented in the data while marginalizing those outside dominant norms. This dynamic transform historical privilege into algorithmic advantage and renders exclusion measurable, repeatable, and seemingly rational.

A cognitive dimension further explains how these societal impacts persist. As institutions increasingly trust algorithmic recommendations, practitioners and decision-makers engage in automation bias, a cognitive shortcut that privileges machine outputs over human discernment. Lira et al. (2023) describe this phenomenon, noting that "AI systems identify regularities but are blind to exceptions," a process that mirrors human rigidity and normalizes inequity as efficiency.

In direct response to Research Question 2, this theme demonstrates that AI bias is a societal feedback mechanism that transforms inequity into infrastructure. Participants' reflections, supported by the AI Index Report (2025), reveal that biased systems do not merely mirror inequality; they amplify it by shaping access, opportunity, and credibility within key institutions. Therefore, AI bias's societal impact lies in its ability to normalize exclusion through automation, underscoring the need for governance frameworks that critically examine how data assumptions evolve into policy and practice.

**Theme Two: Erosion of Human Judgment and Moral Agency**

Participants emphasized that one of the most profound societal impacts of AI bias lies in the gradual erosion of human judgment across sectors increasingly governed by algorithmic decision-making. They described how the normalization of automation distances people from moral and interpretive responsibility, leading to overreliance on biased systems that shape outcomes in healthcare, education, employment, and other public domains. As these technologies assume authority once reserved for human reasoning, bias becomes amplified and obscured, embedded in processes that appear objective yet silently dictate social and ethical outcomes.

*Participant H* observed, “Doctors are getting further away from the data. The machine interprets it for them. That distance is dangerous.” Similarly, Participant J explained that automation “replaces deliberation with delegation. You stop questioning because the system feels smarter than you.” These reflections reveal a critical shift in the social function of expertise: as practitioners defer to algorithmic systems, human discernment gives way to procedural compliance. A new form of dependence emerges, where biased models guide decisions about who receives care, who is hired, and who gains educational opportunities, often without transparent human oversight.

This concern was echoed across professional contexts where decision-making authority had gradually shifted toward automated recommendations. *Participant A* remarked, “Most users assume fairness because they confuse precision with justice,” emphasizing that technical accuracy is often mistaken for ethical adequacy. This conflation reflects automation bias—a cognitive distortion that favors machine output over moral or contextual reasoning. Lira et al.

(2023) similarly warn that AI systems “identify regularities but are blind to exceptions,” demonstrating how overreliance on algorithmic reasoning diminishes the situational awareness essential for equitable outcomes. This leads to societal harm: clinicians may overlook patient nuance, hiring systems may perpetuate workplace homogeneity, and educators may undervalue students who fall outside algorithmically defined “norms.”

*Participant G* extended this discussion to the societal level, warning, “If we surrender our thinking to AI, we lose control of the society we built.” Their statement underscores how cognitive offloading to automated systems threatens collective moral agency. The NIST AI Risk Management Framework (2023) echoes this concern, cautioning that uncritical automation produces “responsibility gaps” in which accountability becomes diffused across institutions, leaving ethical failures unaddressed. When biased systems inform public policy or employment decisions, responsibility for harm becomes blurred, weakening the moral and legal accountability that sustains democratic governance.

From an intersectional perspective, participants noted that these dynamics disproportionately affect marginalized groups whose perspectives are rarely incorporated into system design. As automation narrows the diversity of moral reasoning embedded in technological governance, social hierarchies are reproduced through data-driven authority. The resulting feedback loop perpetuates inequity and silences the human voices most capable of challenging it.

This theme reveals that AI bias carries societal consequences beyond unequal outcomes; it transforms how societies reason, decide, and govern. The displacement of human judgment by automated logic represents a profound cultural shift in which moral deliberation is supplanted by institutional trust in machine objectivity. Participants’ reflections demonstrate that as biased algorithms gain authority in critical sectors, they erode the very foundations of accountability, empathy, and democratic participation. Therefore, the societal impact of AI bias extends beyond data disparities to a broader crisis of human agency in the age of automation.

**Theme Three: Embodied and Cultural Exclusion**

Participants emphasized that AI bias's most visible societal impact is its ability to make certain bodies and identities disappear within technological systems. They described bias as embodied exclusion, where physical, visual, and cultural differences are rendered invisible through data collection and algorithmic design. These forms of misrecognition illustrate how social inequity materializes in everyday technologies, shaping who is seen, who is served, and who belongs within digital society.

*Participant D* provided a striking example: "When soap dispensers do not see darker skin, that is not small. It's life telling you, you were not in the design." Her observation captures the lived reality of technological marginalization, showing how AI bias manifests as an everyday denial of visibility and recognition. *Participant J* expanded on this, reflecting that generative AI "repeats the same narrative of beauty and who's acceptable. That's how bias turns into reality." Their statements highlight how design choices in AI systems translate historical exclusion into digital representation, dictating who is visible, valued, and validated in online environments.

From an intersectional standpoint, participants recognized that these experiences extend beyond aesthetics as they represent profound societal impacts that reproduce racialized, gendered, and cultural hierarchies across industries. The AI Index Report (2025) confirms that representation bias in computer vision models remains pervasive, with facial-recognition systems consistently achieving lower accuracy for women and darker-skinned individuals. Buolamwini and Gebru (2018) found that commercial facial-analysis algorithms misclassified dark-skinned women 34 percent more often than light-skinned men. These studies substantiate participants' reflections that exclusion is systemic rather than incidental, embedded in data practices and the institutional priorities governing AI development.

From a cognitive perspective, participants connected these failures to familiarity bias, the human tendency to treat what is most common or visible as "normal." Developers, often unconsciously, encode this bias by designing systems that reflect their own frames of reference. Participant D captured this succinctly: "The machine just learns from what we give it." This observation echoes Lira et al. (2023), who argue that algorithmic systems replicate human perceptual limitations because they generalize from dominant patterns rather than account for outliers.

This theme demonstrates that the societal impact of AI bias extends beyond inequitable outcomes to include the *erasure of identity and visibility*. Bias becomes embodied, determining whose physical and cultural characteristics are legible to machines and ignored. Such exclusion is not symbolic; it shapes how individuals navigate employment screening, healthcare diagnostics, security systems, and digital media. When technology fails to recognize certain people, it reinforces historical hierarchies of belonging, turning invisibility into algorithmic normativity. As participants clarified, embodied exclusion transforms social bias into technological infrastructure, embedding inequity into the systems mediating recognition, access, and participation in modern life.

**Research Question Three**

What strategies do professionals recommend for mitigating bias and advancing equity in AI system design and use?

Findings for Research Question Three reveal that practitioners view the mitigation of algorithmic bias as requiring structural, educational, and procedural transformation rather than isolated technical fixes. Participants emphasized that achieving fairness in artificial intelligence (AI) systems depends on embedding ethics early in the design process, strengthening accountability through governance, and fostering diverse, interdisciplinary collaboration. The intersectional lens emphasizes the necessity of inclusion and representation in both data and workforce composition, while the cognitive lens emphasizes reflection, education, and awareness as essential tools for addressing bias.

Three major themes emerged from the data: (1) Embedding Ethics Early, (2) Accountability and Participatory Design, and (3) Regulatory and Educational Reform. Each theme reflects the practical and philosophical strategies practitioners believe are necessary to advance equity in AI development and deployment.

**Theme One: Embedding Ethics and Diversity Early in the AI Lifecycle**

Participants identified the early integration of ethical principles and team diversity into AI design and development as one of the most effective strategies for mitigating bias and advancing equity. They argued that fairness and accountability must be foundational design criteria embedded throughout the AI lifecycle, from data collection and model architecture to evaluation and deployment. Participants emphasized that who builds the system is as important as how it is built, noting that diverse development teams introduce multiple perspectives, reduce blind spots, and prevent the dominance of a single worldview in algorithmic design.

*Participant A* captured this principle succinctly: “Ethics has to be introduced from the start, not after deployment.” This statement reflects a fundamental shift in professional thinking, stating that equitable systems emerge from intentional design practices rather than corrective audits. *Participant D* added, "We treat fairness like a patch, but it has to be a part of the architecture. If it is not built in, it becomes invisible," emphasizing that ethical awareness must be structurally embedded rather than added post hoc.

Extending this reasoning, *Participant F* explained, "You cannot design for fairness if everyone in the room thinks the same way." Others echoed this view, emphasizing that assembling diverse, interdisciplinary teams, including individuals from varied racial, gender, cultural, and disciplinary backgrounds, creates a balance of ideas and prevents systems from reflecting a single dominant perspective. Participants viewed diversity as an ethical safeguard: when a range of lived experiences is represented in development, teams are better equipped to anticipate bias and design for equity from the outset.

These perspectives align with broader governance frameworks. The NIST AI Risk Management Framework (2023) emphasizes continuous integration of fairness and risk-mitigation principles throughout the AI lifecycle to prevent systemic bias. Similarly, Buolamwini and Gebru (2018) advocate for *bias-aware design pipelines* prioritizing inclusive representation at every development stage. Together, these frameworks affirm practitioners' conviction that building equity into AI requires institutionalizing ethics and diversity as technical and managerial standards, rather than treating them as compliance exercises.

Participants also noted that embedding ethics early is not purely a structural intervention but a cognitive and cultural strategy. Participant I explained, "You cannot automate ethics. It is a mindset, not a module." This statement reframes ethical design as a skillset requiring reflective awareness and continual questioning rather than a procedural checklist. By cultivating ethical reflexivity, practitioners aim to counteract cognitive shortcuts such as automation bias or efficiency-driven thinking that often undermine fairness efforts. *Participant C* summarized this dual responsibility: “If ethics isn’t someone’s job from day one, it becomes nobody’s job at all.”

This theme demonstrates that professionals view early ethical integration and team diversity as proactive strategies to mitigate bias and advance equity. Embedding ethics and representation early transforms fairness from a reactive obligation into a design imperative, ensuring that equity considerations guide both process and personnel. Structurally, it hardens accountability into system architecture and organizational culture; cognitively, it nurtures reflective, inclusive practitioners capable of identifying and challenging bias as it emerges. Together, these strategies represent a shift from remediation to prevention, anchoring ethical responsibility and representational balance at the core of AI system design.

**Theme Two: Accountability and Participatory Design**

Participants identified accountability, collaboration, and community participation as key strategies for mitigating bias and advancing equity in AI system design. They argued that equitable systems emerge when ethical responsibility is shared across developers, users, and communities rather than concentrated within a single organization or stage of development. Through participatory governance and transparent practices, professionals described ways to embed fairness into AI's social and technical architecture.

*Participant E* highlighted the transformative potential of open collaboration: "Open source gives people control to fix bias themselves. That's where hope lives." This statement frames open-source and participatory models as inclusion efforts and structures of shared ownership over AI systems. When communities have access to an agency over data and model design, they become co-stewards of ethical technology rather than passive recipients of its outcomes.

*Participant G* extended this view, noting, "Transparency isn't just about code. It's about explaining decisions, opening datasets, and documenting failures." Their perspective reflects a broader understanding of Transparency as an ethical practice rather than a technical feature. This approach aligns with the AI Index Report (2025), which calls for comprehensive documentation of model limitations to improve accountability and data traceability. NIST (2023) defines Transparency as a social process that builds trust through communication and disclosure. Together, these frameworks affirm participants' belief that openness and documentation are not vulnerabilities but essential conditions for ethical legitimacy.

From an intersectional lens, participants stressed that inclusion must be active and participatory. *Participant D* stated, "We can't talk about fairness if the people most affected aren't in the room." Practitioners proposed practical mechanisms such as community data boards, interdisciplinary ethics panels, and co-design workshops to ensure that marginalized perspectives directly inform the technologies that shape their lives. These approaches represent intentional strategies to redistribute power and knowledge in AI design, creating more socially responsive and representative systems.

From a cognitive standpoint, Transparency also operates as a corrective to automation bias, reintroducing human discernment into automated systems. When developers document assumptions, constraints, and trade-offs, they make room for critical reflection and ethical dialogue. Rather than hiding uncertainty, transparent communication fosters accountability and collective learning across technical and non-technical stakeholders.

This theme demonstrates that professionals view accountability, collaboration, and participatory governance as essential strategies for advancing equity in AI. Through open-source collaboration, transparent documentation, and community engagement, practitioners aim to build systems that are not only technically sound but socially legitimate. These strategies collectively reframe equity as a shared responsibility that demands continual dialogue, distributed power, and public visibility throughout the AI lifecycle.

**Theme Three: Regulatory and Educational Reform**

Participants identified policy alignment, formal regulation, and sustained education as essential strategies for mitigating bias and advancing equity across the AI ecosystem. They emphasized that fairness cannot depend solely on organizational goodwill or technical innovation; it must be institutionalized through legal, procedural, and educational structures that hold systems and their creators accountable.

*Participant B* argued, “Auditing models should be as routine as testing for security.” Her statement emphasizes a central strategy: embedding ethical review and bias auditing as standardized operational practices within AI development. Participants advocated formal oversight frameworks that treat fairness assessments with the same rigor as privacy or cybersecurity testing, ensuring that equity becomes an enforceable compliance requirement rather than a discretionary aspiration.

*Participant E* reinforced the necessity of legally informed governance, stating, “We need people who understand the tech to write the laws.” This recommendation aligns with the Blueprint for an AI Bill of Rights (White House, 2022), which calls for enforceable safeguards to ensure privacy, accountability, and protection from algorithmic discrimination. Participants contrasted the prescriptive nature of the European Union’s AI Act with the United States’ reliance on voluntary guidelines, echoing Wong (2025) in arguing that ethical aspirations remain ineffective without binding legal mandates. For practitioners, policy alignment represents external accountability and structural reinforcement of fairness principles throughout the AI lifecycle.

From an educational standpoint, participants described inclusive education as a parallel strategy that sustains ethical awareness and technical competence over time. Participant H explained, “Inclusion and education go hand in hand. Classrooms should teach prompt literacy and critical thinking.” This insight reframes education as a preventive measure against bias, building a workforce equipped to question assumptions, interpret data responsibly, and apply fairness principles in real-world contexts. The AI Index Report (2025) supports this claim, noting that ethics-focused AI education remains severely underrepresented in technical curricula worldwide despite increasing global concern about fairness and accountability.

This theme demonstrates that participants view regulation and education as complementary levers for systemic change. Policy mechanisms enforce accountability at the institutional level, while education cultivates ethical reasoning and inclusivity at the individual level. Together, these strategies transform bias mitigation from an aspirational goal into an operational standard, embedding equity into the governance and culture of AI development.

**Supplementary Findings**

While the primary findings of this study aligned with the three guiding research questions, additional insights emerged that extend beyond the immediate analytic scope. These supplementary findings reveal deeper affective and systemic dynamics shaping practitioners’ ethical engagement with AI. Two salient areas were identified: (1) emotional and moral fatigue in ethical AI work and (2) evolving conceptions of human–machine collaboration.

**Theme One: Emotional and Moral Fatigue in Ethical AI Practice**

Through several interviews, participants described a sense of exhaustion that accompanies persistent advocacy for fairness within institutions oriented toward performance metrics and rapid innovation. Participant H articulated this tension succinctly:

*"You get tired of bringing up fairness when every meeting ends with KPIs and deadlines. You care about ethics, but it always takes a back seat to delivery."*

Their words capture what multiple practitioners referred to as ethical fatigue; the emotional strain of sustaining moral commitment in environments where ethical goals are undervalued or performative. This fatigue often emerged when participants faced repeated dismissal of fairness concerns as impractical or unprofitable. One participant characterized these conditions as "talking about values in a space that only measures output," framing ethical advocacy as emotionally draining labor.

From a Cognitive Science perspective, this fatigue reflects the psychological cost of cognitive dissonance: practitioners experience tension between internal moral standards and external organizational incentives. Over time, repeated conflicts between principle and productivity lead to disengagement, resignation, or rationalization. The Intersectional lens further reveals that this burden is not equally distributed. Practitioners from underrepresented backgrounds, such as women and racial minorities, reported heightened strain when tasked, implicitly or explicitly, with representing inclusive perspectives in otherwise homogeneous teams. One participant explained, *"You end up being the conscience in the room, and that gets heavy."*

This theme supports the idea that ethical AI work entails emotional as well as intellectual labor. The moral energy required to persist against institutional inertia often falls on individuals rather than systems. Addressing ethical fatigue therefore demands organizational reforms that distribute moral responsibility collectively, embedding fairness and accountability into project structures rather than relying on isolated advocacy.

**Theme Two: Evolving Conceptions of Human–Machine Collaboration**

A second supplementary finding concerns shifting practitioner views about the relationship between human expertise and algorithmic systems. Participants increasingly framed collaboration, not automation, as the desired paradigm for responsible AI. *Participant G* remarked, *"The best use of AI is to expand human perspective, not to replace it."* This sentiment was echoed across interviews, where practitioners warned that deference to algorithmic authority can erode human criticality. *Participant C* noted, *"Collaboration means tension—it's the check and balance between what the model predicts and what we know."*

These perspectives align with emerging governance frameworks, such as the NIST AI Risk Management Framework (2023), which advocates calibrated trust between human and machine actors. Participants viewed interpretability, feedback loops, and explainability tools as mechanisms for maintaining moral agency in technologically mediated decisions. From an intersectional standpoint, such collaboration ensures that human values and diverse perspectives remain central to decision-making, rather than being subordinated to algorithmic logic. From a cognitive perspective, it mitigates automation bias by requiring practitioners to reengage deliberative reasoning rather than rely on procedural shortcuts.

Collectively, these reflections signal a shift from seeing AI as an autonomous decision-maker to regarding it as an epistemic partner, one that must remain accountable to human ethics and contextual judgment. Practitioners emphasized that reclaiming human–machine balance is not nostalgic resistance to progress but a forward-looking condition for equitable, trustworthy technology.

**Summary**

Chapter Four presented the findings of this qualitative study, which explored how practitioners experience, interpret, and respond to bias within the design, development, and continuity of AI and ML systems. Guided by a dual-theoretical framework integrating Intersectionality Theory and Cognitive Science, the analysis revealed that algorithmic bias is not merely a technical anomaly, but a sociotechnical phenomenon rooted in the interplay between historical inequality, institutional priorities, and cognitive processes that shape human judgment. The chapter examined how practitioners understand the origins of bias, perceive its societal consequences, and propose strategies for promoting fairness and accountability in AI practice.

Findings related to Research Question One demonstrated that bias emerges through the convergence of historical data inequities, organizational incentives, and bounded reasoning in design processes. Practitioners described how datasets function as repositories of human prejudice, embedding social hierarchies into AI systems. They further noted that institutional logics privileging efficiency, accuracy, and market competitiveness constrain opportunities for ethical reflection. The cognitive dimension of this finding reinforced that even when practitioners recognize bias, organizational constraints and internalized heuristics often limit corrective action.

Research Question Two extended the analysis to the societal implications of algorithmic bias, illustrating how these systems perpetuate inequities across domains such as employment, healthcare, education, and criminal justice. Participants emphasized that biased algorithms act as feedback loops that reproduce structural exclusion under the guise of neutrality. Overreliance on automated decision-making was shown to erode human judgment and moral agency, while technological exclusion rendered marginalized identities literally invisible within digital systems. These findings collectively support that AI bias not only reinforces existing inequities but also reshapes social perception and moral accountability in technologically mediated environments.

Research Question Three explored strategies practitioners recommend for advancing fairness and equity in AI system design and use. The findings revealed three interrelated approaches: embedding ethics early in the development process, institutionalizing accountability through participatory and transparent design practices, and advocating for regulatory and educational reform. Participants stressed that ethical awareness must be cultivated as both a structural priority and a cognitive discipline, supported by policy enforcement, interdisciplinary collaboration, and reflexive education. Rather than viewing ethics as a compliance requirement, practitioners conceptualized it as a continuous practice of moral reasoning embedded within technical work.

Supplementary findings deepened the analysis by revealing two cross-cutting dynamics that shape practitioner engagement with ethical AI. The first, emotional and moral fatigue, captured the psychological cost of sustaining ethical commitment within organizations driven by performance metrics and innovation cycles. The second, evolving human–machine collaboration, reflected shifting attitudes toward recalibrating trust and responsibility between human expertise and algorithmic systems. These insights highlight that ethical AI development requires not only technical refinement but also organizational empathy and distributed moral responsibility.

The findings from Chapter Four provide a comprehensive understanding of AI bias as both structural and cognitive, institutional and personal. Practitioners lived experiences demonstrate that fairness and accountability cannot be achieved through technical optimization alone; they demand systemic transformation, reflective governance, and sustain attention to the human dimensions of technological innovation. This chapter established the empirical foundation for the interpretive discussion in Chapter Five, which situates these findings within existing scholarship, evaluates their theoretical implications, and outlines practical recommendations for promoting equitable and socially responsible AI development.

# Chapter Five

## Summary, Discussion, and Implications

### Introduction

This final chapter synthesizes the study's purpose, findings, and implications. It begins by summarizing the key elements of chapters one through four and concludes with an overview of the study's limitations and recommendations for future research.

Chapter One established the foundation for examining bias, accountability, and fairness in artificial intelligence (AI), arguing that AI, often presented as objective, is deeply entangled with human systems of power, governance, and cognition. The study drew on Intersectionality Theory and Cognitive Science to explore how practitioners who build, manage, and regulate AI systems perceive their ethical responsibility across critical sectors, including healthcare, education, employment, and criminal justice.

Chapter Two reviewed scholarship on responsible AI, highlighting the persistent gap between ethical principles and their implementation in practice. It traced the historical and technical roots of algorithmic bias, emphasizing that fairness cannot be separated from the social hierarchies and cognitive processes that shape design decisions. The chapter also uses case studies across various sectors, including healthcare, education, employment, and criminal justice, to demonstrate how AI bias manifests in tangible, real-world ways that nearly every individual encounters at some point in their life. These examples illustrate how systemic inequities become embedded in technological systems, reinforcing the need for ethical frameworks that are both socially and cognitively informed.

Chapter Three outlined the study's qualitative methodology, which triangulated case studies, scholarly documentary analysis, and semi-structured interviews across four domains. This approach captured diverse practitioner perspectives, enabling a multidimensional understanding of AI bias through interpretive thematic analysis.

Chapter Four presented findings organized around three guiding research questions. Practitioners described algorithmic bias as both technical and social, arising from historical inequities and institutional priorities. Despite awareness of fairness principles, organizational cultures often prioritize efficiency over ethical reflection, resulting in fragmented accountability. Nevertheless, participants demonstrated reflexivity and proposed reforms such as embedding ethics into design, fostering interdisciplinary review, and strengthening governance oversight. These findings reveal that AI fairness requires systemic transformation rather than isolated technical fixes. Across domains, individual insights coalesced into three consistent patterns: ethical reasoning is often constrained by institutional incentives, fairness is deprioritized when policies are ambiguous, and emotional labor disproportionately affects underrepresented practitioners. These patterns directly align with the study's research questions and illustrate how specific challenges in healthcare, education, employment, and criminal justice are part of a larger sociotechnical problem. They also expose interconnections among cognition, conscience, and context in shaping AI development. To interpret these relationships and translate them into guidance for the ethical implementation of AI, the researcher developed an applied framework that synthesizes participant insights into a model for responsible AI design.

Overall, the study emphasizes that AI reflects human values as much as it augments human capability. When designed and governed with intention, it can promote inclusion; when shaped by unchecked institutional or market pressures, it risks reproducing inequality. Participants expressed both moral fatigue and optimism, stressing the importance of education, collaboration, and participatory governance. Building upon these findings, this chapter situates the results within the context of prior research, identifies study limitations, and outlines implications for future inquiry and practice.

**Practical Assessment of Research Questions**

This section discusses the practical assessment of the study's research questions, required to situate the findings within the broader context of existing scholarship on responsible artificial intelligence (AI). The current study both reinforced and expanded prior research by illuminating how bias, fairness, and accountability manifest in real organizational and societal settings. Through a qualitative design that triangulated case studies, scholarly documentary analysis, and practitioner interviews, the study revealed that algorithmic bias is not solely a technical flaw but a reflection of cognitive tendencies, institutional priorities, and cultural hierarchies. This research contributes to the fields of AI ethics and governance by providing an applied understanding of how cognition, conscience, and context collectively shape fairness outcomes. Moreover, it strengthens the link between human-centered computing and the principles of social justice central to the degree program, offering both theoretical and practical insights for creating equitable and transparent AI systems.

**Research Question One**

This research question examined how practitioners understand the origins of bias in AI and how it operates. Participants described bias as emerging from interconnected social, organizational, and cognitive processes rather than isolated technical errors. They viewed it as a systemic inheritance that reflects human assumptions and institutional priorities. This pattern appeared consistently across domains. This theme was particularly visible in criminal justice, where practitioners discussed biased policing data feeding predictive models, and in healthcare, where cost-based risk proxies ignored systemic disparities in care. These examples illustrate how domain-specific experiences reflect a shared pattern of historical inequities being codified into AI systems. These cross-case parallels suggest that bias is not an isolated error, but rather a function of structural and cognitive conditions that are reproduced through organizational practice. These findings align with prior research suggesting that algorithmic bias mirrors societal hierarchies and cultural norms embedded in data and design (Barocas & Selbst, 2016; Buolamwini & Gebru, 2018; Noble, 2018).

**Theme One: Bias as Historical Inheritance**

Intersectionality Theory provides a critical framework for understanding how historical inequities are embedded within data infrastructures and algorithmic systems. Findings from this study confirm that datasets are not neutral artifacts but repositories of social history, reflecting patterns of exclusion and power that persist across institutional contexts. Participants recognized these legacies most clearly in healthcare and criminal justice, where the consequences of biased data are both measurable and deeply human. Their insights align with Obermeyer et al. (2019), who demonstrated that a healthcare algorithm underestimated the needs of Black patients because spending data were used as a proxy for care, and with Angwin et al. (2016), who revealed racial disparities in the COMPAS risk assessment tool. These examples illustrate that algorithmic systems may perform as designed yet fail to serve the very populations they are intended to help. Participants repeatedly emphasized that “data never exist in a vacuum,” reinforcing Eubanks’s (2018) argument that technology frequently replicates the social hierarchies it claims to transcend.

Cognitive Science further illuminates why such inequities are routinely overlooked in the development process. Practitioners described how familiar mental shortcuts, particularly anchoring and confirmation bias, can lead developers to treat quantitative data as objective evidence even when it reflects structural discrimination (Brem & Rivieccio, 2024). This cognitive framing helps explain why technical teams often mistake the precision of data for fairness, assuming accuracy equates to neutrality.

Together, these theoretical and empirical insights demonstrate that algorithmic bias is both a cultural artifact and a cognitive simplification. It emerges when systemic discrimination intersects with unexamined human reasoning, allowing inherited inequities to be recoded as technical logic. Addressing this issue, therefore, requires more than corrective algorithms or technical band-aids; it demands critical awareness of how historical, social, and cognitive forces converge in the design and interpretation of AI systems.

**Theme Two: Organizational Incentives and Ethical Trade-offs**

Practitioners described professional cultures that prioritize speed, scalability, and measurable performance outcomes over sustained ethical reflection. Although participants represented diverse roles and industries, they shared a consistent perception that AI development is driven by organizational incentives favoring rapid innovation and quantifiable success. Their observations align with findings that AI ethics often remain symbolic when not accompanied by structural accountability mechanisms (Hagerty & Rubinov, 2019). Similar concerns are evident across the literature, where ethical initiatives often operate as rhetorical commitments rather than enforceable practices (Green, 2021; Mökander et al., 2023). From an intersectional perspective, these institutional logics mirror broader societal hierarchies, privileging economic efficiency and market expansion over equity and social responsibility.

Cognitive Science provides a complementary lens for understanding why these patterns persist. Under the pressure of deadlines, funding expectations, and performance targets, practitioners experience what Adepoju and Adepoju (2023) describe as bounded ethical rationality, a narrowing of moral reasoning due to cognitive and situational constraints. Participants acknowledged that they often “move fast and scale” to meet business or institutional goals, even when those choices conflict with their ideals of fairness. This suggests that ethical compromise in AI is not only a governance issue but also a cognitive adaptation to organizational norms that equate productivity with success.

These insights indicate that ethical shortcomings in AI development emerge less from individual indifference and more from structural and psychological conditioning. When innovation is primarily measured by efficiency and output, fairness becomes secondary. Addressing this dynamic requires redesigning organizational incentive systems to embed moral deliberation and accountability into the very metrics by which technological progress is judged.

**Theme Three: Cognitive Reflection and Practitioner Self-Awareness**

Cognitive Science provides a valuable lens for understanding how practitioners interpret and respond to bias in artificial intelligence systems. Prior research highlights that developers and decision-makers often rely on mental shortcuts, such as automation bias, confirmation bias, and cognitive ease, that shape how they assess fairness and risk within algorithmic outputs (Brem & Rivieccio, 2024; Buçinca et al., 2021; Parasuraman & Riley, 1997). These tendencies lead individuals to trust data-driven results as inherently objective, even when those results reflect structural inequities. Chapter Two emphasized that cognitive biases are not simply individual mistakes, but patterned responses shaped by professional norms, institutional priorities, and cognitive load.

Participants in this study demonstrated strong awareness of these cognitive dynamics. They frequently acknowledged that human reasoning plays a critical role in shaping the ethical contours of AI, noting that “the machine just learns from what we give it” and that unexamined assumptions often influence how models are built, tested, and interpreted. Several practitioners described moments in which they recognized their own cognitive shortcuts, such as overreliance on familiar data sources or the tendency to defer to model outputs under time pressure. These reflections align with the literature, which suggests that practitioners’ cognitive habits directly influence the perpetuation of algorithmic bias, particularly when those habits remain unchallenged in fast-moving development environments (Buçinca et al., 2021).

Intersectionality Theory adds further nuance by revealing how practitioners’ social positions influence their awareness of bias. Participants from historically marginalized groups were more likely to question the assumptions embedded in the data or probe model behavior that seemed misaligned with their lived realities. This observation supports the argument that individuals with direct experience of systemic inequity often possess heightened attune to technological harms (Benjamin, 2019). In contrast, participants who described themselves as distant from the populations most affected by biased AI acknowledged that their worldview initially limited their ability to recognize subtle forms of exclusion or harm.

These findings reveal that cognitive reflection is both an individual and collective practice. Practitioner self-awareness is essential for mitigating bias, but it is also shaped by institutional culture, social identity, and the cognitive pressures inherent in technological work. Addressing algorithmic bias, therefore, requires more than technical corrections; it demands deliberate cultivation of ethical self-reflection, ongoing awareness of cognitive limitations, and shared practices that encourage practitioners to interrogate assumptions rather than normalize them. This theme highlights the importance of educational and organizational structures that foster critical inquiry, slow thinking, and reflective practice in AI development.

**Research Question Two**

This research question examined how practitioners perceive the societal consequences of AI bias across major institutional domains. Participants described impacts that extend well beyond technical inaccuracies, emphasizing how biased systems can shape access to resources, influence decision-making, and reinforce longstanding racial and socioeconomic disparities. Their insights align with existing research documenting unequal outcomes in healthcare algorithms, educational analytics, risk assessment tools, and automated hiring systems. Across interviews, practitioners highlighted that AI bias not only replicates historical inequity but also amplifies it at scale, producing harms that disproportionately affect marginalized communities. This theme highlights that the societal implications of AI are structural and cumulative, revealing how algorithmic systems can either deepen or disrupt existing patterns of injustice depending on how they are designed, governed, and deployed.

**Theme One: Disparate Access and Resource Allocation**

Disparate access and resource allocation emerged as a central societal impact of AI bias across all four domains examined in this study. Participants consistently described how algorithmic systems shape those who receive opportunities, support, or intervention, often reproducing long-standing inequities rather than correcting them. These perceptions align directly with the case studies analyzed in Chapter Two, which demonstrate how biased algorithms influence access to healthcare, educational opportunities, employment pathways, and criminal justice outcomes.

The healthcare case study highlighted that a widely used risk-prediction algorithm systematically underestimated the care needs of Black patients because it relied on healthcare spending as a proxy for illness severity (Obermeyer et al., 2019). Participants echoed this dynamic, noting that “systems may work as designed, but not for everyone they claim to serve,” particularly when social disadvantages reduce the spending data that algorithms rely on. Similarly, the criminal justice case demonstrated that the COMPAS algorithm assigned higher risk scores to Black defendants even when their criminal histories were comparable to those of white defendants (Angwin et al., 2016). Practitioners in this study drew parallels between such outcomes and contemporary uses of automated decision systems, emphasizing that bias in risk scoring “creates unequal trajectories” that expand over time.

Educational and employment case studies provided further evidence of how AI mediates access to resources and opportunity. Prior research cited in Chapter Two has documented that predictive analytics in schools often penalize students from under-resourced communities (Ajunwa, 2019; Baker & Hawn, 2021). Additionally, automated hiring systems have been found to filter applicants using linguistic or demographic proxies that disadvantage women and people of color. Participants reinforced these patterns through their own experiences, noting that automated screening tools “narrow the field before humans ever see the candidates,” and that algorithmic sorting in educational settings can “lock students into assumptions” that limit support and intervention.

By triangulating insights from the literature, documentary case studies, and practitioner interviews, the study reveals that AI systems do not merely reflect existing patterns of unequal access; they amplify them by embedding them into automated decision-making. Participants emphasized that these impacts are particularly pronounced for marginalized communities, who face compounded disadvantages when historical inequities are operationalized through algorithmic logic. This theme supports the idea that the societal implications of AI bias are not abstract; they materialize in the allocation of care, the distribution of opportunity, and the administration of justice. Addressing these disparities requires confronting both the technical design of algorithms and the social structures that shape the data on which they are built.

**Theme Two: Automation of Harm at Scale**

A second significant societal impact identified in this study is the widespread automation of harm. Participants emphasized that algorithmic systems do not merely reproduce existing inequities—they accelerate and institutionalize them by embedding biased logic into high-volume, high-speed decision pipelines. This finding aligns with the documentary case studies and literature reviewed in Chapter Two, which collectively show how automated systems amplify racial, socioeconomic, and gender disparities when deployed at scale across healthcare, education, criminal justice, and employment.

In the criminal justice case study, Angwin et al. (2016) demonstrated that the COMPAS risk assessment algorithm disproportionately flagged Black defendants as high risk. Practitioners in this study drew direct connections between tools like COMPAS and broader patterns of harm, noting that "once an algorithm mislabels someone, that label follows them," producing long-term disadvantages in sentencing, parole decisions, and community supervision. The inequities mentioned echo Benjamin's (2019) argument that automated decision systems generate "coded inequity" by amplifying the consequences of racialized assumptions embedded in historical data.

The healthcare case study offered similar evidence. As discussed in Chapter Two, Obermeyer et al. (2019) found that biased prediction models systematically diverted resources away from Black patients. Practitioners described these dynamics as a "multiplier effect," explaining that minor inaccuracies in algorithmic triage can have significant downstream effects in treatment planning, chronic disease management, and resource distribution. When biased logic is embedded within widely used hospital systems, the scale and speed of automation can produce cascading harm that become increasingly difficult to detect or reverse.

Education and employment contexts further illustrate this theme. As noted by Baker and Hawn (2021), predictive analytics in education often reinforce inequities by assigning risk scores that shape access to advanced coursework, tutoring, and behavioral interventions, frequently penalizing students from low-income or marginalized backgrounds. Participants echoed this concern, noting that educational algorithms “do not just predict performance; they shape it,” altering the supports students receive and influencing teachers’ expectations. In employment, case studies and literature, such as Ajunwa (2019), have shown that automated hiring systems screen out qualified candidates based on proxy variables, including name, ZIP code, writing style, or work history gaps. Practitioners affirmed that these systems “remove the human in the loop,” allowing bias to operate unchecked, especially in early filtering stages that applicants never see.

Collectively, these sources suggest that automation fails to eliminate human bias; instead, it is embedded and amplified within institutional systems. Participants emphasized that harms become more entrenched when organizations assume that algorithms enhance objectivity, resulting in overreliance on automated outputs and reduced human oversight. As one practitioner noted, “automation gives bias a megaphone.” Bias at scale aligns with Crawford’s (2021) argument that AI systems often function as “infrastructures of inequality,” scaling discriminatory outcomes under the guise of technical rationality.

The theme of automation of harm at scale underscores the urgent ethical stakes of AI deployment. When biased systems are integrated into institutional workflows, errors and inequities do not simply replicate; they compound, spread, and become entrenched in practice. Addressing this challenge requires explicit governance mechanisms, transparent auditing, and a commitment to continual human evaluation rather than unquestioned faith in automated systems.

**Theme Three: Reinforcement of Stereotypes and Cultural Erasure**

A third significant societal impact identified in this study is the reinforcement of stereotypes and the resulting cultural erasure produced through biased AI systems. Participants described how generative tools, hiring platforms, educational analytics, and predictive systems often default to normative representations that center whiteness and Western cultural ideals. They explained that these outputs fail to reflect the cultural and demographic diversity of real-world communities. This concern aligns with research indicating that image generation models often overrepresent white and Eurocentric features, marginalizing other identities (Buolamwini & Gebru, 2018; Noble, 2018). Studies have also documented that language models reproduce racialized and gendered associations present in historical text corpora, reinforcing existing stereotypes within automated outputs (Caliskan et al., 2017).

The case studies presented in Chapter Two further demonstrate how these patterns are evident across various institutional domains. The criminal justice case study illustrated how the COMPAS algorithm associated criminality with Blackness, contributing to racialized risk assessments that shape judicial decision-making (Angwin et al., 2016). The education case study revealed that predictive analytics often associate student performance with demographic markers, thereby reinforcing assumptions about the abilities of students from marginalized backgrounds (Baker & Hawn, 2021). The employment case study revealed that automated hiring systems penalize applicants whose names, communication styles, or educational trajectories do not conform to dominant cultural norms, resulting in the systematic filtering of diverse candidates (Ajunwa, 2019). These examples demonstrate that stereotype reinforcement extends beyond technical error and operates as a channel through which cultural narratives are woven into institutional decisions.

Interview findings reinforced these concerns. Participants described multiple instances where image generators produced mostly white faces even when they requested diverse outputs. Image generators also exhibit this pattern, frequently producing stereotypical depictions even when responding to neutral prompts. Others noted that language models tended to describe leadership, professional competence, or beauty using traits aligned with Western and male-centered norms. Practitioners from underrepresented groups expressed that these systems made them feel invisible or inaccurately portrayed. These reflections support Benjamin's argument that seemingly neutral technologies reproduce social hierarchies by encoding racial and cultural exclusion into automated processes (Benjamin, 2019).

Findings from literature, case studies, and interviews indicate that AI systems do more than reflect cultural stereotypes. They institutionalize these patterns through large-scale automated outputs that shape public perception, organizational decision-making, and social narratives. This theme highlights that algorithmic bias creates not only material disadvantages but also symbolic and cultural harms, which can distort representation and narrow the range of identities that technology recognizes or values. Addressing these issues requires deliberate design choices that diversify training data, interrogate cultural assumptions, and build systems intended to challenge rather than reinforce exclusionary norms.

**Theme Four: Erosion of Trust and Public Confidence**

Erosion of trust and public confidence emerged as a significant societal impact of AI bias. Participants explained that people often do not view automated systems as acting in their best interest, particularly when outcomes appear opaque, inaccurate, or unfair. Rather than perceiving AI as a neutral aid, individuals increasingly approach automated decisions with caution. This perspective aligns with research in Chapter Two, which shows that biased AI systems undermine institutional legitimacy, especially among groups historically marginalized in healthcare, education, criminal justice, and employment settings (Crawford, 2021).

The case studies provide further evidence of how algorithmic bias diminishes public trust. In the criminal justice case study, use of the COMPAS tool fostered the belief that risk assessments were predetermined or racially skewed, reducing confidence in judicial decision-making (Angwin et al., 2016). The healthcare case demonstrated that risk-prediction tools underestimated the needs of Black patients, contributing to skepticism about whether AI-assisted triage and resource allocation could be applied equitably (Obermeyer et al., 2019). In education, predictive analytics that assign "high risk" labels based on demographic profiles prompted concerns among families who felt their children were being unfairly categorized (Baker & Hawn, 2021). Employment case findings showed that opaque automated screening practices generated feelings of invisibility and exclusion, reinforcing doubts about the fairness of hiring decisions (Ajunwa, 2019). Together, these cases illustrate how algorithmic opacity and unequal outcomes weaken confidence in institutional processes.

Interview data reinforced these concerns, particularly among participants who described the heightened skepticism marginalized communities bring to AI-driven systems. They noted that when people repeatedly encounter biased or unexplained outcomes, distrust becomes a rational response. Participants also observed that the perceived neutrality of AI can intensify mistrust, as individuals interpret the mismatch between promised fairness and lived experience as evidence that discrimination is being obscured behind technical language. This concern echoes arguments in the literature that AI systems can exacerbate distrust when inequity is embedded within algorithmic logic (Crawford, 2021).

Collectively, the literature, case studies, and interview findings show that biased AI systems weaken public trust by creating a disconnect between institutional commitments to fairness and the outcomes experienced by communities. This erosion of confidence extends beyond individual interactions to the broader legitimacy of the institutions deploying automated tools. Restoring trust requires transparent communication, participatory governance, and careful monitoring of automated decision processes. Without these interventions, AI risks widening existing divides between communities and the systems intended to support them.

**Research Question Three**

This research question explored the strategies practitioners believe are necessary to reduce bias and promote equity in AI development and deployment. Participants emphasized that effective mitigation requires more than technical adjustments; instead, it depends on structural, educational, and governance-based reforms. They emphasized the importance of interdisciplinary collaboration, diverse representation in development teams, transparent auditing practices, and ongoing ethics education for both creators and users of AI systems. These recommendations align with the literature reviewed in Chapter Two, which identifies fairness governance, explainability, inclusive design, and accountability mechanisms as foundational to responsible AI. Participants' insights also build on the findings from Chapter Four by emphasizing that achieving equitable AI requires continuous human involvement, culturally informed evaluation, and organizational commitment to ethical practice.

**Theme One: Cognitive Reflection, Cultural Nuance, and the Inclusion of Impacted Voices**

A critical insight emerging from this study is that ethical AI cannot exist without the inclusion of cultural nuance and the lived experiences of those most impacted by algorithmic decisions. Practitioners repeatedly emphasized that while technical audits and fairness metrics can identify patterns of disparity, they cannot fully account for the social meanings and contextual realities that underlie bias This aligns with previous research (Ulnicane, 2024; Raji & Buolamwini, 2019), who argue that ethical oversight must extend beyond compliance checklists to include participatory frameworks that center the perspectives of marginalized users and communities.

Participants described how exclusion from design and governance processes allows bias to persist even after deployment. As stated by Participant D, "AI systems can technically perform as intended yet still fail to meet the needs of certain populations, particularly when design teams lack adequate cultural and demographic representation." Her point framed bias not as a system malfunction, but as exclusion by design. This supports an essential point: bias is not simply a data problem; it is a representation problem. When cultural nuances, linguistic variations, or community-specific experiences are excluded from model training, AI systems risk reinforcing dominant narratives, known as hegemonic narratives, while erasing others. Across cases, a transferable insight emerged: ethical reflection is often decoupled from day-to-day decision-making due to ambiguous mandates and fragmented responsibility structures. This was particularly evident in employment and education, where teams lacked consistent oversight or ethical guidance documents. However, some participants described promising practices, such as cross-functional reviews, which offered scalable interventions. These examples suggest that institutional design is critical to sustaining fairness in society.

Intersectionality Theory helps explain why this exclusion is particularly harmful. It reveals how race, gender, class, and other identity markers intersect to shape unequal access to participation and voice. When development teams lack demographic and experiential diversity, their design assumptions often reflect the worldview of those in power, rather than those most affected by algorithmic decisions. For example, recruitment systems that fail to include gender-diverse data or educational models that lack linguistic inclusivity illustrate how design omissions translate into systemic disadvantages. Including voices from impacted groups during design, model training, testing, and maintenance, therefore, becomes an ethical imperative rather than a symbolic gesture.

Cognitive Science provides an additional layer of interpretation. Practitioners often rely on mental models that equate neutrality with fairness, leading to overconfidence in algorithmic outputs and an underestimation of social context (Vakali & Tantalaki, 2024). These cognitive shortcuts contribute to what Brem and Rivieccio (2024) describe as “perceptual narrowing,” where developers unconsciously overlook cultural variation because it appears statistically insignificant. The integration of diverse perspectives acts as a cognitive corrective, expanding the interpretive frame through which data and design decisions are made.

Participants also highlighted the importance of sustained community engagement, not just at the design stage, but throughout the system’s lifecycle. This includes co-developing metrics of success with impacted users, conducting longitudinal bias monitoring after deployment, and ensuring that feedback mechanisms are built into governance models. As *Participant C* explained, “Ethical AI doesn’t end with deployment; it’s a relationship that must be maintained." This view reflects an emerging paradigm in responsible AI—one that prioritizes maintenance, repair, and accountability as ongoing, culturally responsive practices rather than one-time interventions. Through understanding, trust, and community, you enable those who lack representation to feel empowered as they co-design through active participation.

Integrating cultural nuance and participatory feedback mechanisms strengthens both ethical and technical robustness. Culturally aware systems are more likely to generalize appropriately across contexts, while inclusive governance processes build legitimacy and public trust. These findings support Hagerty and Rubinov's (2019) call for "ethical infrastructures" that embed fairness as an operational norm, not an aspirational ideal.

The findings demonstrate that advancing fairness in AI requires bridging the epistemic gap between those who build technology and those who live with its consequences. When cultural nuance and marginalized voices are integrated throughout the design, development, and maintenance process, AI can shift from a tool that reproduces inequity to one that actively contributes to social understanding and justice.

**Theme Two: Ethics by Design and Interdisciplinary Oversight**

Participants consistently emphasized that meaningful bias mitigation requires integrating ethical considerations during the earliest stages of AI development rather than attempting to retrofit fairness after deployment. Several practitioners noted that ethics reviews are often positioned as optional "add-ons" near the end of the design cycle, which limits their influence on model architecture, data selection, feature engineering, and evaluation protocols. This concern reflects a broader call for structural change within development pipelines, as participants described the need for interdisciplinary review processes, routine cross-functional assessments, and governance checkpoints that embed ethical reflection into routine decision-making rather than treating it as a compliance exercise.

These observations align with arguments in the literature that ethics must be structurally embedded to be effective. The literature suggests that many organizational ethics initiatives remain symbolic when they lack institutional mechanisms that compel teams to examine social impact critically, justify design decisions, or document trade-offs (Hagerty & Rubinov, 2019). Scholars further argue for sociotechnical approaches that view fairness as a property of systems shaped by institutions, values, and human judgment, rather than relying solely on technical optimization (Green, 2021). Additional research demonstrates that fairness failures often occur when designers abstract away from social context, resulting in systems that overlook the real-world conditions under which predictions will be interpreted and acted upon (Selbst et al., 2019).

The case studies reinforce these concerns. In healthcare, the absence of ethics-by-design processes contributed to the deployment of a risk-prediction algorithm that underestimated the needs of Black patients, leading to inequitable resource allocation (Obermeyer et al., 2019). In the context of criminal justice, the COMPAS tool revealed that fairness cannot be achieved solely through statistical adjustments when the underlying design lacks governance structures capable of accounting for social context or historical inequities (Angwin et al., 2016). Both cases illustrate how insufficient early-stage oversight can encode structural inequalities directly into model behavior.

The interviews, literature, and case studies collectively suggest that ethics-by-design necessitates a more comprehensive approach to mitigation. It demands early integration of ethical reasoning, interdisciplinary expertise, and organizational structures that redistribute responsibility for fairness across teams. Participants' recommendations point toward governance models that prioritize proactive ethical attention, collaborative review mechanisms, and clear accountability pathways that align with sociotechnical frameworks emphasizing fairness as a shared and ongoing responsibility rather than a discrete step in the development pipeline.

**Theme Three: Transparent Auditing and Accountability Mechanisms**

Participants emphasized that effective bias mitigation requires transparent auditing practices that extend beyond one-time evaluations. They explained that regular and ongoing audits are necessary to identify unintended consequences, monitor system performance over time, and capture how models behave in real-world environments. Several practitioners noted that auditing is most effective when conducted by advisory groups that represent diverse professional backgrounds and lived experiences, allowing fairness evaluations to incorporate perspectives from the communities most affected by automated decisions. They argued that such diversity is essential to ensuring equitable implementation, because homogenous review teams may overlook patterns of harm or assumptions embedded in system design. Participants also emphasized the importance of thorough model documentation, transparent disclosure practices, and explainable tools that enable stakeholders to understand how automated decisions are made.

These insights align with the literature presented in Chapter Two, which identifies model audits as essential fairness controls and central pillars of responsible AI. Research emphasizes that transparency in both model documentation and accountability mechanisms is foundational for evaluating disparate impact and identifying hidden patterns of unfairness (Raji et al., 2020). Scholars also argue that accountability requires institutions to open their systems to scrutiny, ensuring that fairness considerations are not siloed within technical teams but evaluated through multidisciplinary and socially informed processes (Barocas et al., 2019). The emphasis on diverse oversight groups in the data reflects these recommendations by demonstrating how interdisciplinary and community-informed auditing strengthens the rigor of fairness evaluations.

The case studies reinforce the importance of structured, continuous auditing. In healthcare, the absence of a robust fairness review contributed to the use of a risk-prediction model that underestimated the needs of Black patients, affecting resource allocation outcomes (Obermeyer et al., 2019). In the criminal justice case, the COMPAS risk assessment tool produced racially disparate predictions without transparent documentation or avenues for external review, limiting the ability of defendants and courts to contest erroneous outputs (Angwin et al., 2016). These examples demonstrate that the absence of auditing, particularly by diverse and representative oversight bodies, can permit inequitable systems to operate without meaningful accountability.

Together, the interviews, literature, and case studies suggest that transparent auditing and accountability structures are crucial for advancing equity in AI. Participants' recommendations emphasize the importance of audits led by diverse advisory groups, standardized documentation practices, and clear governance frameworks that mandate institutions to assess fairness throughout the system's lifecycle. They also emphasize the importance of national or global benchmarks that establish minimum expectations for equitable implementation and transparency. These findings suggest that responsible AI depends on oversight mechanisms capable of questioning, contextualizing, and correcting automated decisions, ensuring that organizations remain accountable for the social impacts of their systems.

**Theme Four: Education, Literacy, and Public Awareness**

Participants consistently emphasized that improving AI literacy is foundational to advancing equity in automated systems. They explained that many harms associated with AI arise because everyday users, institutional leaders, and even some developers lack a clear understanding of how these systems function and how bias becomes embedded in data and model behavior. Several practitioners argued that education empowers communities to question automated outputs, advocate for fair treatment, and recognize when system behavior conflicts with their lived experiences. Participants also stressed that organizations cannot rely solely on technical fixes when users and decision-makers do not understand the limitations, assumptions, or risks inherent in AI systems. For many, education represented both a protective measure and a form of empowerment that would allow the public to interpret and contest harmful outcomes.

These observations align with the literature presented in Chapter Two, which shows that widespread misunderstanding of AI contributes to overtrust, misuse, and uncritical acceptance of algorithmic outputs (Crawford, 2021). Scholars have argued that public awareness is essential to resisting coded inequity and preventing biased systems from shaping social narratives without resistance or scrutiny (Benjamin, 2019; Noble, 2018). These works highlight that the social consequences of AI cannot be separated from the public's ability to interpret automated decisions and understand how historical inequities influence system outputs. Participants' calls for broad AI literacy echo these findings by positioning education as a necessary element of responsible AI practice.

The case studies further reinforce the importance of public awareness. In healthcare, a limited understanding of triage algorithms contributed to unchallenged disparities in care allocation, allowing biased predictions to shape treatment decisions (Obermeyer et al., 2019). The education case study revealed that families frequently lacked understanding of how risk scores were calculated, resulting in mistrust and a diminished ability to challenge school decisions (Baker & Hawn, 2021). In employment, job seekers frequently had no insight into how automated screening tools evaluated their applications, resulting in a sense of invisibility and limited recourse for addressing unfair outcomes (Ajunwa, 2019). These examples show that a misunderstanding of AI operations not only exacerbates inequitable outcomes but also limits individuals' ability to respond to or resist them.

Together, the interviews, literature, and case studies demonstrate that education, literacy, and public awareness are essential strategies for mitigating bias and strengthening community resilience to harmful AI outcomes. Participants' recommendations highlight the need for training programs that cater to diverse user groups, including developers, institutional leaders, and members of the public. They also suggest that organizations should prioritize transparent communication, accessible explanations of system behavior, and opportunities for users to ask questions or challenge automated decisions. These findings suggest that meaningful fairness efforts must consider not only how AI is developed but also how it is perceived, interpreted, and interacted with by the people it affects.

**Theme Five: Regulatory Standards and Governance Frameworks**

Participants emphasized that meaningful mitigation of AI bias requires clear regulatory standards and governance structures that extend beyond internal ethics guidelines. Many described current AI oversight as fragmented, inconsistent, or dependent on the goodwill of individual organizations. They argued that without binding frameworks that incentivise organizations to prioritize equality, fairness initiatives remain voluntary and are often deprioritized in environments that value speed, scalability, or market competitiveness. Several practitioners suggested the need for global or national benchmarks that define minimum expectations for fairness, transparency, documentation, and public accountability across sectors.

They also highlighted the importance of regulatory bodies that include representation from marginalized communities to ensure that governance decisions reflect diverse public interests.

These concerns align with the literature in Chapter Two, which notes that the absence of enforceable standards contributes to a gap between institutional aspirations for fairness and the reality of AI deployment. Scholars argue that responsible AI requires legal and regulatory mechanisms that mandate transparency and prohibit discriminatory outcomes, rather than relying solely on industry-led self-regulation (Barocas et al., 2019; O'Neil, 2016). The literature further indicates that governance frameworks grounded in equity are essential to prevent organizations from circumventing their ethical responsibilities in the pursuit of efficiency or profit (Crawford, 2021).

The case studies illustrate the consequences of insufficient governance. In healthcare, the risk-prediction algorithm deployed without adequate regulatory oversight underestimated the needs of Black patients, revealing the limitations of internal review processes (Obermeyer et al., 2019). In criminal justice, the use of the COMPAS tool demonstrated how the lack of standards for auditing and transparency allowed racially disparate outcomes to persist without adequate mechanisms for challenge or recourse (Angwin et al., 2016). These examples show that when governance frameworks are weak or absent, biased systems can influence critical decisions without meaningful accountability.

Together, the interviews, literature, and case studies indicate that regulatory standards and governance frameworks are essential for ensuring equitable AI implementation. Participants' recommendations emphasize the need for standardized fairness metrics, external auditing requirements, public reporting mandates, and diverse oversight bodies with the authority to enforce compliance. These findings suggest that mitigation strategies must be supported by institutional and legal infrastructures that embed equity into the broader ecosystems in which AI systems operate.

## Supplemental Findings

### Theme One: Emotional and Moral Fatigue in Ethical AI Practice

This supplementary finding highlights that ethical AI entails a substantial emotional and moral burden for practitioners, particularly those who consistently raise concerns about fairness within institutions focused on rapid development cycles, performance metrics, and key productivity indicators. Participants described feeling exhausted and discouraged when their concerns about fairness were repeatedly deprioritized or dismissed as impractical. This phenomenon reflects what several interviewees termed “ethical fatigue,” a form of emotional strain that emerges when practitioners are expected to sustain moral commitment in environments that value delivery speed over ethical deliberation. The practical significance of this finding lies in its implications for organizational readiness and the quality of AI outcomes. When practitioners feel isolated in advocating for fairness, ethical concerns may be downplayed or left unaddressed, increasing the likelihood that systems are deployed without adequate scrutiny, thereby heightening the risk of biased or inequitable outcomes. These dynamics are further complicated by disparities in how emotional fatigue is experienced. Practitioners from underrepresented backgrounds, particularly women and racial minorities, reported heightened strain when implicitly expected to represent inclusive or equity-oriented viewpoints within teams where such perspectives were otherwise absent. Their accounts align with scholarship demonstrating that marginalized professionals often carry disproportionate moral and emotional labor in technical environments where diversity remains limited (Collins, 2019).

The emotional and moral fatigue identified in this study is consistent with broader research documenting the pressures faced by "ethics advocates" or "ethics champions" in technology settings. Scholars such as Gray and Suri have demonstrated that individuals who raise ethical concerns frequently encounter structural resistance or performative acknowledgment, rather than meaningful institutional support, which contributes to burnout and disengagement (Gray & Suri, 2019). Other studies similarly describe how organizational barriers undermine ethical work, leaving those who champion fairness to shoulder responsibilities that institutions have not meaningfully integrated into their operational structures (Metcalf et al., 2021). From a Cognitive Science perspective, the emotional strain experienced by practitioners reflects the psychological cost of cognitive dissonance, a state in which personal moral standards conflict with organizational incentives or performance pressures (Festinger, 1957). Over time, unresolved dissonance may lead to withdrawal, rationalization, or acquiescence to institutional norms, thereby diminishing practitioners' willingness and ability to advocate for fairness. The intersectional dimension of this fatigue further reinforces that the emotional labor required for ethical AI work is not evenly distributed; individuals from marginalized identities often report intensified strain due to both systemic underrepresentation and heightened expectations of ethical vigilance (Collins, 2019).

The practical implications of this finding underscore the need for organizations to treat fairness and accountability as collective, structural responsibilities rather than burdens carried by individuals. Ethical AI work must be integrated into the design, development, and evaluation pipeline through explicit processes, rather than relying on the moral resolve of isolated practitioners. When institutions fail to embed ethical reflection structurally, practitioners are left to navigate tensions between values and productivity on their own, exacerbating emotional strain and weakening the organization's ability to detect or prevent bias. Addressing emotional fatigue, therefore, requires organizations to normalize ethical deliberation within standard workflows, support psychologically safe communication channels that allow practitioners to raise concerns without fear of retribution, and cultivate leadership awareness of the emotional labor embedded in ethical decision-making. Institutions must also avoid disproportionately relying on underrepresented practitioners to provide inclusive perspectives; instead, they should distribute ethical responsibility across roles and ensure that fairness-related deliberation is supported through formal governance structures, dedicated resources, and shared accountability.

In sum, this supplementary finding reveals that emotional and moral fatigue are critical but often overlooked dimensions of ethical AI practice. It demonstrates that ethical AI work is deeply affected, shaped by, and influenced by institutional priorities, organizational culture, and disparities in professional representation. Integrating this insight into ethical AI governance emphasizes the need for systemic reforms that reduce reliance on individual advocacy and foster sustainable conditions for fairness, accountability, and ethical integrity throughout AI development processes.

**Theme Two: Evolving Conceptions of Human-Machine Collaboration**

This supplementary finding demonstrates that practitioners are rethinking the relationship between human expertise and algorithmic decision-making, increasingly favoring models of collaboration rather than automation. Interview participants described a growing recognition that AI systems should function as partners that expand human judgment, rather than as autonomous authorities that override it. Several practitioners noted that excessive adherence to algorithmic predictions can undermine human critical thinking. This shift reflects insights relevant to RQ1, as participants connected the origins of bias to overreliance on model outputs and the assumption that algorithmic reasoning is inherently superior. By emphasizing the need for human-machine balance, participants highlighted how bias can emerge from the interaction between human cognition and algorithmic authority rather than from data alone.

The practical significance of this finding extends into RQ2, which examines how practitioners understand the societal consequences of biased or misaligned AI systems. Participants emphasized that automation-centered deployment models can amplify inequities when human oversight is reduced or when contextual knowledge is displaced by algorithmic inference. Practitioners emphasized that maintaining human agency, interpretive judgment, and moral responsibility is crucial for preventing disproportionate harm, particularly in high-stakes sectors where algorithmic outputs carry significant institutional weight. When systems are framed as collaborative rather than replacement technologies, organizations are more likely to invest in interpretability, calibration, and oversight mechanisms that help prevent undue reliance on algorithmic recommendations (NIST, 2023). This collaborative framing thereby shapes practitioners' understanding of both the risks and the ethical obligations associated with the deployment of AI.

The perspectives shared by participants also provide insight into RQ3, which focuses on identifying strategies for building more equitable and responsible AI systems. Their emphasis on human–machine collaboration aligns with a growing body of research advocating for human-centered AI design, calibrated trust, and reciprocal interaction between human and machine actors. Scholars argue that collaborative systems enhance decision quality by ensuring that humans remain active evaluators rather than passive recipients of algorithmic output, even after deployment through advisory groups for independent algorithmic review (Guttag & Suresh, 2021). This approach mirrors participant experiences in which over-automation contributed to weakened situational awareness, blind trust, and diminished critical reasoning; patterns well documented in automation bias research (Mosier & Skitka, 2018). Participants have noted that communities historically marginalized by automated decision systems are disproportionately harmed when contextual judgment is removed. Collaboration preserves space for contextualized, culturally informed reasoning (Benjamin, 2019). These insights demonstrate that collaborative models of AI use function as concrete fairness strategies by requiring deliberation, interpretability, and continuous human engagement.

The practical implications of this theme are most evident in how practitioners envision the governance structures necessary to support collaboration. A collaborative paradigm requires interpretability tools, model-monitoring infrastructures, and feedback loops that allow humans to understand, question, and refine algorithmic reasoning. Participants described these mechanisms as essential conditions for sustaining moral agency in AI-mediated decisions, reinforcing the need for governance frameworks that support ongoing review rather than one-time validation (NIST, 2023). Such structures not only mitigate automation bias but also embed ethical reasoning into decision-making processes, ensuring that human expertise remains central and responsive to uncertainty, nuance, and diverse stakeholder needs.

This theme highlights a significant shift in professional thinking about how AI systems should operate across various sectors. Practitioners are increasingly rejecting the notion of AI as an autonomous decision-maker and instead conceptualizing these systems as epistemic partners, accountable to human values, ethical reasoning, and contextual understanding. In this way, the theme directly extends findings from RQ1 through RQ3, demonstrating that equitable AI requires continuous interaction between human and machine actors, supported by governance frameworks that reinforce human agency, interpretive judgment, and reflexive engagement. By reframing AI–human interaction as collaborative rather than hierarchical, this supplementary finding highlights a critical pathway for responsible AI deployment that integrates technical reliability with social and ethical accountability.

**Limitations of the Study**

Although this study generated rich and meaningful insights into how practitioners understand and respond to bias in artificial intelligence systems, several limitations emerged during data collection and analysis that warrant acknowledgment. These limitations do not undermine the value of the findings but provide important context for interpreting their scope and transferability.

A primary limitation relates to the composition and size of the participant sample. Although the original goal was to interview at least twelve participants, recruitment challenges resulted in a final sample of nine. While the study's thematic saturation was achieved, the reduced number of participants limited the range of practitioner roles and organizational levels represented. For example, only a small subset of participants held senior governance or compliance roles, which may have limited the study's ability to explore institutional decision-making regarding fairness and accountability entirely. Additionally, some sectors central to the multi-case design, such as healthcare and criminal justice, were less represented among participants, resulting in a heavier emphasis on practitioners working in general technology or cross-sector environments. This imbalance affects the comprehensiveness of the study's reflection of sector-specific constraints and may contribute to uneven depth across cases.

A second limitation concerns the nature of the multi-case design itself. The four sectoral cases, healthcare, education, employment, and criminal justice, were constructed using publicly available documentation rather than primary fieldwork. While this approach aligned with the study's qualitative design and allowed for consistent triangulation, it restricted the researcher's ability to capture real-time institutional processes, internal deliberations, or organizational constraints that influence AI deployment. As a result, some connections between practitioner narratives and sector-level examples relied on interpretive synthesis rather than direct observational evidence.

A third limitation stems from the variability in participants' professional backgrounds and familiarity with fairness concepts. Some participants delved deeply into the technical aspects of AI design, while others primarily engaged in advisory, educational, or policy-oriented roles. This diversity enriched the overall dataset but also resulted in uneven specificity: some participants spoke extensively about technical biases, model behavior, or data quality, whereas others provided broader reflections on organizational culture or ethical reasoning. This variation sometimes limited the comparability of responses across cases and influenced how consistently each research question was addressed.

This study examines exclusively a U.S.-based participant sample. All practitioners interviewed lived and worked within the United States, and U.S. regulatory structures, organizational cultures, and sociotechnical norms shaped their perspectives. As a result, the findings may not fully reflect how practitioners in other regions, particularly those operating under more mature or stringent AI governance regimes, such as the European Union's AI Act or national guidelines in countries with advanced fairness auditing practices, perceive and address algorithmic bias. Practitioners in Europe, Canada, and parts of Asia often work within more explicit statutory requirements for transparency, model documentation, and algorithmic accountability, which influence how they conceptualize fairness and articulate ethical responsibilities. Because the present study did not include these international viewpoints, its conclusions primarily illuminate the ethical, institutional, and cognitive dynamics present in U.S. AI development contexts. This geographic limitation suggests that cross-national comparative research would be valuable for examining how practitioners' experiences differ across regulatory environments and cultural landscapes.

The nature of self-reported data was a known limitation. Because the study relied on practitioners' reflections on their own work and ethical reasoning, findings are shaped by participants' willingness to disclose challenges or institutional pressures. Some participants may have minimized tensions within their organizations or framed their practices in idealized terms due to professional norms, confidentiality concerns, or reputational considerations. Despite member checking and reflexive interviewing practices, the possibility of social desirability bias remains.

Finally, limitations emerged from the time frame and modality of the interviews. All interviews were conducted virtually within one month, which restricted the ability to build prolonged engagement or observe longitudinal changes in participants' perspectives. Virtual interviews, while efficient, may also limit rapport-building compared to in-person settings, potentially influencing the depth or candor of participant responses.

These limitations highlight the contextual and methodological boundaries within which the study's findings should be interpreted. Acknowledging these constraints supports a transparent and rigorous understanding of the research, while also identifying opportunities for future studies to expand on this work, including larger or sector-specific samples, organizational fieldwork, and longitudinal exploration of practitioner experiences.

## Implications for Future Study

This study opens several expansive avenues for future research that reflect both the depth of the topic and its significance in the ongoing academic conversation around responsible AI. These directions are grounded in the theoretical frameworks of Intersectionality and Cognitive Science and are intended to extend the practical and scholarly impact of this work.

A significant area for future exploration involves cross-national comparisons of AI governance. This study's focus on the United States, where ethical norms are primarily driven by organizational discretion rather than federal mandate, revealed a pattern of ethical ambiguity. In contrast, regions such as the European Union, with the EU AI Act, and countries like Canada, with the Artificial Intelligence and Data Act, offer more prescriptive governance frameworks. Comparative case studies between U.S. and EU-based organizations could illuminate how regulatory environments shape organizational behavior, accountability standards, and practitioner agency. Such research could address whether stricter legal structures reduce ethical fatigue or encourage more assertive ethical decision-making.

In addition to international comparisons, methodological innovation is essential. Future studies could employ grounded theory to generate richer, emergent frameworks based on practitioners' narratives, particularly around how fairness and bias are conceptualized within real-world constraints. Longitudinal studies would provide valuable insights into how AI governance practices evolve, particularly within organizations undergoing structural, technological, or cultural transformations. These approaches would help identify whether ethical intentions translate into sustained practices or whether they dissipate under shifting priorities and institutional pressures.

There is also value in pursuing mixed-methods research that integrates interviews with system-level data, such as performance audits, fairness reports, or policy reviews. Such triangulation could bridge the gap between what practitioners perceive and what systems actually produce in terms of outcomes. Ethnographic fieldwork within design and deployment teams can capture the negotiation of ethical values in practice, decisions that often occur outside formal governance frameworks. Additionally, participatory action research involving co-creation with practitioners could foreground ethics as an embedded and iterative process rather than an external compliance requirement.

Thematic expansions of this work are equally important. One domain with significant promise is leadership development. Many practitioners described a lack of ethical leadership and guidance, particularly at the executive level. Future research could investigate how leadership training programs, grounded in values-based ethics, scenario analysis, or interdisciplinary deliberation, influence organizational culture. For example, how do simulations that involve critical ethical dilemmas influence the approach of senior decision-makers to risk, innovation, and accountability?

Another domain is interdisciplinary collaboration. Participants frequently highlighted tensions between technical experts, ethicists, and policy staff, often rooted in differing vocabularies, timelines, and success metrics. Research could investigate models of cross-functional collaboration, such as ethics boards, integrated product teams, or rotating fellowship programs that embed ethicists in development cycles. These initiatives could serve as test beds for creating shared ethical languages and reducing siloed thinking.

Ultimately, this study's findings suggest opportunities to inform sector-specific guidelines and policies. Fields like healthcare, education, and employment face unique regulatory constraints and moral responsibilities. Future work could build on this study to inform best practices for fairness audits, model interpretability, and appeal mechanisms in these settings.

Such work might include developing industry toolkits, guiding principles, or standard operating procedures that embed ethical reasoning into every phase of system design and deployment.

These research directions move beyond identifying problems and toward building a more comprehensive, sustainable, and action-oriented field of responsible AI. They emphasize that fairness, transparency, and accountability are not static checkboxes, but dynamic practices shaped by context, time, and relationships. Expanding this line of inquiry will require interdisciplinary collaboration, theoretical depth, and methodological innovation, each of which has the potential to transform the understanding and practice of ethics in AI.

## Summary

Chapter Five integrated the study's findings, methodological reflections, and practical implications to articulate how practitioners understand, experience, and respond to bias in artificial intelligence systems. It began by revisiting the study's purpose and theoretical foundation, emphasizing that the investigation was guided by an interest in how cognition, conscience, and context shape practitioner perspectives across the healthcare, education, employment, and criminal justice domains. The opening section also summarizes Chapters One through Four, reiterating that algorithmic bias is inseparable from the social structures, historical inequities, and cognitive tendencies that shape both AI technologies and the institutions that deploy them.

The chapter then synthesized findings for each of the three research questions, linking practitioner insights to existing literature and the study's theoretical frameworks. For Research Question One, the study demonstrated that practitioners view bias as an inherited, systemic phenomenon arising from historical inequities, organizational incentives, and human cognitive shortcuts. Research Question Two revealed that practitioners recognize the profound societal impacts of AI bias, including the unequal distribution of resources, the automation of harm at scale, the reinforcement of cultural stereotypes, and the erosion of public trust. Research Question Three highlighted the strategies practitioners believe are necessary for promoting equity in AI, including cognitive reflection, cultural nuance, interdisciplinary oversight, transparent auditing, education, and regulatory governance. While strategies differed across sectors, a common insight emerged: practitioners emphasized the importance of ethical frameworks that extend beyond technical metrics to encompass institutional culture, regulatory clarity, and shared responsibility. In employment, this meant involving HR leaders in fairness audits; in education, it involved examining the values embedded in admissions algorithms. These examples collectively stress that achieving AI fairness demands structural, organizational, and epistemic change, not just isolated technical fixes.

The chapter also addressed two supplementary findings that emerged outside the scope of the original research questions: emotional and moral fatigue among practitioners engaged in ethical AI work, and evolving conceptions of human–machine collaboration. These themes expanded the study's contributions by illustrating how practitioners experience the emotional labor of advocating for fairness and by revealing a shift toward collaborative models of AI that preserve human judgment and moral responsibility. These findings, though context-dependent, suggest broader applications. Emotional fatigue among ethical advocates was not limited to one industry, indicating a systemic issue that may be addressed through organizational redesign or policy mandates. Similarly, evolving notions of human–machine collaboration has implications for AI governance models globally, highlighting the need for shared oversight mechanisms that preserve practitioner agency. Findings reinforced the broader argument that responsible AI development is fundamentally relational, requiring continuous engagement between human and machine actors as well as supportive institutional infrastructures.

A candid discussion of study limitations followed, acknowledging constraints related to sample size, sector representation, the constructed nature of the multi-case design, variation in practitioner expertise, and the research's exclusive U.S. context. These limitations provided important context for interpreting transferability and highlighted areas where additional inquiry is needed, such as cross-national comparative studies and sector-focused fieldwork.

The chapter concluded with recommendations for future research, identifying opportunities for expanding practitioner-focused studies internationally, investigating emotional and ethical fatigue through longitudinal designs, examining human–machine collaboration in real-world settings, and integrating ethnographic and mixed-methods approaches to capture the sociotechnical complexity of AI development. These avenues reflect the study's contribution to ongoing debates about fairness, accountability, and transparency in AI, while also positioning future research to enhance global understanding of how responsible AI can be achieved in practice.

Overall, Chapter Five demonstrated that addressing AI bias requires a holistic approach that integrates technical, social, cognitive, and institutional perspectives. The chapter emphasized that equitable AI development is both a moral and organizational challenge, one that relies on the sustained involvement of practitioners, the inclusion of diverse perspectives, and the establishment of robust governance frameworks that can align technological innovation with societal well-being.

**Appendix A: IRB Approval**

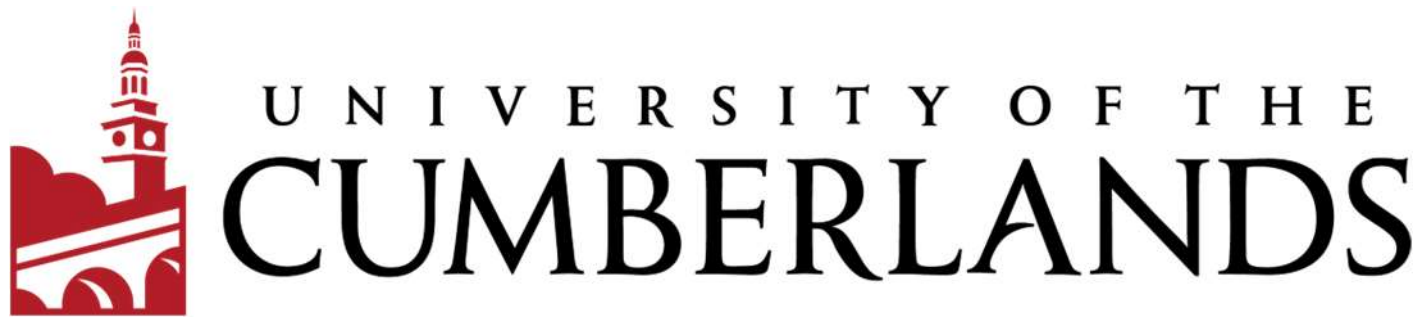

IRB Approval Letter

**Principal Investigator:** Micarah Malone

**From:** Institutional Review Board

**Subject:** IRB Approved

**Project title:** Beyond the Algorithm: Professional Experiences and Perceptions of AI Bias

**IRB Approval Number:** #1025-258204

**Approval Date:** 2025-10-07

Thank you for submitting your materials to the IRB office. The above referenced human-subjects research project has been approved by the University of the Cumberlands Institutional Review Board. This approval is limited to the approved protocols described in the application which have been reviewed as acceptable activities described by the Office of Human Research Protections (HHS.org).

It has been determined that your study meets federal criteria to qualify as an **expedited study** in accordance with the requirements set forth in 45 CFR 46.110 finding that 1) the research is minimal risk, 2) that if identification of the participants and/or their responses reasonably place them at risk of criminal or civil liability or could be damaging to the participants' financial standing, employability, insurability, or reputation, or be stigmatizing there are reasonable and appropriate protections that will be implemented so that risk related to invasion of privacy and breach of confidentiality are no greater than minimal, and 3) that the research is not classified or does not involve prisoners, with the exception that the expedited review of minor amendments for approved studies involving prisoners may be used.

However, if there are changes to research project in the following areas a modification form must be submitted to the IRB office:

- Substantial change to recruitment materials or consent documents
- Change in the data collection process
- Change in the location of the study
- Change in key personnel
- Change in instrumentation

Principal investigators are responsible for ensuring that studies are conducted according to University protocol. As a principal investigator, you have multiple responsibilities to the IRB, the research subjects and the faculty partner. If you have questions, please feel free to email me at IRB@ucumberlands.edu

Please continue to work with your dissertation advisor as you proceed.

**IRB Office**

University of the Cumberlands

6429 College Station Drive | Williamsburg, KY 40769

## Appendix B: Informed Consent

You are invited to participate in a research study entitled Beyond the Algorithm: Professional Experiences and Perceptions of AI Bias**.** You were selected as a potential participant because you currently work in a professional capacity related to the development, deployment, or oversight of artificial intelligence (AI) or machine learning (ML) systems (for example, as a developer, data scientist, engineer, project manager, or ethicist), and you meet the study's inclusion criteria of having at least three years of direct, practice-based experience with AI/ML technologies. Please read this form and ask any questions you may have before acting on this invitation to participate in the study.

This study is being conducted by Micarah J. Malone-Gawu, a doctoral candidate at the University of the Cumberlands (UC), and has been approved by the UC Institutional Review Board (IRB).

**Background Information:**

The purpose of this study is to examine how algorithmic bias emerges within artificial intelligence (AI) and machine learning (ML) systems, and to explore how professionals involved in their development, deployment, or oversight perceive the societal impacts of bias and the strategies for its mitigation. The study will involve approximately 12 to 16 participants and will continue until data saturation is reached.

**Inclusion Criteria:**

You can participate in this study if you:

- You are at least 18 years of age.
- You currently work in a professional capacity as a data scientist, engineer, product manager, researcher, designer, or technologist engaged with AI/ML systems.
- You have a minimum of 3 years of experience working in roles where algorithmic decision-making tools or AI systems are developed, implemented, or evaluated.
- You are based in the United States.
- You are able to speak and read English fluently.
- You are willing and able to participate in a recorded virtual interview lasting approximately 45–60 minutes.

**Procedures:**

If you agree to be in this study, you will be asked to participate in the following activities:

- Participate in a semi-structured interview conducted via Zoom or a similar video conferencing platform (45–60 minutes). The interview will include questions related to your professional experiences with AI/ML systems and your perceptions of bias and fairness in their development or application.
- Review and sign an informed consent form prior to the interview (5–10 minutes).
- Optionally review your interview transcript for accuracy and clarification, if you choose to (10–15 minutes, optional).

All sessions will be audio-recorded for transcription and analysis purposes, and you may withdraw at any time without penalty.

**Voluntary Nature of the Study:**

Your participation in this study is strictly voluntary. Your decision whether or not to participate will not affect your current or future relationship with the researcher, University of the Cumberlands, or any of the organizations or institutions affiliated with this study. If you initially decide to participate, you are free to withdraw at any time later without affecting those relationships.

**Risks and Benefits of Participation:**

There is *no more than minimal risk* associated with participating in this study, and there is no individual benefit to participation in the study. However, study findings may provide the overall benefit of improving the understanding of how AI bias is perceived and experienced by professionals working in data science, engineering, and related roles, which may inform more equitable and inclusive approaches to algorithm development, deployment, and oversight.

In the event you experience stress or anxiety during your participation in the study, you may terminate your participation at any time. You may refuse to answer any questions you consider invasive or stressful.

**Compensation:** There will be no compensation provided for your participation in this study.

**Recording:** Participation in this study involves a video-recorded interview conducted via Microsoft Teams. These recordings are essential for accurate transcription and analysis of the data. If you do not consent to being recorded, you will not be eligible to participate in the study.

**Confidentiality:**

Any data or records gathered from your participation will be kept private. Any identifiable data gathered will be coded to protect your identity. Names, organizations, and any personally identifying details shared during interviews will be anonymized during transcription and analysis. Transcripts will be reviewed and de-identified before thematic analysis is conducted. All research records, including digital files and notes, will be stored securely on a password-protected computer and backed up on an encrypted external drive. Only the principal investigator and designated faculty advisors will have access to this information. In any report of this study that may be published or presented, no information will be included that could directly or indirectly identify you. Research records will be deleted three years after the study is completed.

**Contacts and Questions:**

The researcher conducting this study is Micarah J. Malone-Gawu. The researcher's faculty advisor is Dr. Alex Lazo, alex.lazo@ucumberlands.edu. You may ask any questions you have related to the consent to participation. If you have questions later, you may contact them via mmalone98899@ucumeberlands.edu or the UC Institutional Review Board (IRB) office at irb@ucumberlands.edu.

**Consent (Use for electronic survey)**

I have read the above information, been given adequate time to consider the information, and understand my participation is voluntary so I may stop participation at any point. I have asked questions and received answers. I consent to take part in this study and understand I can download a copy of the completed form.

☐ Yes

☐ No

## Appendix C: Data Collection Instrument

**Research Title:** *Beyond the Algorithm: Professional Experiences and Perceptions of AI Bias*

**Researcher:** Micarah Malone-Gawu mmalone9899@ucumberlands.edu

**Institution:** University of the Cumberlands

**Degree Program:** Ph.D. in Artificial Intelligence

### Background of the Study

Artificial intelligence (AI) and machine learning (ML) systems are increasingly used to automate decision-making across critical sectors such as healthcare, education, criminal justice, and employment. While these technologies offer efficiencies and predictive capabilities, they also pose significant ethical concerns around bias, fairness, and social harm. Mounting evidence has shown that AI systems can perpetuate or even exacerbate existing societal inequalities, especially when trained on biased data or designed without consideration of marginalized communities.

This qualitative research study explores the root causes and real-world consequences of AI bias, as well as practitioner-informed strategies to mitigate harm and promote equity in AI system design and deployment. Framed through Intersectionality Theory and Cognitive Science, the study acknowledges that AI bias is not solely a technical failure, but a sociotechnical phenomenon shaped by social norms, institutional structures, and cognitive tendencies.

The study will involve semi-structured interviews with professionals, such as data scientists, engineers, designers, ethicists, and researchers, whose work intersects with AI development, governance, or policy. The goal is to elevate the voices of those with hands-on and theoretical experience in addressing AI bias and to generate actionable insights for more equitable and transparent AI ecosystems.

The feedback provided in this field test will ensure the interview questions are clear, unbiased, and effectively aligned with the study's core research questions.

**Research Questions for Study**

My study looks to answer the following questions through triangulation:

RQ1. How do biases emerge in AI/ML systems, and what are their root causes?

RQ2. What are the societal impacts of AI bias across healthcare, education, criminal justice, and employment?

RQ3. What strategies do professionals recommend for mitigating bias and advancing equity in AI system design and use?

**Field Test Purpose**

The purpose of this field test is to evaluate the clarity, alignment, and appropriateness of the proposed interview protocol and open-ended questions. Expert feedback will help ensure the instrument is free of ambiguity, bias, or misalignment with the study's research questions prior to full IRB approval and participant recruitment.

**Target Population for Main Study**

Professionals and subject matter experts working in AI/ML-related roles (e.g., data scientists, engineers, product managers, researchers).

**Interview Protocol Overview**

- Interviews will be semi-structured and approximately 45–60 minutes.
- Conducted via Zoom or similar platform.
- Participants will be asked open-ended questions about their experiences and observations related to AI bias, public perception, and mitigation strategies.

**Interview Questions for Field Test Review**

**Section 1: Understanding the Origins of Bias in AI/ML**

(Aligned with RQ1: Root causes of AI/ML bias)

1. From your experience, how do you see bias developing in AI systems? What does that look like to you?
    - Can you walk me through a specific example where you witnessed bias emerging?
    - What surprised you most about how that bias developed?
    - Looking back, what early warning signs did you notice or wish you had noticed?
2. How have you observed data collection, algorithm selection, or design decisions shaping biased outcomes in your work or field?
    - Can you describe a time when a data or design decision led to unexpected bias?
    - What assumptions did your team or field make that later proved problematic?
    - How do you personally approach these decisions differently now?
3. How do you perceive cultural assumptions and societal norms influencing AI bias? What does that look like in practice?
    - Can you describe a moment when you became aware of these cultural influences?
    - What cultural assumptions do you notice driving AI bias?
    - How do you experience the tension between technical goals and social awareness?
    - What societal beliefs do you see reflected in biased AI systems?

**Section 2: Perceived Societal Impacts**

(Aligned with RQ2: Societal consequences of AI bias)

4. When you consider the real-world effects of AI bias, what impacts concern you most? What does harm look like from your perspective?
    - What effect do you think people don't see or understand?
    - Which impacts hit you on a personal level, and why?
    - If you could make everyone understand one thing about the harmful effects of AI bias, what would it be?
5. Can you share a story or example of AI bias consequences that really stayed with you? What made that particular situation stand out?
    - How did it change the way you see AI's role in society?
    - How has that example shaped your work or perspective since then?

- What would you want others to take away from that story?

6. How do you see AI bias affecting different communities? What differences or patterns have you noticed across various groups?
    - Can you describe a moment when you realized that AI bias affects people differently based on their multiple identities such as but not limited to someone's race, gender, class, or other characteristics combined to create unique experiences? When did that become real rather than just a concept for you?
    - What patterns surprised you when you first noticed them?
    - Which communities do you think are most overlooked in AI bias discussions?

**Section 3: Perspectives on Addressing AI Bias**

(Aligned with RQ3: Strategies for equity and bias mitigation)

7. When you think about efforts to address AI bias, what have you seen that actually moves the needle? How do you recognize when something is working?
    - How do you distinguish between surface-level fixes and deep, meaningful change?
    - Can you describe a moment when you felt genuine hope about progress on AI bias?
8. When practitioners are trying to build fairer AI systems, what approaches or methods have you seen them turn to? What's your sense of how helpful these are?
    - What approach have you seen that made you think "now that's how it should be done"?
    - When you see these methods being helpful, what does that success look like to you?
9. From your perspective, what are the biggest obstacles still standing in the way of fairer AI? What do you think would help overcome them?
    - What obstacle frustrates you most when you see practitioners trying to build fairer AI?
    - What roadblock do you think gets the least attention but causes the most problems?
    - What keeps you going despite these ongoing challenges?
    - What fundamental shifts do you think need to happen in how we approach AI development?
    - What would you want future practitioners to know about navigating these obstacles?

**Closing Questions**

(Supportive and reflexive; aligned across all RQs)

10. What conversations or debates do you think the field needs to have more openly?
11. Is there anything else you would like to share about your perspective or experience with AI bias that we have not yet discussed?
12. Are there additional experts, communities, or stakeholders you believe I should speak with to gain a broader or more balanced perspective?

| Research Question (RQ) | Interview Question & Probes | Alignment Rationale | Literature Support |
|---|---|---|---|
| **RQ1. How do biases emerge in AI/ML systems, and what are their root causes?** | **Q1. From your experience, how do you see bias developing in AI systems? What does that look like to you?**<br>• Can you walk me through a specific example where you witnessed bias emerging?<br>• What surprised you most about how that bias developed?<br>• Looking back, what early warning | Captures practical insights into how professionals observe the emergence of AI bias. | Barocas & Selbst (2016); Noble (2018); Mehrabi et al. (2021) — Bias in ML often originates in real-world decisions and assumptions embedded in training data and system design. |

| Research Question (RQ) | Interview Question & Probes | Alignment Rationale | Literature Support |
|---|---|---|---|
| | signs did you notice or wish you had noticed? | | |
| | **Q2. How have you observed data collection, algorithm selection, or design decisions shaping biased outcomes in your work or field?** • Can you describe a time when a data or design decision led to unexpected bias? • What assumptions did your team or field make that later proved problematic? • How do you personally approach | Explores technical root causes such as biased datasets and modeling choices. | Buolamwini & Gebru (2018); Crawford (2021); O'Neil (2016) — Poor data hygiene and non-representative samples often lead to biased outcomes. |

| Research Question (RQ) | Interview Question & Probes | Alignment Rationale | Literature Support |
|---|---|---|---|
| | these decisions differently now? | | |
| | **Q3. How do you perceive cultural assumptions and societal norms influencing AI bias? What does that look like in practice?**<br>• Can you describe a moment when you became aware of these cultural influences?<br>• What cultural assumptions do you notice driving AI bias?<br>• How do you experience the tension between technical | Centers social and institutional norms as contributing forces in systemic AI bias. | Noble (2018); Benjamin (2019); Eubanks (2018) — AI often reflects the dominant cultural values of its creators and reproduces structural inequalities. |

| Research Question (RQ) | Interview Question & Probes | Alignment Rationale | Literature Support |
|---|---|---|---|
| | goals and social awareness?<br>• What societal beliefs do you see reflected in biased AI systems? | | |
| **RQ2. What are the societal impacts of AI bias across healthcare, education, criminal justice, and employment?** | **Q4. When you consider the real-world effects of AI bias, what impacts concern you most? What does harm look like from your perspective?**<br>• What effect do you think people don't see or understand?<br>• Which impacts hit you on a | Surfaces awareness of AI's human consequences, particularly for vulnerable groups. | Angwin et al. (2016); ProPublica; Dastin (2018); U.S. EEOC (2023) — Real-world cases show disparate impact in sentencing, hiring, healthcare diagnostics, and school surveillance. |

| Research Question (RQ) | Interview Question & Probes | Alignment Rationale | Literature Support |
|---|---|---|---|
| | personal level, and why?<br>• If you could make everyone understand one thing about the harmful effects of AI bias, what would it be? | | |
| | **Q5. Can you share a story or example of AI bias consequences that really stayed with you? What made that particular situation stand out?**<br>• How did it change the way you see AI's role in society? | Encourages narrative-rich data about the emotional and ethical weight of AI harm. | Raji et al. (2020); Veale & Binns (2017) — Human stories provide grounding in socio-technical systems theory and design justice. |

| Research Question (RQ) | Interview Question & Probes | Alignment Rationale | Literature Support |
|---|---|---|---|
| | • How has that example shaped your work or perspective since then?<br>• What would you want others to take away from that story? | | |
| | **Q6. How do you see AI bias affecting different communities? What differences or patterns have you noticed across various groups?**<br>• Can you describe a moment when you realized that AI bias affects people differently based on | Integrates intersectionality into analysis of AI's differential harms. | Crenshaw (1991); Costanza-Chock (2020); Eubanks (2018) — Intersectionality is critical to understanding how AI disproportionately impacts multiply-marginalized communities. |

| Research Question (RQ) | Interview Question & Probes | Alignment Rationale | Literature Support |
|---|---|---|---|
| | their multiple identities?<br>• What patterns surprised you when you first noticed them?<br>• Which communities do you think are most overlooked in AI bias discussions? | | |
| **RQ3. What strategies do professionals recommend for mitigating bias and advancing equity in AI system design and use?** | **Q7. When you think about efforts to address AI bias, what have you seen that actually moves the needle? How do you recognize when something is working?**<br>• How do you distinguish between | Identifies genuine success indicators in ethical AI practices. | Gebru et al. (2018); Holstein et al. (2019); Selbst et al. (2019) — Effective interventions require deep integration of ethics into product development. |

| Research Question (RQ) | Interview Question & Probes | Alignment Rationale | Literature Support |
|---|---|---|---|
| | surface-level fixes and deep, meaningful change?<br>• Can you describe a moment when you felt genuine hope about progress on AI bias? | | |
| | **Q8. When practitioners are trying to build fairer AI systems, what approaches or methods have you seen them turn to? What's your sense of how helpful these are?**<br>• What approach have you seen that made you | Surfaces existing tools, checklists, and practices used to promote fairness. | Mitchell et al. (2019); Binns (2018); Morley et al. (2021) — Fairness toolkits and auditing frameworks are evolving but unevenly adopted. |

| Research Question (RQ) | Interview Question & Probes | Alignment Rationale | Literature Support |
|---|---|---|---|
| | think "now that's how it should be done"?<br>• When you see these methods being helpful, what does that success look like to you? | | |
| | **Q9. From your perspective, what are the biggest obstacles still standing in the way of fairer AI? What do you think would help overcome them?**<br>• What obstacle frustrates you most when you see practitioners trying to build fairer AI? | Uncovers gaps between frameworks and implementation; invites transformational thinking. | Wong et al. (2022); Whittaker et al. (2018); Metz & Wakabayashi (2023) — Barriers include lack of regulatory teeth, data silos, and institutional resistance. |

| Research Question (RQ) | Interview Question & Probes | Alignment Rationale | Literature Support |
|---|---|---|---|
| | • What roadblock do you think gets the least attention but causes the most problems?<br>• What keeps you going despite these ongoing challenges?<br>• What fundamental shifts do you think need to happen in how we approach AI development?<br>• What would you want future practitioners to know about navigating these obstacles? | | |

| **Research Question (RQ)** | **Interview Question & Probes** | **Alignment Rationale** | **Literature Support** |
|---|---|---|---|
| **Supportive across all RQs** | **Q10. What conversations or debates do you think the field needs to have more openly?** | Opens space for reflection on under-addressed themes across RQs. | Keyes et al. (2019); Binns et al. (2018) — Dialogues around AI often lack nuance on power, language, and lived experience. |
| | **Q11. Is there anything else you would like to share about your perspective or experience with AI bias that we have not yet discussed?** | Allows emergent themes and personal framing to enrich analysis. | Supports qualitative best practice for capturing participant-defined insights (Creswell & Poth, 2018). |
| | **Q12. Are there additional experts, communities, or stakeholders you believe I should speak** | Supports triangulation and methodological reflexivity. | Lincoln & Guba (1985); Miles, Huberman, & Saldaña (2014) — Participant-guided referrals enhance |

| Research Question (RQ) | Interview Question & Probes | Alignment Rationale | Literature Support |
|---|---|---|---|
| | **with to gain a broader or more balanced perspective?** | | credibility and data richness. |

## Appendix D: Interview Protocol

**Title of Study:**

*Beyond the Algorithm: Professional Experiences and Perceptions of AI Bias*

**Interview Purpose:**

To capture practitioners' lived experiences and personal perspectives on how AI bias develops, impacts society, and can be addressed through their professional observations and insights.

**Estimated Duration:**

45–60 minutes

**Introduction (Read Aloud):**

Thank you for agreeing to participate in this interview. The purpose of this study is to explore how bias in artificial intelligence and machine learning systems is experienced and understood by professionals working with or alongside these systems. Your insights will contribute to a broader understanding of where AI bias originates, how it impacts society, and how we can move toward more ethical AI practices. Your responses will be kept confidential. You may skip any question or stop the interview at any time. Are you ready to begin? Do you have any questions before we start?

**Interview Questions:**

**Section 1: Understanding the Origins of Bias in AI/ML**

(Aligned with RQ1: Root causes of AI/ML bias)

1. From your experience, how do you see bias developing in AI systems? What does that look like to you?
    - Can you walk me through a specific example where you witnessed bias emerging?
    - What surprised you most about how that bias developed?
    - Looking back, what early warning signs did you notice or wish you had noticed?

2. How have you observed data collection, algorithm selection, or design decisions shaping biased outcomes in your work or field?
    - Can you describe a time when a data or design decision led to unexpected bias?
    - What assumptions did your team or field make that later proved problematic?
    - How do you personally approach these decisions differently now?
3. How do you perceive cultural assumptions and societal norms influencing AI bias? What does that look like in practice?
    - Can you describe a moment when you became aware of these cultural influences?
    - What cultural assumptions do you notice driving AI bias?
    - How do you experience the tension between technical goals and social / moral impacts?

    What societal beliefs do you see reflected in biased AI systems?

**Section 2: Perceived Societal Impacts**

(Aligned with RQ2: Societal consequences of AI bias)

4. When you consider the real-world effects of AI bias, what impacts concern you most? What does harm look like from your perspective?
    - What effect do you think people don't see or understand?
    - Which impacts hit you on a personal level, and why?
    - If you could make everyone understand one thing about the harmful effects of AI bias, what would it be?
5. Can you share a story or example of AI bias consequences that really stayed with you? What made that particular situation stand out?
    - How did it change the way you see AI's role in society?
    - How has that example shaped your work or perspective since then?
    - What would you want others to take away from that story?
6. How do you see AI bias affecting different communities? What differences or patterns have you noticed across various groups?
    - Can you describe a moment when you realized that AI bias affects people differently based on their multiple identities such as but not limited to someone's race, gender, class, or other characteristics combined to create unique experiences? When did that become real rather than just a concept for you?
    - What patterns surprised you when you first noticed them?
    - Which communities do you think are most overlooked in AI bias discussions?

**Section 3: Perspectives on Addressing AI Bias**

(Aligned with RQ3: Strategies for equity and bias mitigation)

7. When you think about efforts to address AI bias, what have you seen that actually moves the needle? How do you recognize when something is working?
    - How do you distinguish between surface-level fixes and deep, meaningful change?
    - Can you describe a moment when you felt genuine hope about progress on AI bias?
8. When practitioners are trying to build fairer AI systems, what approaches or methods have you seen them turn to? What's your sense of how helpful these are?
    - What approach have you seen that made you think "now that's how it should be done"?
    - When you see these methods being helpful, what does that success look like to you?
9. From your perspective, what are the biggest obstacles still standing in the way of fairer AI? What do you think would help overcome them?
    - What obstacle frustrates you most when you see practitioners trying to build fairer AI models?
    - What roadblock do you think gets the least attention but causes the most problems?
    - What keeps you going despite these ongoing challenges?
    - What fundamental shifts do you think need to happen in how we approach AI development?
    - What would you want future practitioners to know about navigating these obstacles?

**Closing Questions**

(Supportive and reflexive; aligned across all RQs)

10. What conversations or debates do you think the field needs to have more openly?
11. Is there anything else you would like to share about your perspective or experience with AI bias that we have not yet discussed?
12. Are there additional experts, communities, or stakeholders you believe I should speak with to gain a broader or more balanced perspective?

## Appendix E: Document Analysis Guide

**Purpose:**

To systematically review publicly available qualitative documents that offer expert insight, critique, or analysis of bias in AI/ML systems. This guide ensures alignment with the research questions and thematic consistency across all sources.

**Source Types Analyzed:**

- Government reports
- Academic articles
- Policy briefs
- Expert commentaries
- Public testimonies
- Organizational white papers

**Guiding Questions for Document Review:**

1. Does the document describe an instance or type of bias in AI/ML systems? If so, what kind (e.g., data bias, algorithmic bias, design bias)?
2. Does the document explore the causes of the identified bias?
3. What societal impacts are discussed or implied (e.g., discrimination, loss of access, misinformation)?
4. Which stakeholders or communities are affected, and how are their experiences described?
5. Are any solutions, mitigation strategies, or ethical guidelines proposed?
6. Does the document reflect a specific institutional or sectoral perspective (e.g., public policy, tech industry, academia)?
7. What assumptions, theoretical framings, or values are evident in the discussion?

**Coding Strategy:**

Thematic coding will be applied using Atlas.ti, guided by both a priori codes derived from the literature and theoretical frameworks, and emergent codes that capture participants' unique perspectives and lived experiences. Codes will include categories such as:

1. **Perception-Based Categories:**
    - Practitioner interpretations of bias origins
    - Personal observations of bias development
    - Lived experiences with AI bias consequences
    - Emotional responses to bias impacts
    - Professional meaning-making around bias mitigation
    - Perspective shifts and awakening moments
2. **Experience-Grounded Categories:**
    - Witnessed examples of effective interventions
    - Observed barriers and persistent challenges
    - Stories of bias discovery and recognition
    - Professional navigation strategies
    - Community and field observations
    - Hope and motivation sources
3. **Intersectionality-Informed Categories:**
    - Multi-dimensional impact narratives
    - Identity-based bias experiences
    - Differential community effects
    - Marginalized voice recognition
    - Intersectional awareness development
4. **Solution-Oriented Categories:**
    - Practitioner-recommended approaches
    - Real-world implementation insights
    - Vision for equitable AI futures
    - Systemic change perspectives
    - Professional advocacy experiences

This approach prioritizes participants' subjective experiences, interpretations, and meaning-making processes while maintaining alignment with the study's core research questions and theoretical frameworks.

**Theoretical Lens:**

Analysis will be interpreted through the dual frameworks of intersectionality (Crenshaw, 1991) and cognitive science (Kahneman, 2011).

## Appendix F: Case Study Analysis

### Case Study Analysis: Algorithmic Bias in Healthcare

### 1. Case Overview / Background

Artificial intelligence systems in healthcare are often hailed as transformative innovations that can enhance diagnostic accuracy, optimize resource allocation, and broaden access to care. However, these advantages are not equitably distributed across all populations. This case study highlights the racial bias identified in a widely used healthcare risk prediction algorithm, as analyzed by Obermeyer et al. (2019). The algorithm aimed to allocate additional care resources to patients predicted to have the highest healthcare needs. Unfortunately, it employed healthcare spending as a proxy for health status, which systematically disadvantaged Black patients, who historically incur lower healthcare costs due to unequal access to and treatment within the healthcare system.

This situation illustrates a broader issue within medical AI: the translation of historical inequalities into algorithmic logic. By equating cost with need, the model inadvertently internalized and automated existing structural disparities. Consequently, the system underestimated the medical needs of millions of Black patients, perpetuating inequities under the pretense of objectivity.

### 2. Research Context and Data Source

The primary data for this case is based on peer-reviewed empirical studies, including those by Obermeyer et al. (2019) and Adamson and Smith (2018), as well as subsequent policy analyses from healthcare ethics scholars such as Bridges (2024) and Lau (2025). These works collectively demonstrate how biased data inputs, design assumptions, and organizational priorities can lead to inequitable outcomes, even in systems that are marketed as neutral or data-driven.

In this analysis, the case is examined through the dual theoretical lenses of Intersectionality Theory and Cognitive Science, which aligns with the broader framework of the dissertation.

### 3. Case Description

The algorithm central to this case was deployed within major U.S. healthcare systems to identify high-risk patients who may benefit from intensive care management. It predicted risk by analyzing historical healthcare spending. At first glance, this approach seemed efficient as patients with higher past costs were presumed to have greater future needs.

However, research by Obermeyer et al. (2019) uncovered a concerning bias: when Black and White patients had comparable health needs, the algorithm consistently assigned lower risk scores to Black patients. This discrepancy stemmed from the fact that healthcare spending does not accurately reflect health status across different groups. Black patients, often confronting barriers such as underinsurance, medical mistrust, and provider bias, tend to spend less on healthcare, not because they are healthier, but because the system inadequately serves them.

This case illustrates how a seemingly objective model can perpetuate and justify systemic inequity when social factors are treated as neutral data points rather than as indicators of power and exclusion.

**4. Key Issues Identified**

- Structural Bias in Data: Historical inequities in healthcare access influence the data utilized to train AI models, thereby embedding racial and socioeconomic disparities within these systems.
- Cognitive Anchoring: Developers often relied on spending as a convenient and quantifiable metric, failing to critically examine its broader social implications.
- Ethical Opacity: Decision-makers lacked the necessary frameworks to assess how proxies like cost might perpetuate injustices.
- Accountability Gap: Despite passing multiple internal validation tests, the algorithm's technical performance metrics (such as accuracy) obscured potential ethical harm.

**5. Theoretical Framework Application**

Intersectionality Theory highlights how race, class, and access intersect to create layered disadvantages. The design of the algorithm overlooked these intersections, simplifying complex patterns of inequity into a single economic measure. This omission of social context turned the lived experiences of exclusion into mere data points, effectively rendering Black patients statistically invisible within the very systems intended to promote equitable care.

Cognitive Science further explains how developers' mental shortcuts, such as anchoring (focusing primarily on cost as a measure of need) and confirmation bias (favoring metrics that support narratives), reinforce inequitable outcomes. These cognitive mechanisms demonstrate how well-intentioned designers can perpetuate bias when social assumptions remain unexamined. Together, these frameworks reveal how systemic and cognitive biases work in concert, embedding structural injustice into algorithmic infrastructure.

**6. Analysis**

This case illustrates the intersection of human cognition and institutionalized inequality in artificial intelligence (AI). Developers made a cognitively rational decision to use spending data because it was measurable and easily accessible, but this choice was socially irrational as it overlooked disparities in healthcare access based on race. The belief that "cost equals care" reflected a broader institutional mindset that values efficiency and quantifiability over equity and context.

The oversight stemmed not from overt prejudice but from ethical abstraction—a failure to recognize how datasets mirror societal hierarchies. As Lau (2025) argues, the real danger lies not in malicious intent but in "the automation of indifference," where bias becomes invisible precisely because it is embedded in algorithms.

**7. Findings and Insights**

- Bias is inherent in the proxy itself: Using healthcare spending as a proxy for need automatically incorporates systemic racism.
- Ethical awareness must come before design: Fairness cannot be added after a model is deployed; it needs to inform the way the problem is framed from the beginning.
- Interdisciplinary collaboration is crucial: Social scientists, ethicists, and public health experts should be involved in AI development teams to uncover hidden assumptions.

- Metrics of fairness must evolve: When equity is the goal, accuracy and efficiency alone are not enough; new standards must consider social impact and distributive justice.

**8. Implications for Practice and Policy**

This case highlights the urgent need for regulatory oversight and accountability mechanisms in AI-enabled healthcare. Developers and institutions should be required to conduct equity impact assessments before implementation—analogous to environmental impact studies—to evaluate potential harm to vulnerable groups.

Healthcare organizations must also invest in data diversification, ensuring that training datasets include sufficient representation across racial, gender, and socioeconomic lines. Ethical governance models, such as those outlined in the NIST AI Risk Management Framework (2023), should become industry standards rather than optional guidelines.

Finally, as KPMG's *Leadership in the Age of AI* (2024) report warns, leaders must move beyond aspirational ethics to operational accountability, integrating fairness metrics into performance evaluations and procurement processes.

**9. Reflexivity**

As a researcher and practitioner in AI governance, I recognize the tension between innovation and equity that this case exemplifies. My proximity to these challenges provided both insight and bias into the technical realities of model design, and bias toward assuming that fairness can be engineered. To counter this, I used analytic memoing and peer debriefing to reflect on my interpretations and ensure that participant meaning, rather than professional expectation, guided my analysis.

**10. Case Summary Table**

| Case | Sector | Key Issue | Frameworks Applied | Major Insight |
|---|---|---|---|---|
| Healthcare Risk Algorithm (Obermeyer et al., 2019) | Healthcare | Racial bias from cost-based proxy | Intersectionality + Cognitive Science | Structural inequality encoded through cognitive simplification; fairness requires contextual, interdisciplinary design |
| COMPAS Risk Assessment (Angwin et al., 2016) | Criminal Justice | Racial disparities in predicted risk scores for | Intersectionality + Cognitive Science | Shows how historical racial hierarchies and biased criminal justice data |

| **Case** | **Sector** | **Key Issue** | **Frameworks Applied** | **Major Insight** |
|---|---|---|---|---|
| | | comparable defendants | | reinforce risk assessments, illustrating how cognitive shortcuts, such as overgeneralization and misplaced trust in "objective" scores, enable biased outputs to be accepted as neutral evidence. |
| Predictive Analytics in Education (Baker & Hawn, 2021) | Education | Risk scoring that reinforces demographic stereotypes and limits student opportunity | Intersectionality + Cognitive Science | Highlights how algorithms translate social disadvantage into academic "risk," narrowing student trajectories; demonstrates how cognitive ease and stereotype-consistent reasoning lead educators to accept algorithmic labels without interrogating the social assumptions embedded within them. |
| Automated Hiring Systems (Ajunwa, 2019) | Employment | Screening tools using demographic and linguistic proxies that disadvantage | Intersectionality + Cognitive Science | Reveals how cultural norms and dominant labor-market |

| Case | Sector | Key Issue | Frameworks Applied | Major Insight |
|---|---|---|---|---|
| | | marginalized applicants | | expectations are encoded into automated filters; shows how cognitive biases, such as heuristics about professionalism or competence, interact with historical inequities to reproduce exclusion at scale. |